\documentclass[traditabstract,fleqn]{aa}
\usepackage{graphicx}
\usepackage{amssymb}
\usepackage{amsmath}
\usepackage{natbib}
\bibpunct{(}{)}{;}{a}{}{,} 
\usepackage{subfigure}
\usepackage{txfonts}

\begin{document}

\title{Binary black holes in nuclei of extragalactic radio sources}

\author{J. Roland\inst{1}, S. Britzen\inst{2}, A. Caproni\inst{3}, C. Fromm\inst{2}, C. Gl\"{u}ck\inst{4} and A. Zensus\inst{2}}

\authorrunning{Roland \& al.}
\titlerunning{Modeling compact radio sources}

\institute{Institut d'Astrophysique, UPMC Univ Paris 06,
           CNRS, UMR 7095, 98\,bis Bd Arago ,
           75014 Paris, France
           \and
           Max-Planck-Institut f\"ur Radioastronomie,
           Auf dem H\"ugel 69, Bonn 53121, Germany
           \and
           N\'ucleo de Astrof\'isica Te\'orica, Universidade Cruzeiro do Sul, R. Galv\~ao Bueno 868, Liberdade, 01506-000 S\~ao Paulo, SP, Brazil
           \and
           I. Physikalisches Institut, Universit\"at zu K\"oln, Z\"ulpicher Str. 77, 50937 K\"oln, Germany}

\offprints{J. Roland, \email{roland@iap.fr}}

\date{Received 05/03/2012 / Accepted 05/06/2013}

\abstract{If we assume that nuclei of extragalactic radio sources contain  
binary black hole systems, the two black holes can eject VLBI components 
 in which case two families of different VLBI trajectories will be observed. 
Another important consequence of a binary black hole system is that 
 the VLBI core is associated with one black hole, and if a VLBI component 
is ejected by the second black hole, one expects to be able to detect the offset 
of the origin of the VLBI component ejected by the black hole that is not associated 
with the VLBI core. The ejection of VLBI components is perturbed by the precession 
of the accretion disk and the motion of the black holes around the center of gravity 
of the binary black hole system. We modeled the ejection of the component taking 
into account the two pertubations and present a method to fit the coordinates 
of a VLBI component and to deduce the characteristics of the binary black hole system. 
Specifically, this is the ratio $T_{p}/T_{b}$ where $T_{p}$ is the precession period of the 
accretion disk and $T_{b}$ is the orbital period of the binary black hole system, 
the mass ratio $M_{1}/M_{2}$, and the radius of the Binary Black Hole system $R_{bin}$. From  
 the variations of the coordinates as a function of time of the ejected 
VLBI component, we estimated the inclination angle $i_{o}$ and the bulk Lorentz 
factor $\gamma$ of the modeled component. We applied the method to component 
S1 of 1823+568 and to component C5 of 3C 279, which presents a large offset of 
the space origin from the VLBI core. We found that 1823+568 contains a binary black hole 
system whose size is $R_{bin} \approx 60$ $\mu as$ ($\mu as$ is a microarcsecond) 
and 3C 279 contains a binary black hole system whose size is $R_{bin} \approx 420$ $\mu as$. 
We calculated the separation of the two black holes and the coordinates 
of the second black hole from the VLBI core. This information will be important 
to link the radio reference-frame system obtained from VLBI observations 
and the optical reference-frame system obtained from GAIA.

}

\maketitle

\section{Introduction}
\label{sec:intro}

\begin{figure}[ht]
\centerline{
\includegraphics[scale=0.5, bb =100 380 540 520,clip=true]{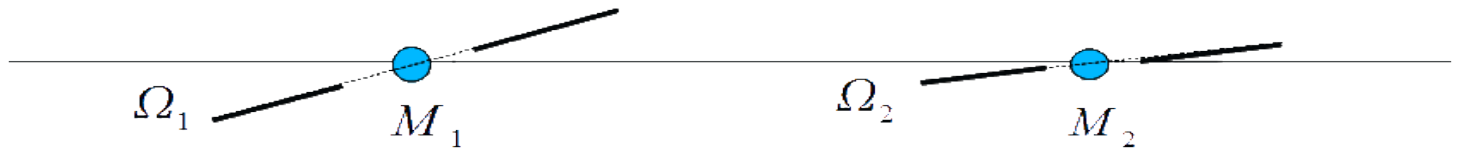}}
\caption{BBH system model. The two black holes can have an accretion disk 
and can eject VLBI components. If it is the case, we observe two different 
families of trajectories and an offset between the VLBI core and the origin 
of the VLBI component if it is ejected by the black hole that is not associated with 
the VLBI core. The angles $\Omega_{1}$ and $\Omega_{2}$ between the accretion 
disks and the rotation plane of the BBH system can be different.}
\label{fig:BBH_MODEL}
\end{figure}

VLBI observations of compact radio sources show that the ejection 
of VLBI components does not follow a straight line, but undulates. 
These observations suggest a precession of the accretion disk. 
To explain the precession of the accretion disk, we assumed that 
the nuclei of radio sources contain binary black hole systems 
(BBH system, see Figure \ref{fig:BBH_MODEL}) .

A BBH system produces three pertubations of the VLBI ejection due to
\begin{enumerate}
    \item the precession of the accretion disk,
    \item the motion of the two black holes around the center of gravity 
    of the BBH system, and
    \item the motion of the BBH system in the galaxy.
\end{enumerate}

In this article, we do not take into account the possible third 
pertubation due to the motion of the BBH system in the galaxy.

\begin{figure}[ht]
\centerline{
\includegraphics[scale=0.5, width=8cm,height=6cm]{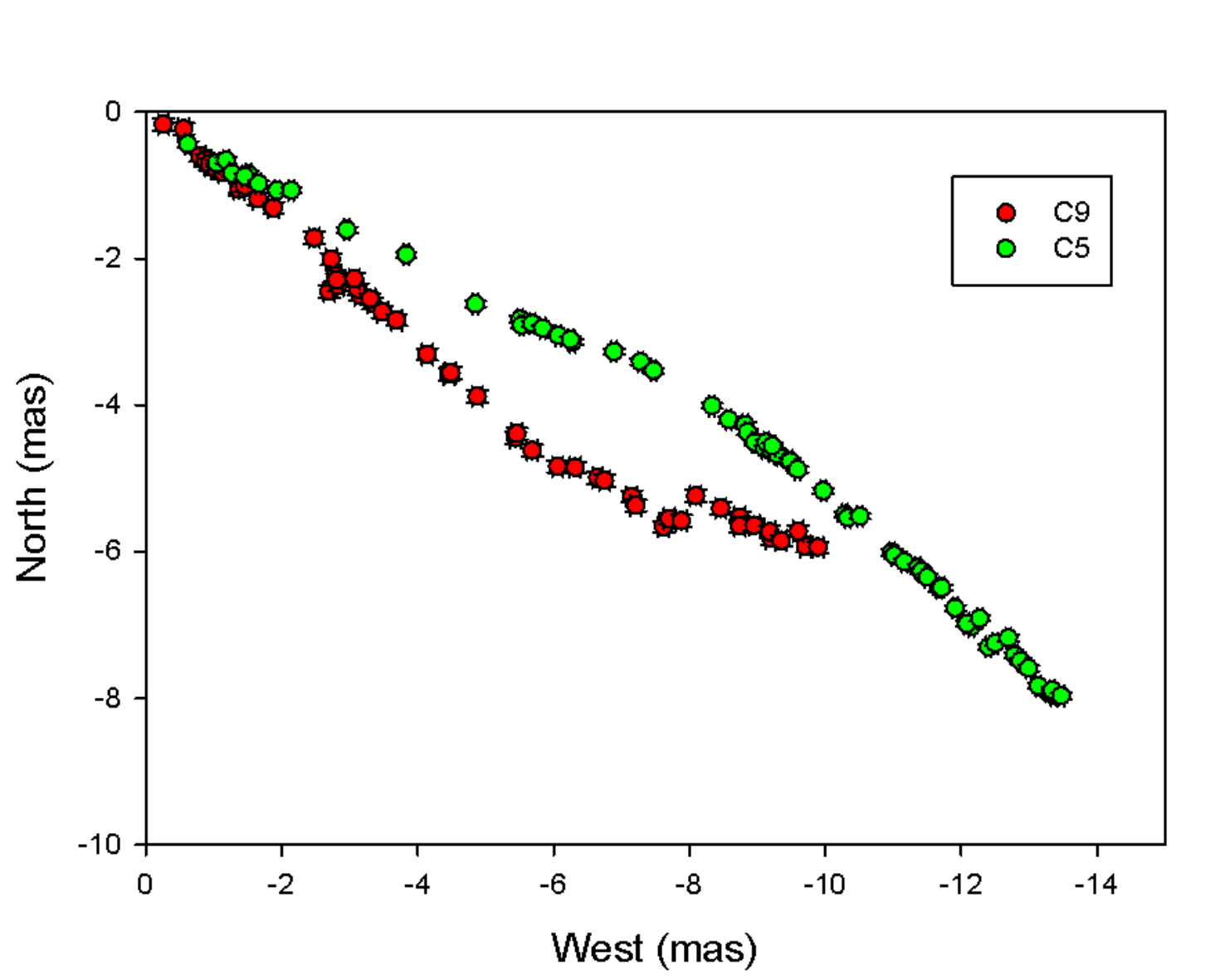}}
\caption{Trajectories of the VLBI components C5 and C9 of 3C 273 using MOJAVE data 
\citep{LiCo+:09}. 
We observe two different types of trajectories, suggesting that they are ejected from 
two different black holes.}
\label{fig:C5+C9_Traj_Mojave_data_New}
\end{figure}

A BBH system induces several consequences, which are that
\begin{enumerate}
    \item even if the angle between the accretion disk and the plane 
    of rotation of the BBH system is zero, the ejection does not follow 
    a straight line (due to the rotation of the black holes around the 
    center of gravity of the BBH system),
    \item the two black holes can have accretion disks with different 
    angles with the plane of rotation of the BBH system and can eject 
    VLBI components; in that case we observe two different families 
    of trajectories; a good example of a source with two families of trajectories 
    is 3C 273 whose components C5 and C9 follow two different types of trajectories 
    (see Figure \ref{fig:C5+C9_Traj_Mojave_data_New}), and
    \item if the VLBI core is associated with one black hole, and if  
     the VLBI component is ejected by the second black hole, 
    there will be an offset between the VLBI core and the origin of the 
    ejection of the VLBI component; this offset will correspond to the radius 
    of the BBH system.
\end{enumerate}

The precession of the accretion disk can be explained 
using a single rotating black hole (Lense-Thirring effect) or by the magnetically 
driven precession \citep{CaLi+:06}. However, a 
single black hole and a BBH system have completely different consequences. 
In the case of a BBH system, one has an extra perturbation of the ejected component 
due to the motions of the black holes around the center of gravity of the BBH system. 
One can expect to observe two different families 
of trajectories (if the two black holes eject VLBI components) and an offset 
of the origin of the ejected component if it is ejected by the black hole 
that is not associated with the VLBI core.\\

We modeled the ejection of the VLBI component using a geometrical model that 
takes into account the two main perturbations due to the BBH system, i.e.
\begin{enumerate}
    \item the precession of the accretion disk and
    \item the motion of the two black holes around the center of gravity 
    of the BBH system.
\end{enumerate}

Modeling the ejection of VLBI components using a BBH system has been 
developed in previous articles, for instance \citet{BrRo+:01} modeled 
0420-014, \citet{LoRo:05} modeled 3C 345, 
and \citet{RoBr+:08} modeled 1803+784. Observationnal VLBI studies have been 
performed to directly detect BBH systems in active galactic nuclei \citep{BuSp:11, TiWa:11}.

In section \ref{sec:model} we recall the main lines of the model. 
The details of the model can be found in \citet{RoBr+:08}.

We determined the free parameters of the model by comparing the observed 
coordinates of the VLBI component with the calculated coordinates of the 
model.

This method requires knowing of the variations of the two 
coordinates of the VLBI component as a function of time. Because these 
observations contain the kinematical information, we will be able
to estimate the inclination angle of the source and the bulk Lorentz
factor of the ejected component.

In this article we present a method to 
solve this problem, either for a precession model or 
for a BBH system model, based on understanding the 
space of the solutions.

Practically, two different cases can occur when we try to solve 
this problem.
\begin{enumerate}
    \item Either the VLBI component is ejected from the VLBI 
    core, or the offset is smaller than or on the order of the smallest error bars of the 
		VLBI positions of the ejected component (case I),
    \item or  the VLBI component is ejected with an offset larger 
    than the smallest error bars of the 
		VLBI positions of the ejected component (case II).
\end{enumerate}

Case II is much more complicated to solve than case I, because the observed 
coordinates contain an unknown offset that is larger than the error bars. 
Therefore, we first have to find the offset, then correct the VLBI data from the 
offset, and finally find the solution corresponding to the corrected data.

We present the method for solving the problem in section \ref{sec:method}.
To illustrate case I, we solve the fit of component S1 of 1823+568 
using MOJAVE data in section \ref{sec:case_I}.
To illustrate case II, we solve the fit of component C5 of 3C 279 
using MOJAVE data in section \ref{sec:case_II}.

\section{Model}
\label{sec:model}

\subsection{Introduction: Two-fluid model}
\begin{figure}[ht]
\centerline{
\includegraphics[scale=0.3,  bb =0 -50 800 500,clip=true]{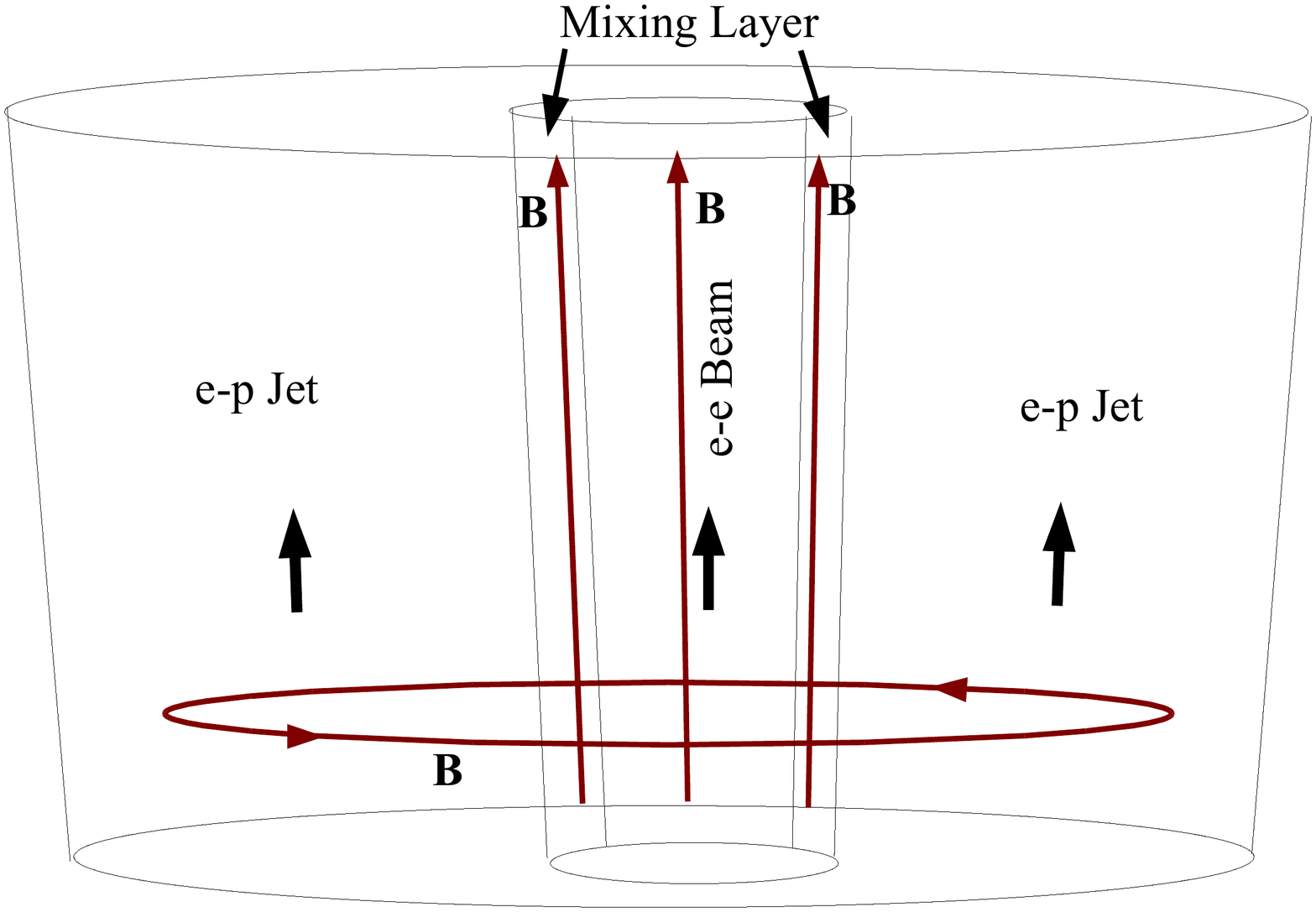}}
\caption{Two-fluid model. The outflow consists of an $e^{-}-e^{+}$ plasma, 
 \textit{the beam}, moving at a highly relativistic speed, surrounded 
by an $e^{-}-p$ plasma, and of \textit{the jet}, moving at a mildly relativistic speed. 
The magnetic field lines are parallel to the flow in the 
beam and the mixing layer, and are toroidal in the jet.}
\label{fig:TF_Model}
\end{figure}

We describe the ejection of a VLBI component in the framework of the
two-fluid model \citep{SoPe+:89,PeRo:89,PeRo:90,PeSo:92}. 
The two-fluid description of the outflow is
adopted with the following assumptions:
\begin{enumerate}
    \item The outflow consists of an $e^{-}-e^{+}$ plasma (hereafter
    \textit{the beam}) moving at a highly relativistic speed (with
    corresponding Lorentz factor\footnote{The bulk Lorentz factor 
		is limited to 30 to ensure the 	propagation	stability  
		of the relativistic beam in the subrelativistc jet.} 
		$\gamma_{b}\leq 30$) surrounded by an $e^{-}-p$ plasma 
		(hereafter \textit{the jet}) moving at a mildly relativistic 
		speed of $v_{j}\leq 0.4\times c$.
    \item The magnetic field lines are parallel to the flow in the
    beam and the mixing layer, and are toroidal in the jet 
    (see Figure \ref{fig:TF_Model}).
\end{enumerate}

\citet{MuPe+:88} and \citet{RoPe+:88} found that the Cygnus A hot spots 
could be explained by a an $e^{-}-p$ plasma 
moving at a mildly relativistic speed, i.e. $v_{j}\leq 0.4 \: \times \: c$. 
Consequently, the two-fluid model was introduced to explain superluminal radio 
sources observed in the nuclei of radio sources.

The $e^{-}-p$ jet carries most of the mass and the kinetic energy
ejected by the nucleus. It is responsible for the formation of
kpc-jets, hot spots, and extended lobes \citep{RoHe:96}. 
The relativistic $e^{\pm}$ beam moves in a channel through 
the mildly relativistic jet and is responsible for the formation 
of superluminal sources and their $\gamma$-ray emission 
\citep{RoTe+:94}. The relativistic beam can
propagate when the magnetic field $B$ is parallel to the flow in the
beam and in the mixing layer between the beam and the jet, and when it
is greater than a critical value \citep{PeSo+:88,AcSc:93}. 
The magnetic field in the jet becomes rapidly 
toroidal as a function of distance from the core \citep{PeRo:90}.

The observational evidence for the two-fluid model has been 
discussed by e.g. \citet{RoHe:96}. Observational evidence for 
relativistic ejection of an $e^{\pm}$ beam comes from the $\gamma$-ray 
observations of MeV sources \citep{RoHe:95,SkDe+:97}
and from VLBI polarization observations \citep{AtRo+:99}.

The formation of X-ray and $\gamma$-ray spectra, assuming relativistic
ejection of $e^{\pm}$ beams, has been investigated by \citet{MaHe+:95,MaHe+:98} 
for Centaurus A.

The possible existence of
VLBI components with two different apparent speeds has been pointed
out for the radio galaxies Centaurus A \citep{TiJa+:98}, 
Virgo A \citep{BiSp+:99} and 3C 120 \citep{GoMa+:01}.
If the relativistic beam transfers some energy and/or relativistic particles
to the jet, the relativistic particles in the jet will radiate and
a new VLBI component with a mildly relativistic speed will be observed
(3C 120 is a good example of a source showing this effect).

\subsection{Geometry of the model}

\begin{figure}[ht]
\centerline{
\includegraphics[scale=0.45, bb = -100 250 700 700,clip=true]{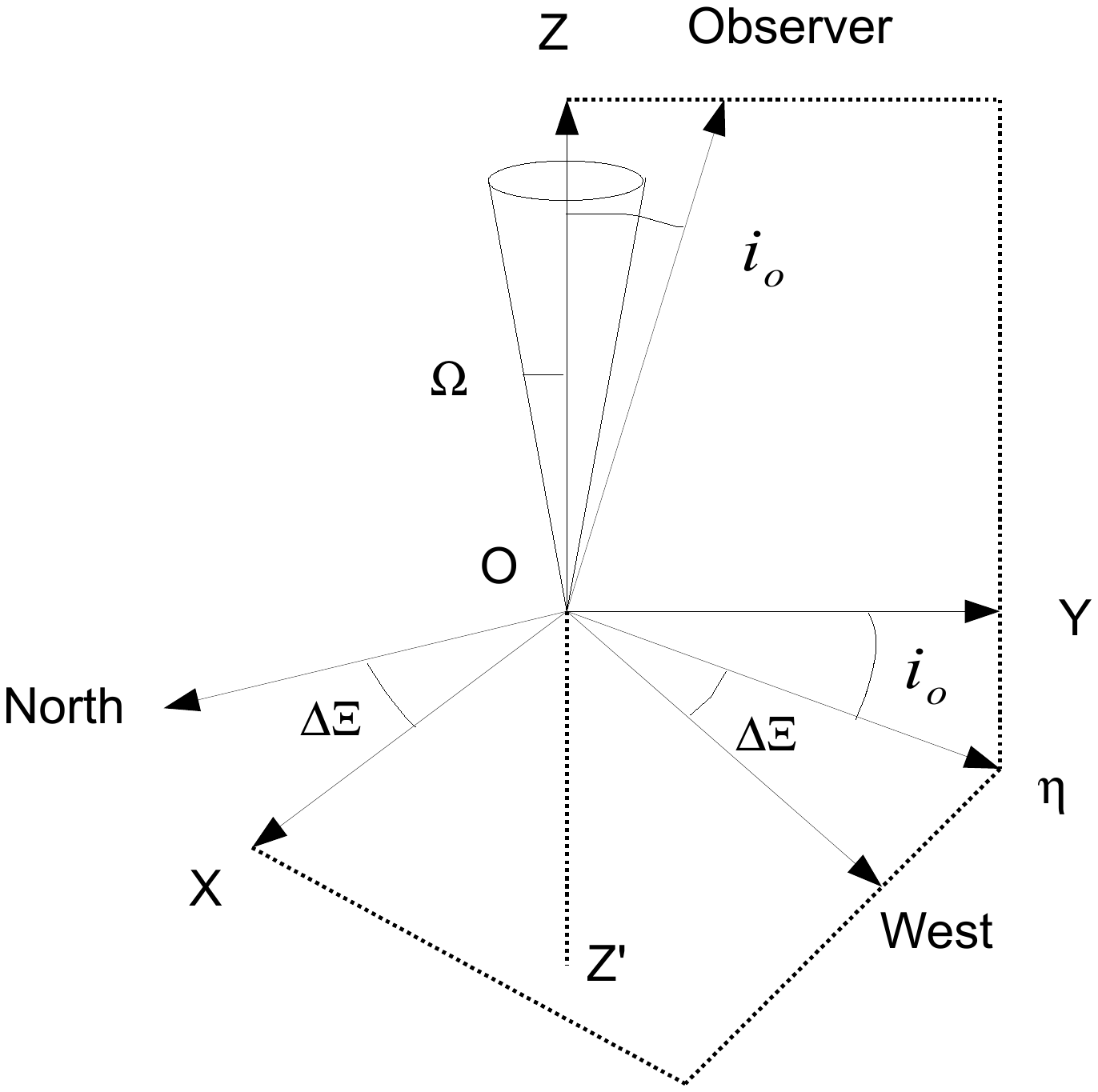}}
\caption{Geometry of the problem. The planes $X$ - $\eta$ and west - north
are perpendicular to the line of sight. In the west - north plane, the axis $\eta$ 
corresponds to the mean ejection direction of the VLBI component. $\Omega$ 
is the opening angle of the precession cone.}
\label{fig:Refence_System}
\end{figure}

We call $\Omega$ the angle between the accretion disk and the orbital 
plane ($XOY$) of the BBH system. The component is ejected 
 on a cone (the precession cone) with its axis in the $Z'OZ$ plane and of opening angle 
$\Omega$. We assumed that the line of sight is in
the plane ($YOZ$) and forms an angle $i_{o}$ with the axis $Z'OZ$
(see Figure \ref{fig:Refence_System}). The axis $\eta$ corresponds to the mean 
ejection direction of the VLBI component projected in a plane perpendicular 
to the line of sight, so the plane perpendicular to the line of
sight is the plane ($\eta OX$). We call $\Delta\Xi$ the rotation angle  
 in the plane perpendicular to the line of sight to transform
the coordinates $\eta$ and $X$ into coordinates $N$ (north) and $W$ (west), 
which are directly comparable with the VLBI observations. We have

\begin{equation}
W  =  - x \cos(\Delta\Xi) + (z \sin(i_{o}) + y \cos(i_{o})) \sin(\Delta\Xi)\ ,
\label{eq:West}
\end{equation}

\begin{equation}
N  =  x \sin(\Delta\Xi) + (z \sin(i_{o}) + y \cos(i_{o})) \cos(\Delta\Xi)\ .
\label{eq:North}
\end{equation}

The sign of the coordinate W was changed from \citet{RoBr+:08} 
 to use the same definition as VLBI observations.

\subsection{General perturbation of the VLBI ejection}

For VLBI observations, the origin of the coordinates is 
black hole~1, i.e. the black hole ejecting the VLBI components. 
For the sake of simplicity, we assumed that the two black holes have circular orbits, 
i.e. $e=0$. Therefore, the coordinates of the moving components 
\textit{in the frame of reference where black hole 1 is considered the origin} 
are \citep{RoBr+:08}

\begin{eqnarray}
x_{c} & = & [R_{o}(z) \cos(\omega_{p}t-k_{p}z(t)+\phi_{o}) \nonumber \\
        &&{} + x_{1} cos(\omega_{b}t-k_{b}z(t)+\psi_{o}) - x_{1} \cos(\psi_{o})] \nonumber \\
        &&{} exp(-t/T_{d})\ ,
\label{eq:xc}
\end{eqnarray}

\begin{eqnarray}
y_{c} & = & [R_{o}(z) \sin(\omega_{p}t-k_{p}z(t)+\phi_{o}) \nonumber \\
        &&{} + y_{1} \sin(\omega_{b}t-k_{b}z(t)+\psi_{o}) - y_{1} \sin(\psi_{o})] \nonumber \\
        &&{} exp(-t/T_{d})\ ,
\label{eq:yc}
\end{eqnarray}

\begin{equation}
z_{c} = z_{c}(t)\ ,
\label{eq:zc}
\end{equation}
where
\begin{itemize}
  \item $R_{o}(z)$ is the amplitude of the precession perturbation, given by 
     $R_{o}(z) = R_{o}z_{c}(t)/(a +z_{c}(t))$, with $a = R_{o}/(2\;tan\Omega)$,
  \item $\omega_{p}$ is $\omega_{p} = 2 \pi / T_{p}$, where $T_{p}$ is the precession period, and
     $k_{p}$ is defined by $k_{p} = 2 \pi/T_{p}V_{a}$, where $V_{a}$ is the speed of 
     the propagation of the perturbations,
  \item $\omega_{b}$ is $\omega_{b} = 2 \pi/T_{b}$, where $T_{b}$ is the BBH system period
     and $k_{b}$ is defined by $k_{b} = 2 \pi/T_{b} V_{a}$,
  \item $T_{d}$ is the characteristic time of the damping of the perturbation,
  \item $x_{1}$ and $y_{1}$ are given by
     \begin{equation}
        x_{1} = y_{1} = -\frac{ M_{2}}{M_{1}+M_{2}} \times \left[\frac{T_{b}^{2}}{4\pi^{2}}G(M_{1}+M_{2}) \right]^{1/3} \ .
     \end{equation}
\end{itemize}

We define with $R_{bin}$ the distance between the two black holes as 
the size of the BBH system. It is

\begin{eqnarray}
    R_{bin} = \left[\frac{T_{b}^{2}}{4\pi^{2}}G(M_{1}+M_{2}) \right]^{1/3} \ .
\label{eq:Rbin}
\end{eqnarray}

In $mas$ units (milli arc second units), it is 
\begin{eqnarray}
   R_{bin} \approx 2.06 \; 10^{8} \; \left[\frac{T_{b}^{2}}{4\pi^{2}}G(M_{1}+M_{2}) \right]^{1/3} / D_{a} \ ,
\label{eq:Rbin_mas}
\end{eqnarray}
where $D_{a} = D_{l}/(1 + z)^{2}$ is the angular distance, $D_{l}$ is the luminosity 
distance, and $z$ is the redshift of the source.

The differential equation governing the evolution of
$z_{c}(t)$ can be obtained by defining the speed 
of the component, namely

\begin{equation}
    v_{c}^{2} = \left(\frac{dx_{c}(t)}{dt}\right)^{2} + \left(\frac{dy_{c}(t)}{dt}\right)^{2} +
    \left(\frac{dz_{c}(t)}{dt}\right)^{2}\ ,
\label{eq:v2}
\end{equation}
where $v_{c}$ is related to the bulk Lorentz factor by 
$v_{c}/c = \sqrt{(1 - 1/\gamma_{c}^{2})}$.

Using (\ref{eq:xc}), (\ref{eq:yc}) and (\ref{eq:zc}), we find from
(\ref{eq:v2}) that $dz_c/dt$ is the solution of the equation

\begin{equation}
    A\left(\frac{dz_{c}}{dt}\right)^{2} + B\left(\frac{dz_{c}}{dt}\right) + C = 0\ .
\label{eq:dzt}
\end{equation}

The calculation of the coefficients $A$, $B$ and $C$ can be found 
in Appendix  A of \citet{RoBr+:08}.

Equation (\ref{eq:dzt}) admits two solutions corresponding to 
the jet and the counter-jet.

Following \citet{CaKr:92}, if we call 
$\theta$ the angle between the velocity of the component and 
the line of sight, we have

\begin{equation}
    cos(\theta(t))=\left(\frac{dy_{c}}{dt}sin\;i_{o}+\frac{dz_{c}}{dt}cos\;i_{o}\right)/v_{c}\ .
    \label{eq:cos}
\end{equation}

The Doppler beaming factor $\delta$, characterizing the 
anisotropic emission of the moving component, is
\begin{equation}
    \delta_{c}(t) = \frac{1}{\gamma_{c} \left[1 - \beta_{c} cos(\theta(t))\right]}\ ,
    \label{eq:delta}
\end{equation}
where $\beta_{c} = v_{c} /c$. The observed flux density is

\begin{equation}
    S_{c} = \frac{1}{D_{l}^{2}}\delta_{c}(t)^{2+\alpha_{r}}(1+z)^{1-\alpha_{r}}\int_{c}j_{c}dV\ ,
    \label{eq:flux}
\end{equation}
where $D_{l}$ is the luminosity distance of the source, $z$ its redshift, 
$j_{c}$ is the emissivity of the component, and $\alpha_{r}$ 
is the synchrotron spectral index (it is related to the flux density by 
$S\propto\nu^{-\alpha_{r}}$). As 
the component is moving relativistically toward the observer, 
the observed time is shortened and is given by

\begin{equation}
    t_{obs} = \int_{0}^{t}\left[1- \beta_{c} cos(\theta(t'))\right]\left(1+z\right)dt'\ .
    \label{eq:tobs}
\end{equation}

\subsection{Coordinates of the VLBI component}
Solving (\ref{eq:dzt}), we determine the coordinate $z_{c}(t)$ of a
point-source component ejected relativistically in the perturbed beam.
Then, using (\ref{eq:xc}) and (\ref{eq:yc}), we can find the coordinates
$x_{c}(t)$ and $y_{c}(t)$ of the component. In addition, for each point 
of the trajectory, we can calculate the derivatives  $dx_{c}/dt$, $dy_{c}/dt$,
$dz_{c}/dt$ and then deduce $\cos\theta$ from (\ref{eq:cos}), $\delta_{c}$
from (\ref{eq:delta}), $S_{\nu}$ from (\ref{eq:flux}) and $t_{obs}$
from (\ref{eq:tobs}).

After calculating the coordinates $x_{c}(t)$, $y_{c}(t)$ and $z_{c}(t)$, 
they can be transformed to $w_{c}(t)$ (west) and
$n_{c}(t)$ (north) coordinates using (\ref{eq:West}) and
(\ref{eq:North}).

As explained in \citet{BrRo+:01}, \citet{LoRo:05}, and 
\citet{RoBr+:08}, the radio VLBI component has to be described as 
an extended component along the beam.
We call $n_{rad}$ the number of points (or integration steps along the beam)
for which we integrate to model the component. The coordinates
$W_{c}(t)$, $N_{c}(t)$ of the VLBI component are then
\begin{equation}
    W_{c}(t) = \left(\sum_{i=1}^{n_{rad}} w_{ci}(t)\right)/n_{rad} \ 
    \label{eq:West_c}
\end{equation}
and
\begin{equation}
    N_{c}(t) = \left(\sum_{i=1}^{n_{rad}} n_{ci}(t)\right)/n_{rad} \ 
    \label{eq:North_c}
\end{equation}
and can be compared with the observed coordinates of the VLBI component, which 
correpond to the radio peak intensity coordinates provided by model-fitting 
during the VLBI data reduction process. 

When, in addition to the radio, optical observations are available 
that peak in the light curve, this optical emission can be modeled 
as the synchrotron emission of a point source
ejected in the perturbed beam, see \citet{BrRo+:01} and
\citet{LoRo:05}. This short burst of very energetic
relativistic $e^{\pm}$ is followed immediately by a very long burst
of less energetic relativistic $e^{\pm}$. This long burst is 
modeled as an extended structure along the beam and is responsible for
the VLBI radio emission. In that case the origin $t_{o}$ of the VLBI
component is the beginning of the first peak of the optical
light curve and is not a free parameter of the model.

\subsection{Parameters of the model}

In this section, we list the possible free
parameters of the model. They are

\begin{itemize}
  \item $i_{o}$ the inclination angle,
    \item $\phi_{o}$ the phase of the precession at $t=0$,
    \item $\Delta\Xi$ the rotation angle in the plane perpendicular
    to the line of sight (see (\ref{eq:West}) and (\ref{eq:North})),
    \item $\Omega$ the opening angle of the precession cone, 
    \item $R_{o}$ the maximum amplitude of the perturbation, 
    \item $T_{p}$ the precession period of the accretion disk,
    \item $T_{d}$ the characteristic time for the damping of the beam perturbation,
    \item $M_{1}$ the mass of the black hole ejecting the radio jet,
    \item $M_{2}$ the mass of the secondary black hole,
    \item $\gamma_{c}$ the bulk Lorentz factor of the VLBI component,
    \item $\psi_{o}$ the phase of the BBH system at $t=0$,
    \item $T_{b}$ the period of the BBH system,
    \item $t_{o}$ the time of the origin of the ejection of the VLBI component,
    \item $V_{a}$ the propagation speed of the perturbations,
    \item $n_{rad}$ is the number of steps to describe the extension of the VLBI 
    component along the beam,
    \item $\Delta W$ and $\Delta N$ the possible offsets of the origin of the VLBI
    component.
\end{itemize}

We will see that the parameter $V_a$ can be used to study the degeneracy of 
the solutions, so we can keep it constant to find the solution. The range of 
values that we study for parameter $V_a$ is $0.01 \times c \leq V_a \leq 0.45 \times c$ 
\footnote{We limit ourselves to nonrelativistic hydrodynamics in this model.}.

The parameter $n_{rad}$ is known when the size of the VLBI component is known.

This means that, pratically, the problem we have to solve is a 15 free parameter problem.

We have to investigate the different possible scenarios with regard to the 
sense of the rotation of the accretion disk and the sense of the orbital 
rotation of the BBH system. These possibilities correspond to $\pm\:
\omega_{p}(t- z/V_{a})$ and $\pm \:\omega_{b}(t- z/V_{a})$. Because the
sense of the precession is always opposite to the sense of the
orbital motion \citep{Ka:97}, 
we study the two cases denoted by $+-$ and $-+$, 
where we have $\omega_{p}(t- z/V_{a})$, $-\omega_{b}(t- z/V_{a})$ and
$-\omega_{p}(t- z/V_{a})$, $\omega_{b}(t- z/V_{a})$, respectively.

\section{Method for solving the problem}
\label{sec:method}
\subsection{Introduction}

In this section, we explain the method for fitting VLBI observations 
using either a precession model or a BBH system model. The software 
is freely available on request to J Roland (roland@iap.fr).\\

This method is a practical one that provides solutions, but the method is not 
unique and does not guarantee that all possible solutions are found.\\

We calculate the projected trajectory on the plane
of the sky of an ejected component and determine the
parameters of the model to simultaneously produce the best fit 
with the observed west and north coordinates.
The parameters found minimize
\begin{equation}
    \chi^{2}_{t}= \chi^{2}(W_{c}(t)) + \chi^{2}(N_{c}(t)) \ ,
    \label{eq:chi2}
\end{equation}
where $\chi^{2}(W_{c}(t))$  and $\chi^{2}(N_{c}(t))$ 
are the $\chi^{2}$ calculated by comparing the 
VLBI observations with the calculated coordinates $W_{c}(t)$ and 
$N_{c}(t)$ of the component. For instance, to find the inclination 
angle that provides the best fit, we minimize $\chi^{2}_{t}(i_{o})$. 

A good determination of the $1\;\sigma$ (standard deviation) error bar can be 
obtained using the definition

\begin{eqnarray}
    (\Delta i_{o})_{1 \sigma} = | i_{o}(\chi^{2}_{min} + 1) -  i_{o}(\chi^{2}_{min}) | \ ,
    \label{eq:1_sigma}
\end{eqnarray}
which provides two values $(\Delta i_{o})_{1 \sigma+}$ and 
$(\Delta i_{o})_{1 \sigma-}$ (see \citet{LaMa+:76} and 
\citet{HeLe+:02}).

The concave parts of the surface $\chi^{2}(i_{o})$ contain a 
minimum. We can find solutions without a minimum; they 
correspond to the convex parts of the surface $\chi^{2}(i_{o})$ and are 
called \textit{mirage solutions}.

\begin{figure}[ht]
\centerline{
\includegraphics[scale=0.4, bb =-100 200 700 650,clip=true]{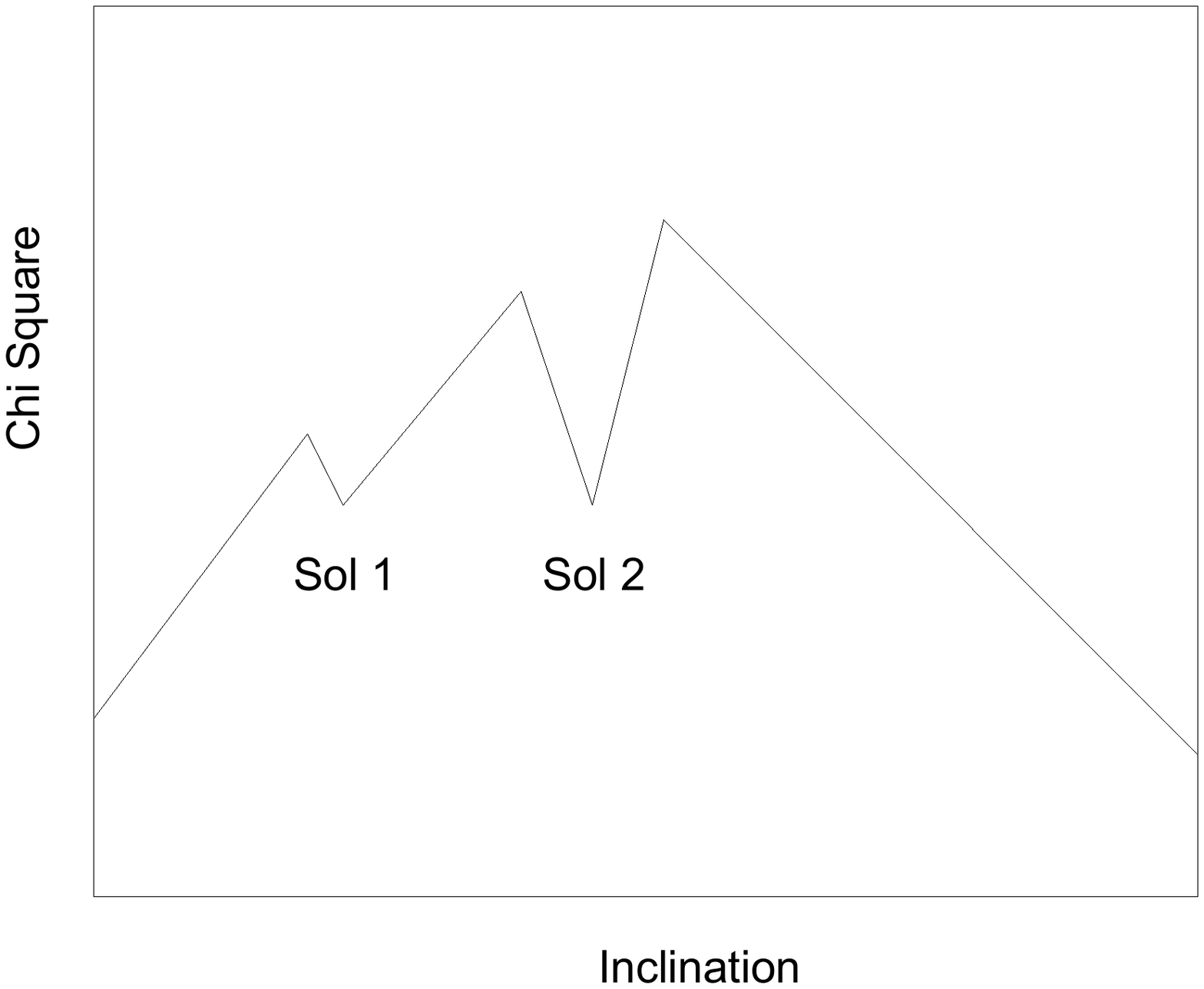}}
\caption{Example of a possible profile of the solution $\chi^{2}(i_{o})$. 
There are two possible solutions for which $\chi^{2}(Sol1) \approx \chi^{2}(Sol2)$. 
They correspond to the concave parts of the surface $\chi^{2}(i_{o})$. 
However, solution 2 is more robust than solution~1, i.e. it is the deepest one, 
and it will be the solution we adopt.}
\label{fig:Chi2_io}
\end{figure}

To illustrate the properties of the surface $\chi^{2}(i_{o})$ we plot in 
Figure \ref{fig:Chi2_io} a possible example of a profile of the solution 
$\chi^{2}(i_{o})$. In Figure \ref{fig:Chi2_io}, there are two possible solutions 
for which $\chi^{2}(Sol1) \approx \chi^{2}(Sol2)$, solution 2 is more 
robust 
than solution 1, i.e. it is the deepest one, and it will be the solution 
we will keep. 

We define the \textbf{\textit{robustness of the solution}} as the square root of the 
difference between the smallest maximum close to the minimum and the minimum of the 
function $\chi^{2}$. A solution of robustness 3 is a 3 $\sigma$ solution, i.e. 
3 $\sigma$ $\Leftrightarrow$ $\Delta \chi^{2} = 9$. 

The main difficulties we have to solve are the following:
\begin{enumerate}
    \item find all possible solutions,
    \item eliminate the mirage solutions,
    \item find the most robust solutions.
\end{enumerate}

For a given inclination angle of the BBH system problem, 
there exists a parameter that allows us
to find the possible solutions. This fundamental parameter is 
the ratio $T_{p}/T_{b}$, where $T_{p}$ and $T_{b}$ are 
the precession period of the accretion disk and the binary period 
of the BBH system respectively (see details in paragraph 2 of section \ref{sec:BBH_solution}, 
section \ref{sec:Chi2TpTb1823}, and section \ref{sec:Chi2TpTb3C279}).

Any minimum of the $\chi^{2}$ function can be a local minimum and not 
a global minimum.
However, because we investigate a 
wide range of the parameter $T_{p}/T_{b}$, namely 
$1 \leq T_{p}/T_{b} \leq 1000$, we expect to be able to find 
all possible solutions (the limit $T_{p}/T_{b} \leq 1000$ is given as an 
indication, in practice a limit of $T_{p}/T_{b} \leq 300$ is enough). 

Note that when the solution is found, it is not unique, 
but there exists a family of solutions. The solution shows a degeneracy 
and we will see that the parameter to fix the degeneragy or to find 
the range of parameters that provide the family of solutions is 
$V_{a}$, the propagation speed of the perturbation along the beam.

Generally, for any value of the parameters, the surface $\chi^{2}(\lambda)$
is convex and does not present a minimum. Moreover, when we are on the 
convex part of the surface $\chi^{2}(\lambda)$, one of the important 
parameters of the problem can diverge. The two important parameters of 
the problem that can diverge are 
\begin{enumerate}
    \item the bulk Lorentz factor of the $e^{\pm}$ beam, which has to be 
    $\gamma_{b}\leq 30$. This limit is imposed by the stability criterion  
     for the propagation of the relativistic beam in 
    the subrelativistic $e^{-}-p$ jet,
    \item the total mass of the BBH system.
\end{enumerate}

The most frequent case of divergence we can find corresponds to 
$\gamma_{b} \rightarrow \infty$. These mirage solutions are catastrophic 
and must be rejected. As we will see, generally, we have to study 
the robustness of the solution in relation to the parameters $T_{p}/T_{b}$, 
$M_{1}/M_{2}$, $\gamma$ and $i_{o}$.

\subsection{Solution of the precession model}
In a first step, we fit a simple precession model 
without a BBH system. This corresponds to the precession induced by a spinning 
BH (Lense-Thirring effect) or by the magnetically 
driven precession \citep{CaLi+:06}. This has the advantage of determining whether the solution 
corresponds to case I or to case II and of  preliminarily determining 
the inclination angle and the bulk Lorentz factor of the ejected 
component.

We have to investigate the different possible scenarios with regard to the 
sense of the rotation of the accretion disk. These possibilities correspond 
to $\pm\: \omega_{p}(t- z/V_{a})$. Accordingly, we study the two cases.

Assuming a simple precession model, these are the steps to fit the coordinates $X(t)$ 
and $Y(t)$ of a VLBI component:
\begin{enumerate}
    \item \textit{Determining the solution $\chi^{2}(i_{o})$ and  
    the time origin of the component ejection.} 
    In this section we assume that $V_{a} = 0.1 \; c$ (as $\chi^{2}(V_{a})$ remains 
		constant when $V_{a}$ varies, any value of $V_{a}$ can be used, see details in 
		the next paragraph). We calculate $\chi^{2}(i_{o})$, i.e., 
		we minimize $\chi^{2}_{t}(i_{o})$ when the inclination angle varies gradually 
		between two values. At each step of $i_{o}$, we determine each free parameter 
		$\lambda$ such that $\partial \chi^{2}_{t}/\partial \lambda = 0$. 
		
		Firstly, the important parameter to determine is the time origin of the ejection of 
		the VLBI component. We compare the times of the observed peak flux with the modeled peak flux.  
		The time origin is obtained when the two peak fluxes occur at the same time. 
		The solutions corresponding to case II show a significant  
		difference between the time origin of the ejection of the VLBI component 
		deduced from the fit of the peak flux and the time origin obtained from 
		the interpolation of the core separation. 
		
		Second, we can make a first determination of the inclination angle 
		and of the bulk Lorentz factor.
    
    \item \textit{Determining the family of solutions.}
		The solution previously found is not unique and shows a degeneracy.
		The parameter $V_a$ can be used to study the degeneracy of the solution. 
		Indeed, if we calculate $\chi^{2}(V_{a})$ when $V_{a}$ varies, we find that 
		$\chi^{2}(V_{a})$ remains constant. For the inclination angle found in the previous 
		section and the parameters of the corresponding solution, we calculate 
		$\chi^{2}(V_{a})$ when $V_{a}$ varies between $0.01 \; c \leq V_{a} \leq 0.45 \; c$ 
		and deduce the range of the precession period.
		
    \item \textit{Determining the possible offset of the origin of the VLBI component.}
    In this section, we keep $V_{a} = 0.1 \; c$ and using the inclination angle previously found 
		and the corresponding solution, we calculate $\chi^{2}(\Delta x, \Delta y)$ when $\Delta x$ 
		and $\Delta y$ vary ($\Delta x$ and $\Delta y$ are the possible offsets of the VLBI origin). 
		Solutions corresponding to case II show a significant offset of the space origin. 
		Note that determining the offsets of the VLBI coordinates does not 
		depend on the value of the inclination angle.
\end{enumerate}

\subsection{Solution of the BBH model}
\label{sec:BBH_solution}

We have to investigate the different possible scenarios with regard to the 
sense of the rotation of the accretion disk and the sense of the orbital 
rotation of the BBH system. Because the sense of the precession is always 
opposite to the sense of the orbital motion, we study the two cases 
where we have $\omega_{p}(t- z/V_{a})$, $-\omega_{b}(t- z/V_{a})$ and
$-\omega_{p}(t- z/V_{a})$, $\omega_{b}(t- z/V_{a})$, respectively.

Assuming a BBH model, this is the method for fitting the coordinates $X(t)$ 
and $Y(t)$ of a VLBI component:
\begin{enumerate}
    \item \textit{Determining the BBH system parameters for various 
		values of $T_{p}/T_{b}$.}
    In this section, we keep the inclination angle previously found  
		and $V_{a} = 0.1 \; c$. We determine the 
		BBH system parameters for different values of $T_{p}/T_{b}$, namely 
		$ T_{p}/T_{b} = 1.01$, $2.2$, $4.6$, $10$, $22$, $46$, $100$, and $220$ 
		for a BBH system with $M_{1} = M_{2}$ (these values of $T_{p}/T_{b}$ are chosen 
		because they are equally spaced on a logarithmic scale). 
		Generally, the BBH systems obtained with a low value of $T_{p}/T_{b}$, namely 
		$ T_{p}/T_{b} = 1.01$, $2.2$, or $4.6$ are systems with a large radius and 
		the BBH systems obtained with a high value of $T_{p}/T_{b}$, namely 
		$ T_{p}/T_{b} = 10$, $22$, $46$, $100$, or $220$ are systems with a small radius. 
		
    \item \textit{Determining the possible solutions: the 
		$\chi^{2}(T_{p}/T_{b})$ - diagram.}
		In this section, we keep the inclination angle previously found, 
		$V_{a} = 0.1 \; c$ and $M_{1} = M_{2}$. The crucial parameter for finding 
		the possible solutions is $T_{p}/T_{b}$, i.e., the ratio of the 
		precession period and the binary period. Starting from the 
		solutions found in the previous section, we calculate 
		$\chi^{2}(T_{p}/T_{b})$ when $T_{p}/T_{b}$ varies between 1 and 300. 
		We find that the possible solutions characterized by a specific value of the 
		ratio $T_{p}/T_{b}$. We note that some of the solutions can be 
		mirage solutions, which have to be detected and excluded.
		
    \item \textit{Determining the possible offset of the space origin.}
		In this section, we keep the inclination angle previously found, 
		$V_{a} = 0.1 \; c$ and $M_{1} = M_{2}$. Starting with the solution found 
		in the previous section, we calculate $\chi^{2}(\Delta x, \Delta y)$ when $\Delta x$ 
		and $\Delta y$ vary ($\Delta x$ and $\Delta y$ are the possible offsets of the VLBI origin). 
		If we find that an offset of the origin is needed, we correct the VLBI coordinates by 
		the offset to continue. Note that determining the offsets of 
		the VLBI coordinates does not depend on the value of the inclination angle.
		
    \item \textit{Determining the range of possible values of $T_{p}/T_{b}$.}
		In this section, we keep $V_{a} = 0.1 \; c$, $M_{1} = M_{2}$.
		Previously, we found a solution characterized by a value of $T_{p}/T_{b}$ 
		for a given inclination. Therefore we calculate $\chi^{2}(i_{o})$ when $i_{o}$ varies 
		with a variable ratio $T_{p}/T_{b}$. We obtain the range of possible values 
		of $T_{p}/T_{b}$ and the range of possible values of $i_{o}$.
		
    \item \textit{Preliminary determination of $i_{o}$, $T_{p}/T_{b}$ and $M_{1}/M_{2}$.}
		In this section, we keep $V_{a} = 0.1 \; c$. This section is the most complicated 
		one and differs for solutions corresponding to case I and case II. We indicate 
		the main method and the main results (the details are provided in section 
		\ref{sec:Mass_ratio_case_I} for the fit of component S1 of 1823+568 solutions and in section 
		\ref{sec:Mass_ratio_case_II} for the fit of component C5 of 3C 279). We calculate $\chi^{2}(i_{o})$ 
		for various values of $T_{p}/T_{b}$ and $M_{1}/M_{2}$. Generally, we find that 
		there exist critical values of the parameters $T_{p}/T_{b}$ and $M_{1}/M_{2}$,  
		which separate the domains for which the solutions exist or become mirage solutions.
		The curves $\chi^{2}(i_{o})$ show a minimum for given values $(i_{o})_{min}$ 
		and if necessary, we study the robustness of the solution in relation to the parameter $\gamma$, 
		therefore we calculate $\chi^{2}(\gamma)$ at $i_{o} = (i_{o})_{min}$ for the corresponding 
		values of $T_{p}/T_{b}$ and $M_{1}/M_{2}$. When these critical values are obtained, 
		we find the domains of $T_{p}/T_{b}$ and $M_{1}/M_{2}$, which produce the solutions 
		whose robustness is greater than 1.7 $\sigma$ and the corresponding inclination angle $i_{o}$. 
		
    \item \textit{Determining a possible new offset correction.}
		Using the solution found in the previous section, we calculate again 
		$\chi^{2}(\Delta x, \Delta y)$ when $\Delta x$ and $\Delta y$ vary. 
		When a new offset of the origin is needed, we correct the VLBI 
		coordinates by the new offset to continue. Note that this new offset correction is 
		smaller than the first one found previously.
		
    \item \textit{Characteristics of the final solution to the fit of the VLBI 
		component.} We are now able to find the BBH system parameters that produce the best 
		solution for the fit with the same method as described in point 5, 
		\textit{Preliminary determination of $i_{o}$, $T_{p}/T_{b}$ and $M_{1}/M_{2}$.}
		
    \item \textit{Determining the family of solutions.}
		The solution previously found is not unique and shows a degeneracy.
		The parameter $V_a$ can be used to study the degeneracy of the solution. 
		Indeed, when we calculate $\chi^{2}(V_{a})$ for varying $V_{a}$, we find that 
		$\chi^{2}(V_{a})$ remains constant. Using the solution found in the previous 
		section and the parameters of the corresponding solution, we calculate 
		$\chi^{2}(V_{a})$ when $V_{a}$ varies between $0.01 \; c \leq V_{a} \leq 0.45 \; c$ 
		and deduce the range of the precession period, the binary period, and the 
		total mass of the BBH system.
		
		\item \textit{Determining the size of the accretion disk.}
		Because we know the parameters of the BBH system, we can deduce the rotation period 
		of the accretion disk and its size.
\end{enumerate}

\section{Method - Case I}
\label{sec:case_I}
\subsection{Introduction: Fitting the component S1 of 1823+568}
\begin{figure}[ht]
\centerline{
\includegraphics[scale=0.42, bb =-100 200 700 650,clip=true]{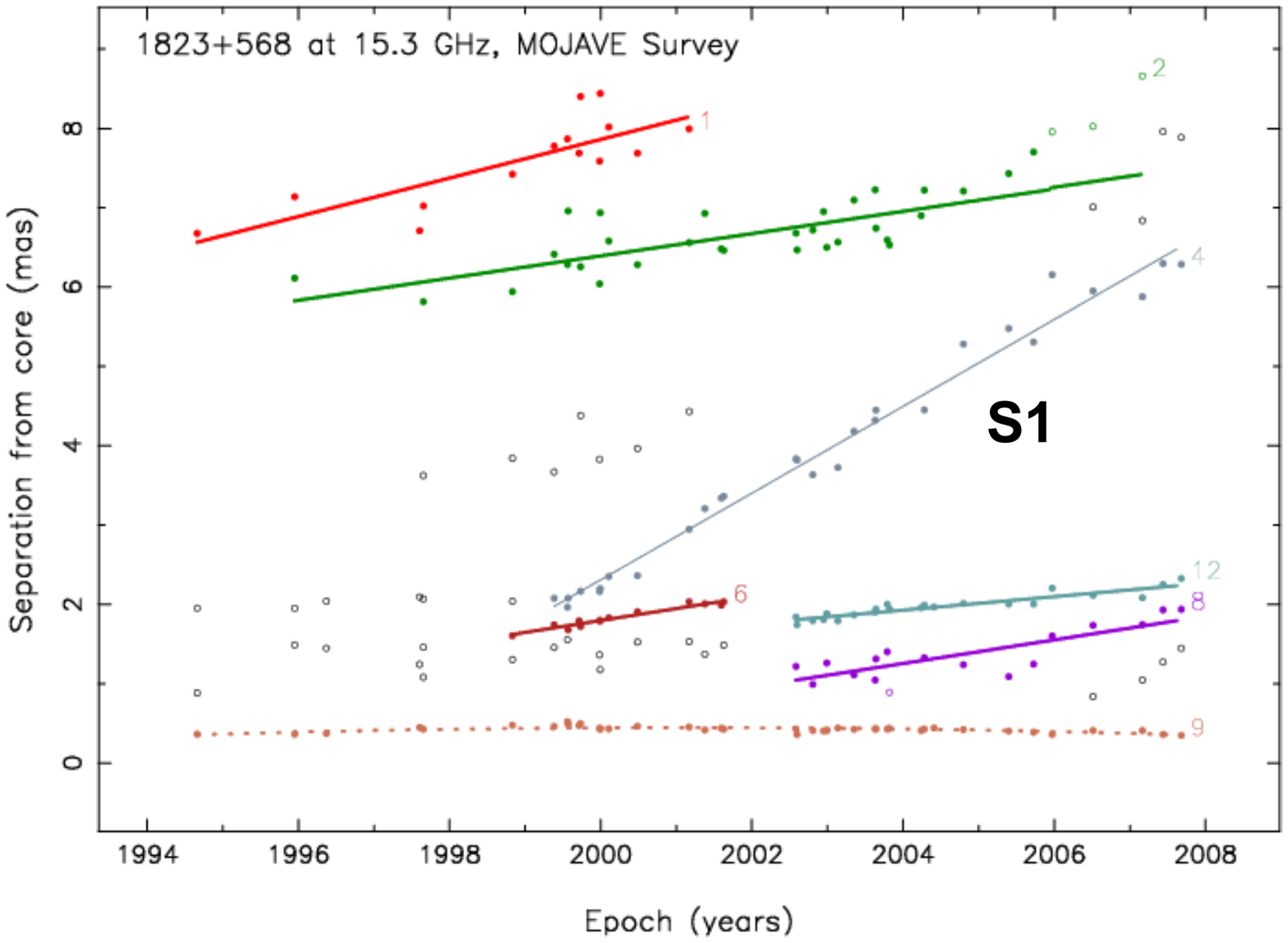}}
\caption{Separation from the core for the different VLBI 
components for the source 1823+568 from MOJAVE data \citep{LiCo+:09}. 
For details concerning the plot and the line fits see \citet{LiCo+:09}.
We  fit  component S1 corresponding to component 4 from 
the MOJAVE survey. Component S1 moves fast, which may indicate 
that two families of VLBI components exist in the case of 1823+568. If this is the case,
the nucleus of 1823+568 could contain a BBH system.}
\label{fig:1823+568_sep_time}
\end{figure}

Case I corresponds to a VLBI component ejected either from 
the VLBI core or to one where the offset of the origin of the ejection is 
smaller than or on the order of the smallest error bars of the VLBI component 
coordinates. It is the simplest case to solve. To illustrate the 
method of solving the problem corresponding to case I, we fit 
the component S1 of the source 1823+568 (Figures \ref{fig:1823+568_map} and \ref{fig:1823+568_sep_time}).

\subsection{VLBI data of 1823+568}

1823+568 is an quasar at a redshift of $0.664 \pm 0.001$ \citep{LaPe+:86}. 
The host galaxy is elliptical according to HST observations \citep{FaUr+:97}. 
The jet morphology on kpc-scales is complex 
- a mirrored S in observations with the MTRLI at 1666 MHz and with the VLA at 2 and 6 cm 
\citep{ODeBa+:88}. The largest extension of 1823+568 is 15'', corresponding 
to 93 kpc. On pc-scales the jet is elongated and points in a southern 
direction from the core \citep{PeRe:88}  - in accordance with the kpc-structure. 
Several components could be identified in the jet, e.g., \citet{GaCa+:89} and \citet{GaMu+:94}, 
\citet{GaCa:96}, \citet{JoMa+:05}. A VSOP Space VLBI image of 1823+568 has been 
obtained by \citet{LiAl+:09}. 
All identified components show strong polarization. The linear polarization is parallel 
to the jet ridge direction. Most of the components show slow apparent superluminal motion. 
The fast component S1 moved with an apparent velocity of about 20 c $\pm$ 2 c until 2005 
and subsequently decreases \citep{Gl:10}. 
Twenty-two VLBA observations obtained at 15 GHz within the 2-cm MOJAVE survey between 1994.67 
and 2010.12 have been re-analyzed and model-fitted to determine the kinematics of the 
individual components. For details of the data reduction and analysis see \citet{Gl:10}.  

\begin{figure}[ht]
\centerline{
\includegraphics[scale=0.25, bb =+000 000 550 820,clip=true]{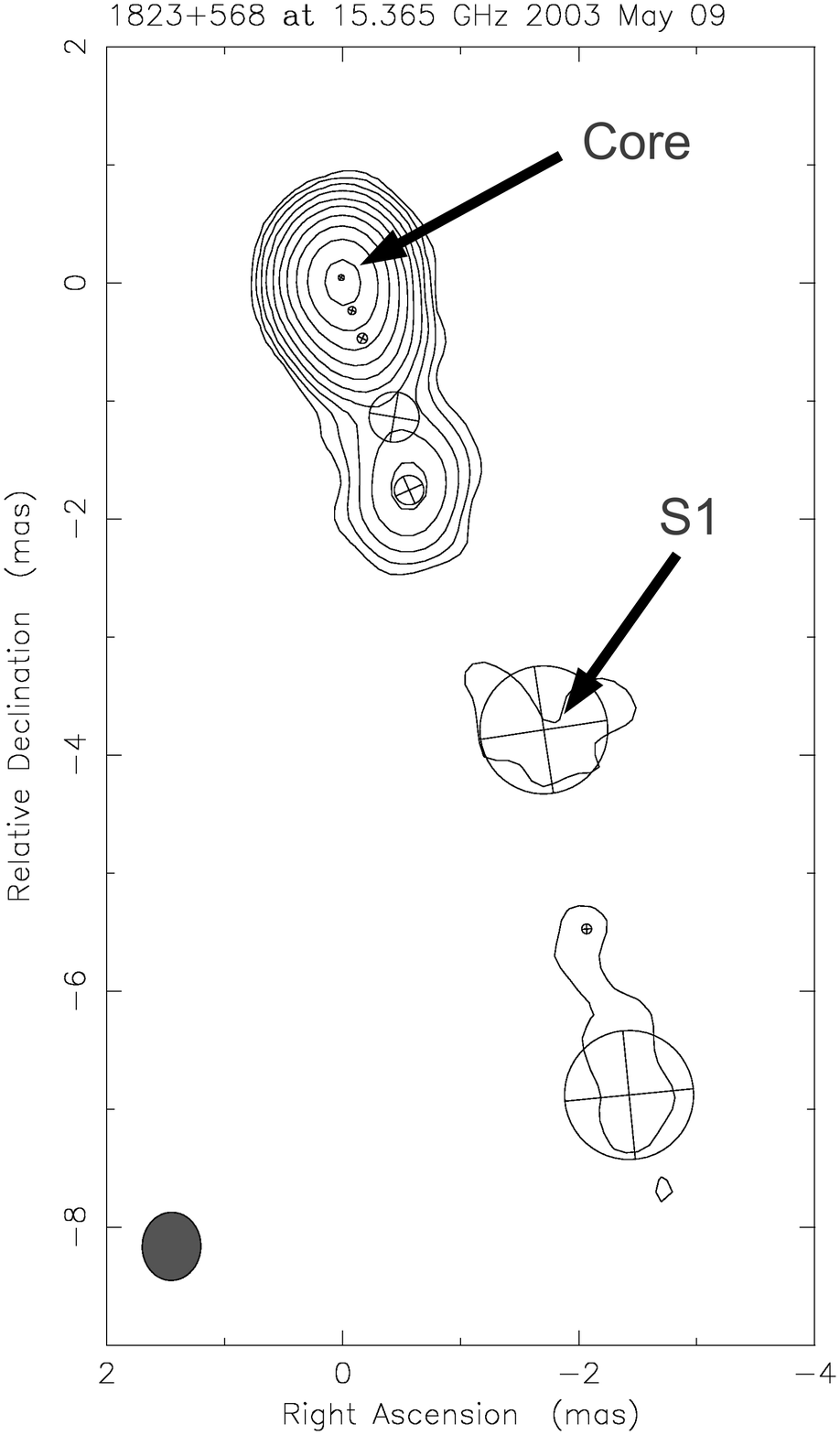}}
\caption{$15\,\mathrm{GHz}$ natural weighted VLBI image of 1823+568 with fitted circular 
Gaussian components observed 9 May 2003 \citep{Gl:10}. The map peak flux density was $1.27\,\mathrm{Jy/beam}$, where 
the convolving beam was $0.58\times0.5\,\mathrm{mas}$ at position angle (P.A.) $-2.09^{\circ}$. 
The contour levels were drawn at 0.15, 0.3, 0.6, 1.2, 2.4, 4.8, 9.6, 19.2, 38.4, and 76.8\,\% of 
the peak flux density.}
\label{fig:1823+568_map}
\end{figure}

The radio map of 1823+568, observed 9 May 2003, is shown in 
Figure \ref{fig:1823+568_map}. The data are taken from \citet{Gl:10}.

\subsection{Preliminary remarks}
\label{sec:Preliminary_remarks_1823+568}

The redshift of the source is $z_{s} \approx 0.664$, 
and using for the Hubble constant $H_{o} \approx 72$ km/s/Mpc, 
the luminosity distance of the source is $D_{l}\approx 3882$ Mpc 
and the angular distance is $D_{a} = D_{l} /(1+z)^2$. 

For details of the values of the data and of their error bars  
see \citet{Gl:10}. At 15 GHz, calling the beam size $Beam$, 
we adopted for the minimum values $\Delta_{min}$ 
of the error bars of the observed VLBI coordinates, the values in the range:
\begin{equation}
    Beam /15 \leq \Delta_{min} \leq Beam /12 \ ,
\label{eq:min_error}
\end{equation}
see Section \ref{sec:appendix_III} for details concerning this choice.

For 1823+568, observations were performed at 15 GHz and the beam size
is mostly circular and equal to $Beam \approx 0.5$ $mas$. We adopted as minimum 
values of the error bars the values $(\Delta W)_{min} \approx Beam/12 \approx 40$ $\mu as$ and 
$(\Delta N)_{min} \approx Beam/12 \approx 40$ $\mu as$ for the west and north coordinates 
of component S1, i.e., when the error bars obtained from the VLBI data reduction were  
smaller than $(\Delta W)_{min}$ or $(\Delta N)_{min}$, they were enlarged to the 
minimum values. The minimum values were chosen empirically, but the adopted values were justified 
a posteriori by comparing the $\chi^{2}$ value of the final solution and the 
number of constraints to make the fit and to 
obtain a reduced $\chi^{2}$ close to 1. For the component S1, we have 
$(\chi^{2})_{final} \approx 51$ for 56 constraints, the reduced $\chi^{2}$ is 
$\chi^{2}_{r} = (\chi^{2})_{final} /56 \approx 0.91$. \citet{LiHo:05} suggested that the positional 
error bars should be about $1/5$ of the beam size. However, if we had chosen 
$(\Delta W)_{min} = (\Delta N)_{min} \approx Beam /5 \approx 100$ $\mu as$, 
we would have $(\chi^{2})_{final} \ll 56$, indicating that the minimum error bars 
would be overestimated (see details in Section \ref{sec:appendix_III}).

To obtain a constant projected trajectory of the VLBI component 
in the plane perpendicular to the line of sight, 
the integration step to solve equation (\ref{eq:dzt}) changes when the 
inclination angle varies, . 
The integration step was $\Delta t = 0.8$ yr 
when $i_{o} = 5^{\circ}$. When $i_{o}$ varied, it was 
$\Delta t = 0.8 (sin(5^{\circ})/sin(i_{o}))$ yr.

The trajectory of component S1 is not long enough to constrain the parameter 
$T_{d}$, i.e., the characteristic time for the damping of the beam perturbation. 
We fit assuming that $T_{d} \leq 2500$ yr; this value produced a good trajectory 
shape.

The time origin of the ejection of the component S1, deduced from the interpolation 
of VLBI data, is $t_{o} \approx 1995.6$ (Figure \ref{fig:1823+568_sep_time}).

Close to the core, the size of S1 is $\approx 0.24$ $mas$, therefore we  
assumed that $n_{rad} = 75$,  where $n_{rad}$ is the number of steps 
to describe the extension of the VLBI component along the beam. 
At $i_{o} = 5^{\circ}$ with an integration step $\Delta t = 0.8$ yr, we calculated 
the length of the trajectory corresponding to each integration step. The size 
of the component is the sum of the first $n_{rad} = 75$ lengths.

\subsection{Final fit of component S1 of 1823+568}
\label{sec:solution_1823+568}

Here we present the solution to the fit of S1, the details for the fit 
can be found in Section \ref{Fit_1823+568}. 

We studied the two cases $\pm \omega_{p}(t - z/V_{a})$. The final solution 
of the fit of component S1 using a BBH system corresponds 
to $+\omega_{p}(t - z/V_{a})$ and $- \omega_{b}(t - z/V_{a})$. \\

The main characteristics of the solution of the BBH system 
associated with 1823+568 are that
\begin{itemize}
    \item the radius of the BBH system is $R_{bin} \approx 60$ $\mu as$ $\approx 0.42$ $pc$,
    \item the VLBI component S1 is not ejected by the VLBI core, and the offsets  
    of the observed coordinates are $\Delta W \approx +5$ $\mu as$ 
    and $\Delta N \approx 60$ $\mu as$,
    \item the ratio $T_{p}/T_{b}$ is $8.88 \leq T_{p}/T_{b} \leq 9.88$, and
    \item the ratio $M_{1}/M_{2}$ is $0.095 \leq M_{1}/M_{2} \leq 0.25$.
\end{itemize}

The results of the fits obtained for $T_{p}/T_{b} = 8.88$ and $T_{p}/T_{b} = 9.88$ are 
given in section \ref{sec:Final_Fit_1823+568}. The solutions found with $T_{p}/T_{b} \approx 8.88$ are 
slightly more robust, but both solutions can be used.

To continue, we arbitrarily adopted  the solution with $T_{p}/T_{b} \approx 8.88$ 
and $M_{1}/ M_{2} \approx 0.17$. We deduced the main parameters of 
the model, which are that
\begin{itemize}
    \item the inclination angle is $i_{o} \approx 3.98^{\circ}$,
    \item the angle between the accretion disk and the rotation plane 
    of the BBH system is $\Omega \approx 0.28^{\circ}$ (this is also the opening 
    angle of the precession cone),
    \item the bulk Lorentz factor of the VLBI component is $\gamma_{c} \approx 17.7$, and
    \item the origin of the ejection of the VLBI component is $t_{o} \approx 1995.7$.
\end{itemize}

\begin{figure}[ht]
\centerline{
\includegraphics[scale=0.5, width=8cm,height=6cm]{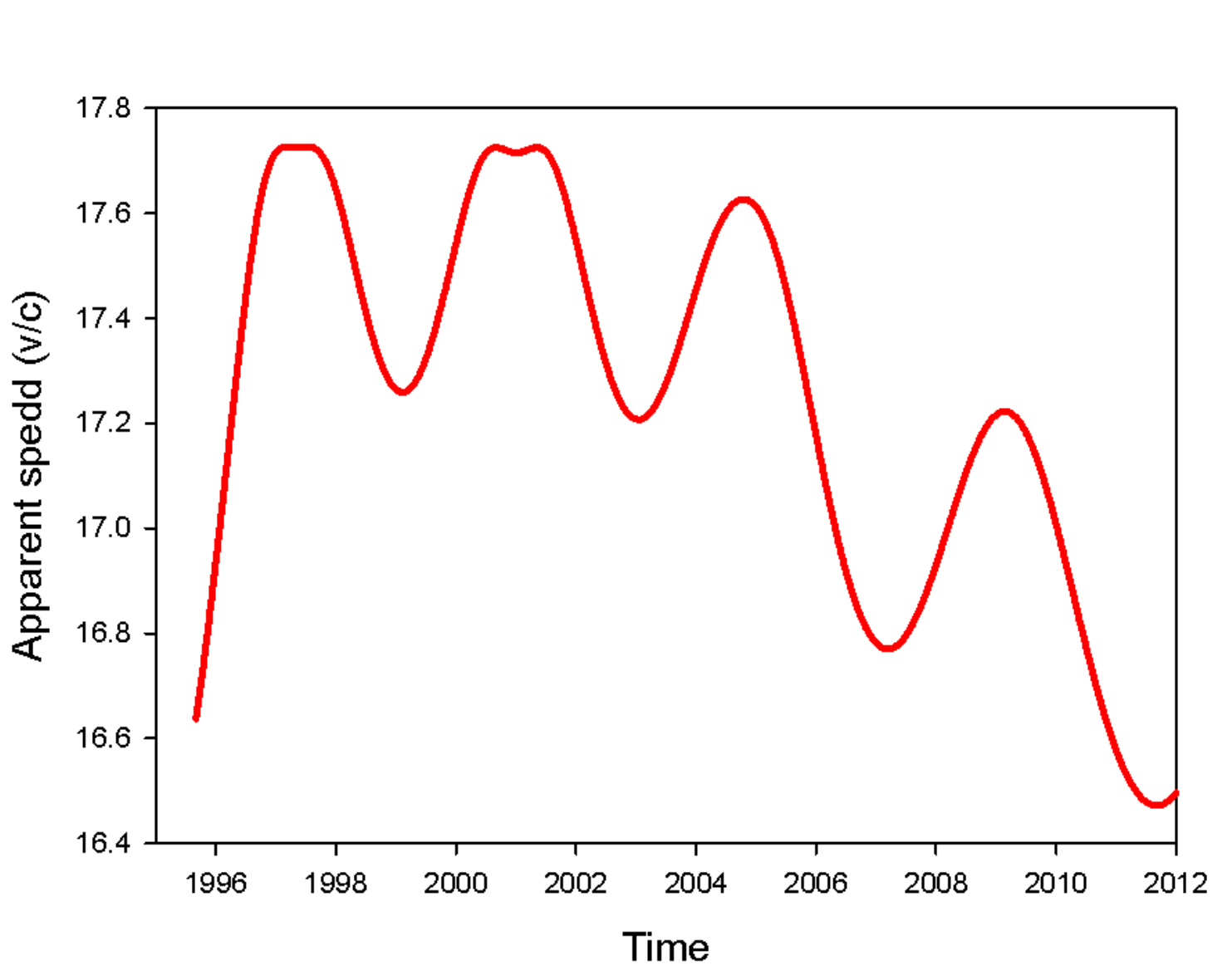}}
\caption{Apparent speed of component S1 increases at the begining, 
then it is $\approx 17.5$ c until 2005, and finally, it decreases slowly assuming 
a constant bulk Lorentz factor $\gamma_{c} \approx 17.7$.}
\label{fig:App_Speed_S1_1823+568}
\end{figure}

The variations of the apparent speed of component S1 are shown 
in Figure \ref{fig:App_Speed_S1_1823+568}.

We can determine the Doppler factor (equation \ref{eq:delta}), and 
consequently, we can estimate the observed flux density (equation \ref{eq:flux}). This was used 
to fit the temporal position of the peak flux and to determine the temporal origin of 
the ejection of the VLBI component (see Section \ref{sec:Prec_1823+568} for the details).

\begin{figure}[ht]
\centerline{
\includegraphics[scale=0.5, bb =-275 0 700 650,clip=true]{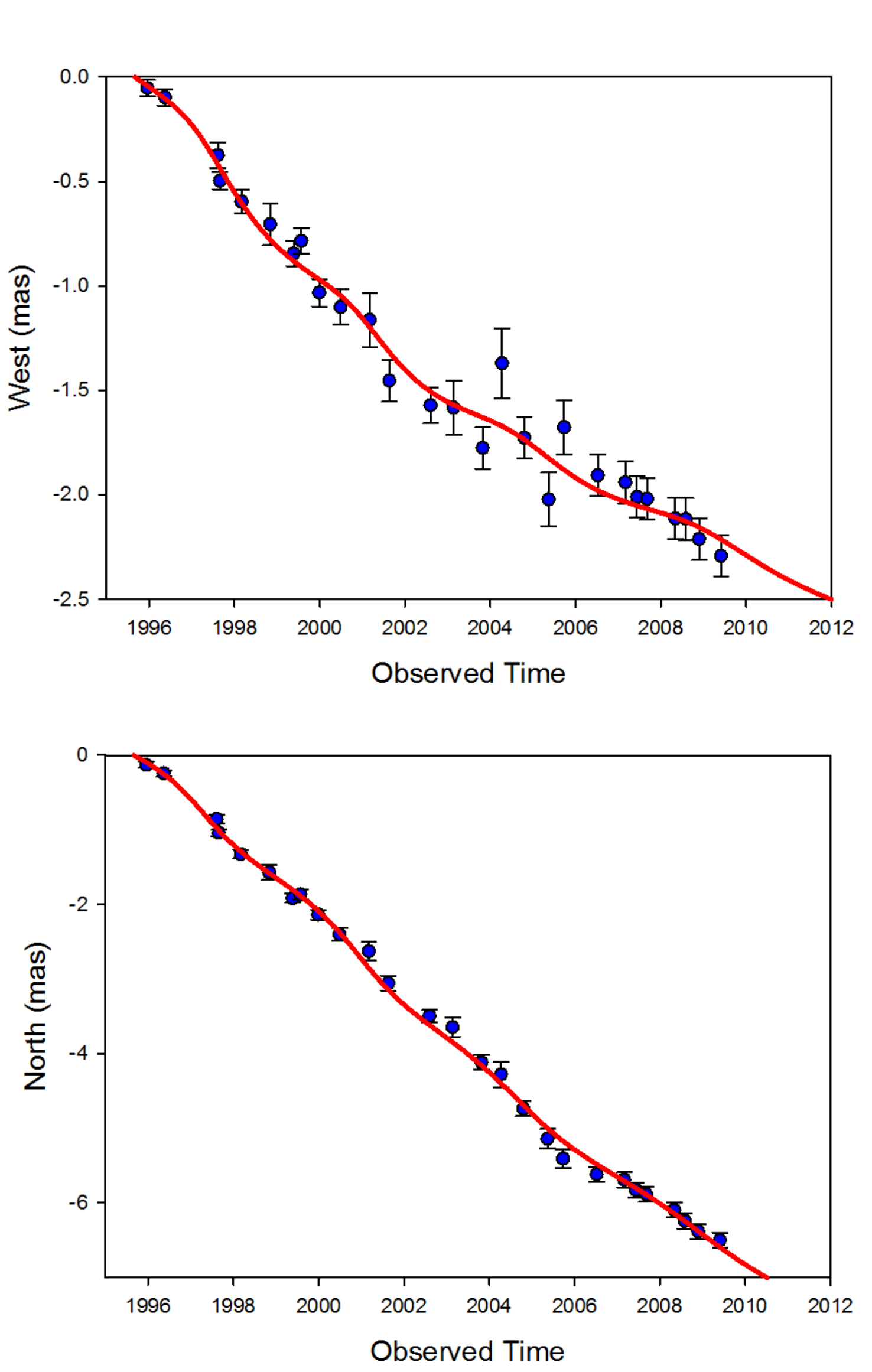}}
\caption{Fit of the two coordinates $W(t)$ and $N(t)$ of component S1 of 1823+568. 
They correspond to the solution with $T_{p}/T_{b} \approx 8.88$, 
$M_{1}/ M_{2} \approx 0.17$, and $i_{o} \approx 3.98^{\circ}$. The points are the observed 
coordinates of component S1 that were corrected by the offsets $\Delta W \approx +5$ $\mu as$ 
and $\Delta N \approx 60$ $\mu as$ (the VLBI coordinates and their error bars are taken from \citet{Gl:10}). 
The red lines are the coordinates of the component trajectory calculated using the BBH model.}
\label{fig:S1_Xt+Yt_4_New_OffbSol2+Fin}
\end{figure}

The fit of the two coordinates $W(t)$ and $N(t)$ of the component S1 of 1823+568 is  
shown in Figure \ref{fig:S1_Xt+Yt_4_New_OffbSol2+Fin}. 
The points are the observed coordinates of component S1 that were corrected 
by the offsets $\Delta W \approx +5$ $\mu as$ and $\Delta N \approx 60$ $\mu as$, and 
the red lines are the coordinates of the component trajectory calculated using the BBH model 
assuming the solution parameters, i.e., $T_{p}/T_{b} \approx 8.88$, $M_{1}/ M_{2} \approx 0.17$, 
$i_{o} \approx 3.98^{\circ}$, etc.

Finally, we compared this solution with the solution obtained using 
the precession model. The $\chi^{2}_{min}(i_{o})$ is about 51 for the fit using 
the BBH system and about 67 for the precession model (see section \ref{sec:Prec_1823+568}),
i.e., the BBH system solution is a 4 $\sigma$ better solution. 
To fit the ejection of component S1 we used 56 observations (the west and north coordinates 
corresponding to the 28 epochs of observation), so the reduced 
$\chi^{2}$ is $\chi^{2}_{r} = 51 /56 \approx 0.91$, indicating that the minimum values 
used for the error bars are correct.

\subsection{Determining the family of solutions}
For the inclination angle previously found, i.e., 
$i_{o}\approx 3.98^{\circ}$, $T_{p}/T_{b} \approx 8.88$, $M_{1}/M_{2} \approx 0.17$, 
and $R_{bin} \approx 60$ $\mu as$, we gradually varied $V_{a}$ between $0.01$ c 
and $0.45$ c. The function $\chi^{2}(V_{a})$ remained constant,  
indicating a degeneracy of the solution. We deduced the 
range of variation of the BBH system parameters. They are given in Table 1.

\begin{center}
Table 1 : Ranges for the BBH system parameters\medskip%

\begin{tabular}
[c]{c||c|c}\hline
$V_{a}$                  & $0.01 \: c$                           & $0.45 \: c$                           \\\hline
$T_{p}(V_{a})$           & $\approx 540000$ yr                   & $\approx 6700$ yr                     \\\hline
$T_{b}(V_{a})$           & $\approx 60600$ yr                    & $\approx 750$ yr                      \\\hline
$(M_{1}+ M_{2})(V_{a})$  & $\approx 1.6 \; 10^{5}$ $M_{\odot}$   & $\approx 1.05 \; 10^{9}$ $M_{\odot}$  \\\hline
\end{tabular}
\end{center}

The period of the BBH system is not obviously 
related to a possible periodicity of the radio or the optical 
light curve.

\subsection{Determining the size of the accretion disk}

From the knowledge of the mass ratio $M_{1}/M_{2} \approx 0.17$ and 
the ratio $T_{p}/T_{b} \approx 8.88$, we calculated in the previous section 
the mass of the ejecting black hole $M_{1}$, the orbital period $T_{b}$, 
and the precession period $T_{p}$ for each value of $V_{a}$.

The rotation period of the accretion disk, $T_{disk}$, is given by 
\citep{BrRo+:01}

\begin{equation}
    T_{disk} \approx \frac{4}{3}\frac{M_{1}+M_{2}}{M_{2}} T_{b} \frac{T_{b}}{T_{p}} \ .
    \label{eq:Tdisk}
\end{equation}

Thus we calculated the rotation period of the accretion disk, and 
assuming that the mass of the accretion disk is $M_{disk} \ll M_{1}$, 
the size of the accretion disk $R_{disk}$ is 
\begin{equation}
    R_{disk} \approx \left(\frac{T_{disk}^{2}}{4\pi^{2}}GM_{1}\right)^{1/3} \ .
    \label{eq:Rdisk}
\end{equation}

We found that the size of the accretion disk does not depend on $V_{a}$ and is 
$R_{disk}  \approx 0.090 \; pc \approx 0.013 \; mas$.

\section{The method - Case II}
\label{sec:case_II}
\subsection{Introduction: Application to component C5 of 3C 279}
\label{sec:IntrocaseII}
Case II corresponds to an ejection of the VLBI component with an offset 
of the origin of the component larger than   
the smallest error bars of the VLBI component coordinates. This is the most 
difficult case to solve because data have to be corrected by an unknown offset 
That is larger than the smallest error bars.

When we apply the precession model, there are two signatures of case II, which are
\begin{enumerate}
    \item the problem of the time origin of the VLBI component, and 
    \item the shape of the curve $\chi^{2}_{t}(i_{o})$.
\end{enumerate}

Using the precession model, we modeled the flux and compared the time position of 
the first peak flux with the time position of the observed peak flux. 
If the origin time deduced from interpolating the VLBI data was very different than 
the origin time deduced from the precesion model, we concluded that there is 
a time origin problem (see Section \ref{sec:Prec_3C279}). We show that 
this origin-time problem is related to the offset of the space origin of the VLBI 
component, i.e., the VLBI component is not ejected by the VLBI core and this offset 
is larger than the smallest error bars (see Section \ref{sec:Offset_prec_3C279}).\\

When the offset of the space origin is larger than the smallest error 
bars of the component positions and the VLBI coordinates are not corrected 
by this offset, the curve $\chi^{2}_{t}(i_{o})$ can have a very characteristic shape:
\begin{enumerate}
    \item the inclination angle is limited to a specific interval, i.e., 
    $i_{min} \leq i_{o} \leq i_{max}$,
    \item when $i_{o} \rightarrow i_{max}$ and when $i_{o} \rightarrow i_{min}$, 
    the bulk Lorentz factor of the VLBI component diverges, i.e., $\gamma_{c} \rightarrow \infty$, and 
    \item the function $\chi^{2}_{t}(i_{o})$ does not have a minimum 
    in the interval $i_{min} \leq i_{o} \leq i_{max}$.
\end{enumerate}
See Figure \ref{fig:Chi2+Gamma_C5_3C279_Precession} corresponding to the precession 
model applied to component C5 of 3C 279.

\subsection{MOJAVE data of 3C 279}

The radio quasar 3C 279 (z = 0.536 \citet{MaSu+:96}) is one of the brightest extragalactic 
radio sources and has been observed and studied in detail for decades.
Superluminal motion in the outflow of the quasar was found by \citet{WhSh+:71} 
and \citet{CoCa+:71}. Thanks to the increasing resolution and sensitivity of modern 
observation techniques, a more complex picture of 3C 279 appeared, including  multiple 
superluminal features moving along different trajectories downstream the jet  \citep{UnCo+:89}. 
The apparent speed of these components span an interval between 4 c and 16 c 
\citep{CoCo+:79,WePi+:01}.

We used the MOJAVE observations of 3C 279 \citep{LiTi+:01a}. 
Seventy-six VLBA observations obtained at 15 GHz within the 2-cm MOJAVE survey between 1999.25  
and 2007.64 were re-analyzed and model-fitted to determine the coordinates of the VLBI components.
We used the NRAO Astronomical Image Processing System (AIPS) to calibrate 
 the data. We performed an amplitude calibration and applied a correction for 
the atmospheric opacity for the high-frequency data $\left(\nu>15\,\mathrm{GHz}\right)$. 
The parallactic angle correction was taken into account before we calibrated the 
phases using the pulse-scale signal and a final fringe fit. The time- and 
frequency-averaged data were imported to DIFMAP \citep{Sh:97}, were we used the CLEAN and 
MODELFIT algorithm for imaging and model fitting, respectively.

The fully calibrated visibilities were fitted in DIFMAP using the algorithm MODELFIT 
and 2D circular Gaussian components. These components were characterized by their 
flux density, $S_{mod}$, position $r_{mod}$, position angle (P.A.), $\theta_{mod}$ 
(measured from north through east), and their full-width at half-maximum (FWHM). Since the number 
of fitted Gaussians was initially not limited, we only then added a new component when 
the $\chi^{2}$ value decreased significantly. This approach led to a minimum number 
of Gaussians that can be regarded as a reliable representation of the source. We modeled 
each epoch separately to avoid biasing effects. The kinematics of the source could thus 
be analyzed by tracking the fitted components. The average beam for 
the $15\,\mathrm{GHz}$ observations is $0.51\,\mathrm{mas}\times	1.34\,\mathrm{mas}$.

\begin{figure}[ht]
\centerline{
\includegraphics[scale=0.5, bb =-100 020 700 550,clip=true]{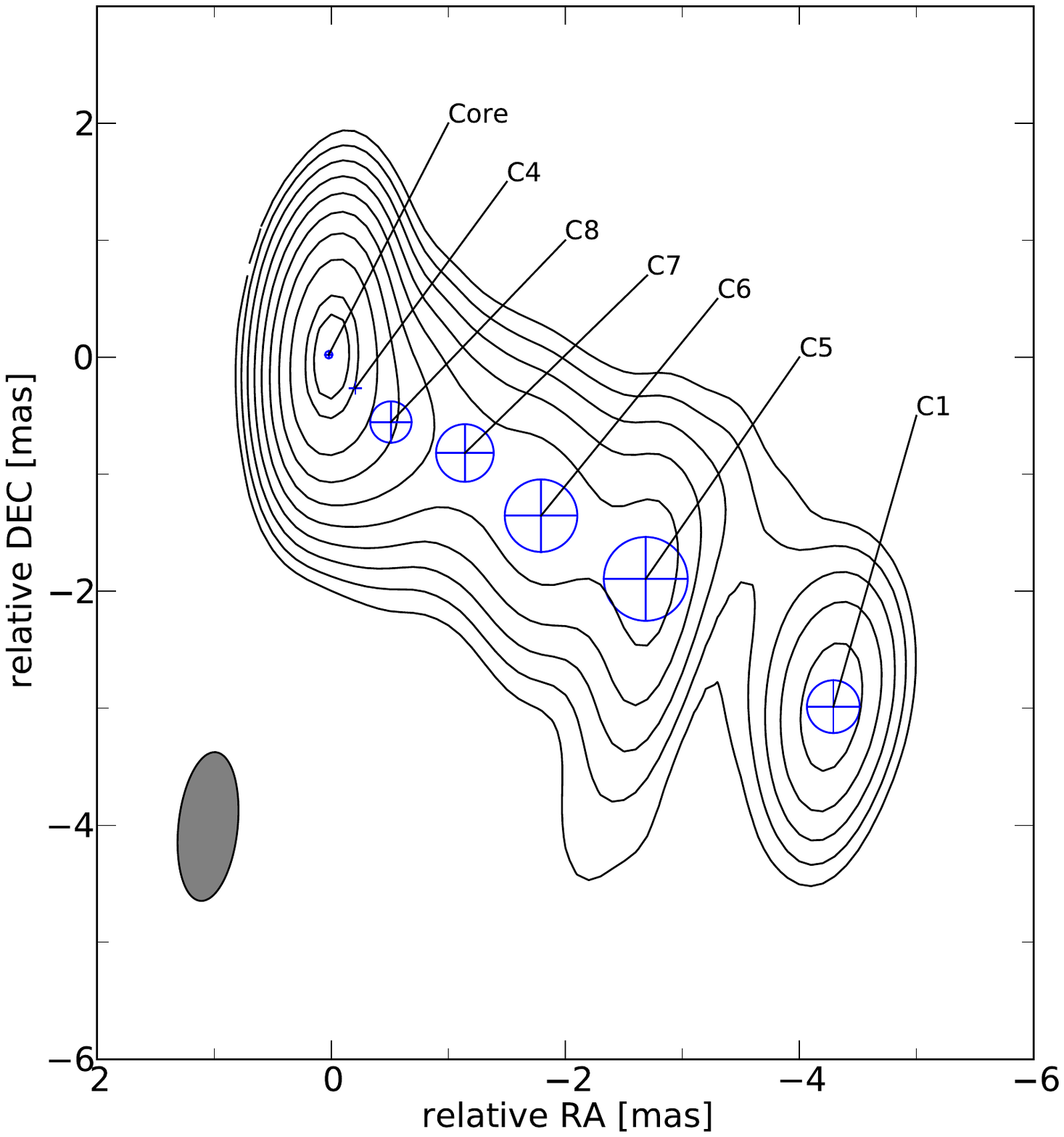}}
\caption{$15\,\mathrm{GHz}$ natural-weighted VLBI image of 3C\,279 with fitted circular 
Gaussian components observed 15 June 2003 \citep{LiAl+:09}. 
The map peak flux density was $8.3\,\mathrm{Jy/beam}$, where 
the convolving beam was $1.3\times0.5\,\mathrm{mas}$ at position angle (P.A.) $-6.0^{\circ}$. 
The contour levels were drawn at 0.2, 0.5, 1.0, 2.0, 4.0, 8.0, 16, 32, 64, and 80\,\% of 
the peak flux density. Component C4 is a \textit{stationary} component 
(see Figure \ref{fig:3C279_sep_time}).}
\label{fig:3C279_Map}
\end{figure}

The radio map of 3C 349, observed 15 june 2003, is shown in 
Figure \ref{fig:3C279_Map}. The data are taken from \citet{LiAl+:09}.

\begin{figure}[ht]
\centerline{
\includegraphics[scale=0.42, bb =-100 150 700 600,clip=true]{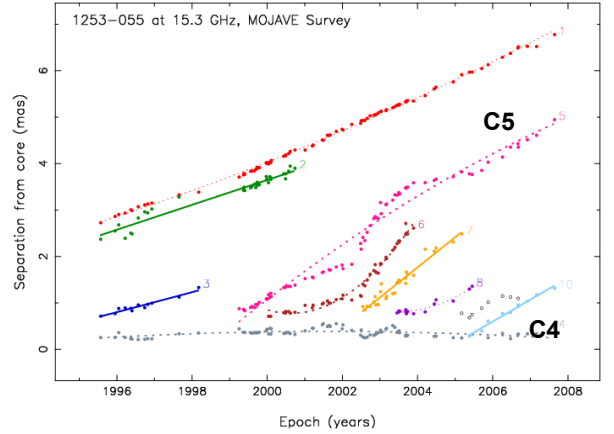}}
\caption{Separation from the core for the different VLBI 
components for the source 3C 279 from MOJAVE data \citep{LiCo+:09}. 
For the obtaining of the plotted line fits see \citet{LiCo+:09}.
We fit component C5. Component C5 is ejected from an origin with a 
large offset from the VLBI core.}
\label{fig:3C279_sep_time}
\end{figure}

\subsection{Preliminary remarks}
The redshift of 3C 279 is $z \approx 0.536$, and using for 
the Hubble constant $H_{o}\approx 72$ km/s/Mpc, the luminosity distance 
of the source is $D_{l}\approx 3070$ Mpc and the angular distance 
is $D_{a} = D_{l} /(1+z)^2$. 

For details of the values of the data   
see \citet{LiAl+:09}. Because the observations were performed at 15 GHz and the beam 
size was $0.51\,\mathrm{mas}\times	1.34\,\mathrm{mas}$, we adopted for the minimum 
values of the error bars the values $(\Delta W)_{min} \approx Beam/15 \approx 34$ $\mu as$ and 
$(\Delta N)_{min} \approx Beam/15 \approx 89$ $\mu as$ for the west and north coordinates 
of component C5. The adopted values were justified a posteriori by comparing  
the $\chi^{2}$ value of the final solution and the number of constraints to make the fit 
and to obtain a reduced $\chi^{2}$ close to 1. For the component C5, we have 
$(\chi^{2})_{final} \approx 150$ for 152 constraints, thus the reduced $\chi^{2}$ is: 
$(\chi^{2})_{r} \approx 0.99$. It has been suggested by \citet{LiHo:05} that 
the positional error should be within 20\% of the convolving beam size, i.e., $\approx Beam/5$. 
See Section \ref{sec:appendix_III} for details concerning the choice adopted in this article 
and the determination of the $\chi^{2}$, the characteristics of the 
solution using minimum erros bars are as large as $\approx Beam/5$.

The integration step used to solve equation (\ref{eq:dzt}) is $\Delta t = 0.8$ yr 
when $i_{o} = 5^{\circ}$. When $i_{o}$ varies, it is $\Delta t = 0.8 (sin(5^{\circ})/sin(i_{o}))$ yr.

The trajectory of component C5 is not long enough to constrain the parameter 
$T_{d}$, i.e., the characteristic time for the damping of the beam perturbation. 
We fit assuming that $T_{d} \leq 2000$ yr.

The time origin of the ejection of the component C5 cannot be deduced easily 
from the interpolation of VLBI data \citep{LiCo+:09}. However, 
we show in Section \ref{sec:Prec_3C279} how, using the precession model, 
it is possible to obtain the minimum time origin of the VLBI component by comparing the time position 
of the calculated first peak flux with the observed time position of the first peak flux.

Close to the core, the size of C5 is $\approx 0.25$ $mas$, therefore we  
assumed that $n_{rad} = 75$,  where $n_{rad}$ is the number of steps 
to describe the extension of the VLBI component along the beam.

\subsection{Final fit of component C5 of 3C 279}
\label{sec:solution_3C279}

Here we present the solution to the fit of C5, the details for the fit 
can be found in Section \ref{Fit_3C279}. The fit of component C5 using a BBH system corresponds 
to $-\omega_{p}(t - z/V_{a})$ and $+ \omega_{b}(t - z/V_{a})$.\\

The main characteristics of the solution of the BBH system 
associated with 3C 279 are that
\begin{itemize}
    \item the radius of the BBH system is $R_{bin} \approx 420$ $\mu as$  $\approx 2.7$ $pc$,
    \item the VLBI component C5 is not ejected by the VLBI core and the offsets  
    of the observed coordinates are $\Delta W \approx +405$ $\mu as$ 
    and $\Delta N \approx +110$ $\mu as$,
    \item the ratio $T_{p}/T_{b}$ is $T_{p}/T_{b} \approx 140$, and
    \item the ratio $M_{1}/M_{2}$ is $M_{1}/M_{2} \approx 2.75$.
\end{itemize}

The results of the fits obtained for $T_{p}/T_{b} \approx 140$ and $M_{1}/M_{2} \approx 2.75$ 
are given in Appendix \ref{sec:Final_fit_3C279}.

Adopting the solution with $T_{p}/T_{b} \approx 140$ 
and $M_{1}/ M_{2} \approx 2.75$, we deduced the main parameters of 
the model. 
\begin{itemize}
    \item The inclination angle is $i_{o} \approx 10.4^{\circ}$.
    \item The angle between the accretion disk and the rotation plane 
    of the BBH system is $\Omega \approx 2.4^{\circ}$ (this is also the opening 
    angle of the precession cone).
    \item The bulk Lorentz factor of the VLBI component is $\gamma_{c} \approx 16.7$.
    \item The origin of the ejection of the VLBI component is $t_{o} \approx 1999.0$.
\end{itemize}

\begin{figure}[ht]
\centerline{
\includegraphics[scale=0.5, width=8cm,height=6cm]{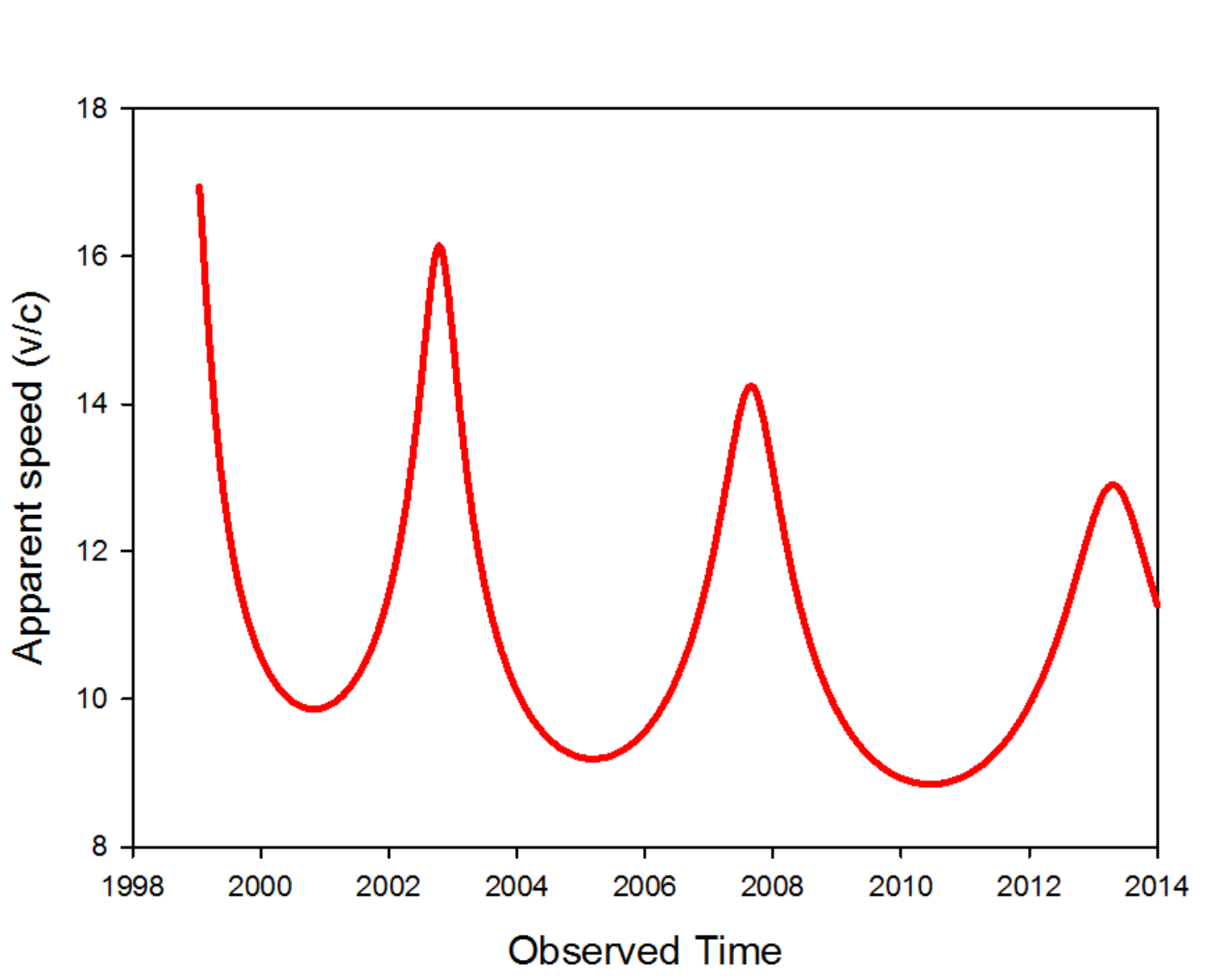}}
\caption{Apparent speed of component C5 varies between 17 c and 9 c assuming 
a constant bulk Lorentz factor $\gamma_{c} \approx 16.7$.}
\label{fig:3C279_C5_Apparent_Speed}
\end{figure}

The variations of the apparent speed of component C5 are shown 
in Figure \ref{fig:3C279_C5_Apparent_Speed}.

\begin{figure}[ht] 
\centerline{
\includegraphics[scale=0.5, width=8cm,height=6cm]{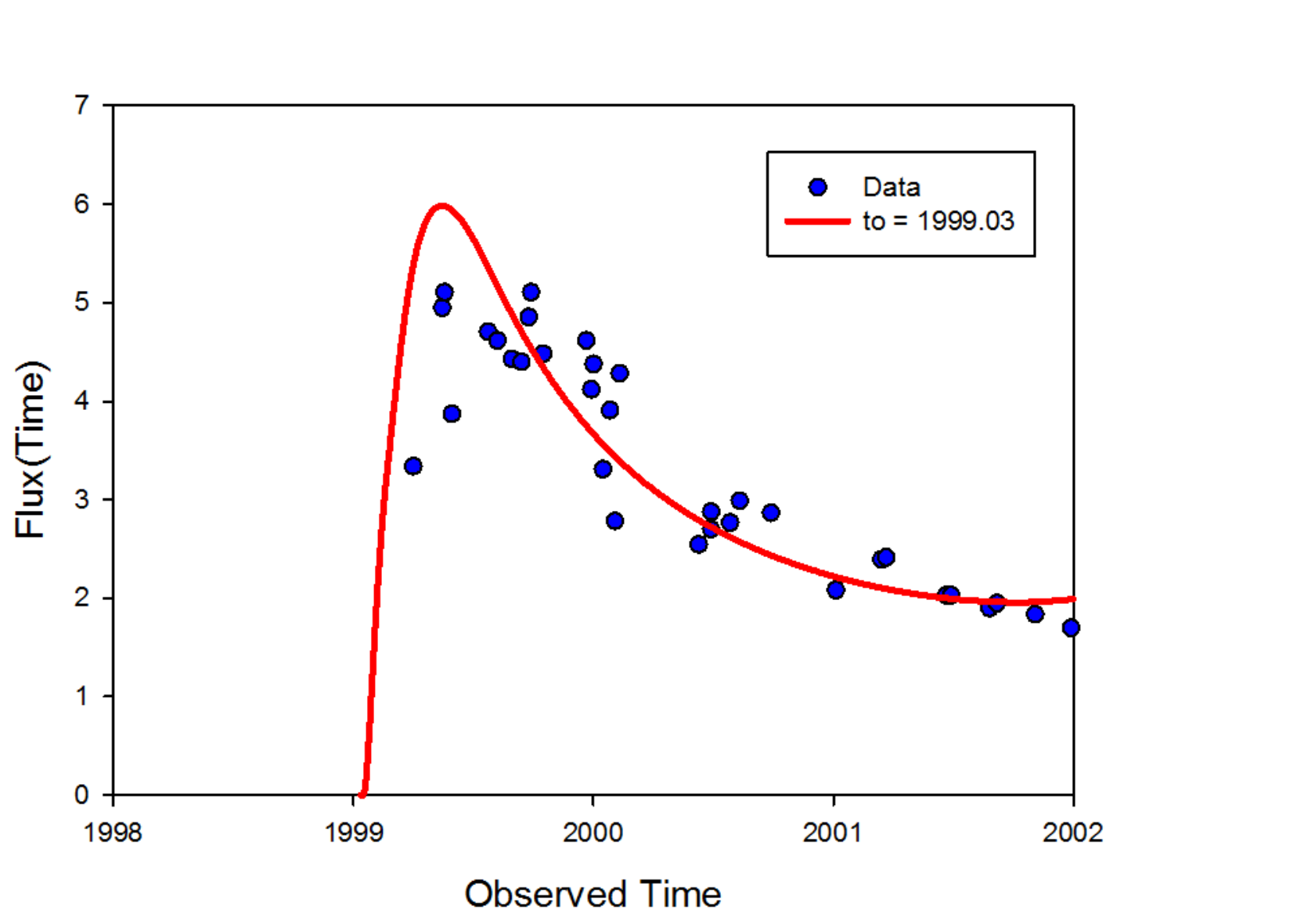}}
\caption{Flux variations of component C5 using the BBH model. The time origin of the ejection 
of C5 is 1999.03.}
\label{fig:3C279_C5_Flux_BBH}
\end{figure}

We can determine the Doppler factor (equation \ref{eq:delta}), and 
consequently we can estimate the observed flux density (equation \ref{eq:flux}). 
Using the precession model, we fitted the temporal position of the peak flux and determined 
the temporal origin of the ejection of the VLBI component 
(see Section \ref{sec:Prec_3C279} for the details). Using the BBH model, we calculated and plotted  
in Figure \ref{fig:3C279_C5_Flux_BBH} the flux variations of C5 using equation (\ref{eq:flux_model}). 
We found that the time origin of the ejection of component C5 is 
$t_{o} \approx 1999.03$. Although equation (\ref{eq:flux_model}) is a rough estimate of the flux 
density variations, it allows us

\begin{itemize}
	\item  to check the time origin of the ejection of the VLBI component found using 
	the BBH model,
	\item to compare the time positon of the modeled first peak flux with the observed first peak flux,
	\item to obtain a good shape of the variation of the flux density during the first few years and 
	explain the difference between the radio and the optical light curves. In some cases, 
	in addition to the radio, optical observations show a light curve with peaks separated 
	by about one year, see for instance the cases of 0420-016 \citep{BrRo+:01} 
	and 3C 345 \citep{LoRo:05}. Using equation (\ref{eq:flux_model}), the optical emission 
	can be modeled as the synchrotron emission of a point source ejected in the perturbed 
	beam \citep{BrRo+:01,LoRo:05}. This short burst of very energetic relativistic $e^{\pm}$ 
	is followed immediately by a very long burst of less energetic relativistic $e^{\pm}$. 
	This long burst is modeled as an extended structure along the beam and is responsible 
	for the VLBI radio emission.
\end{itemize}

\begin{figure}[ht]
\centerline{
\includegraphics[scale=0.5, bb =-275 0 700 650,clip=true]{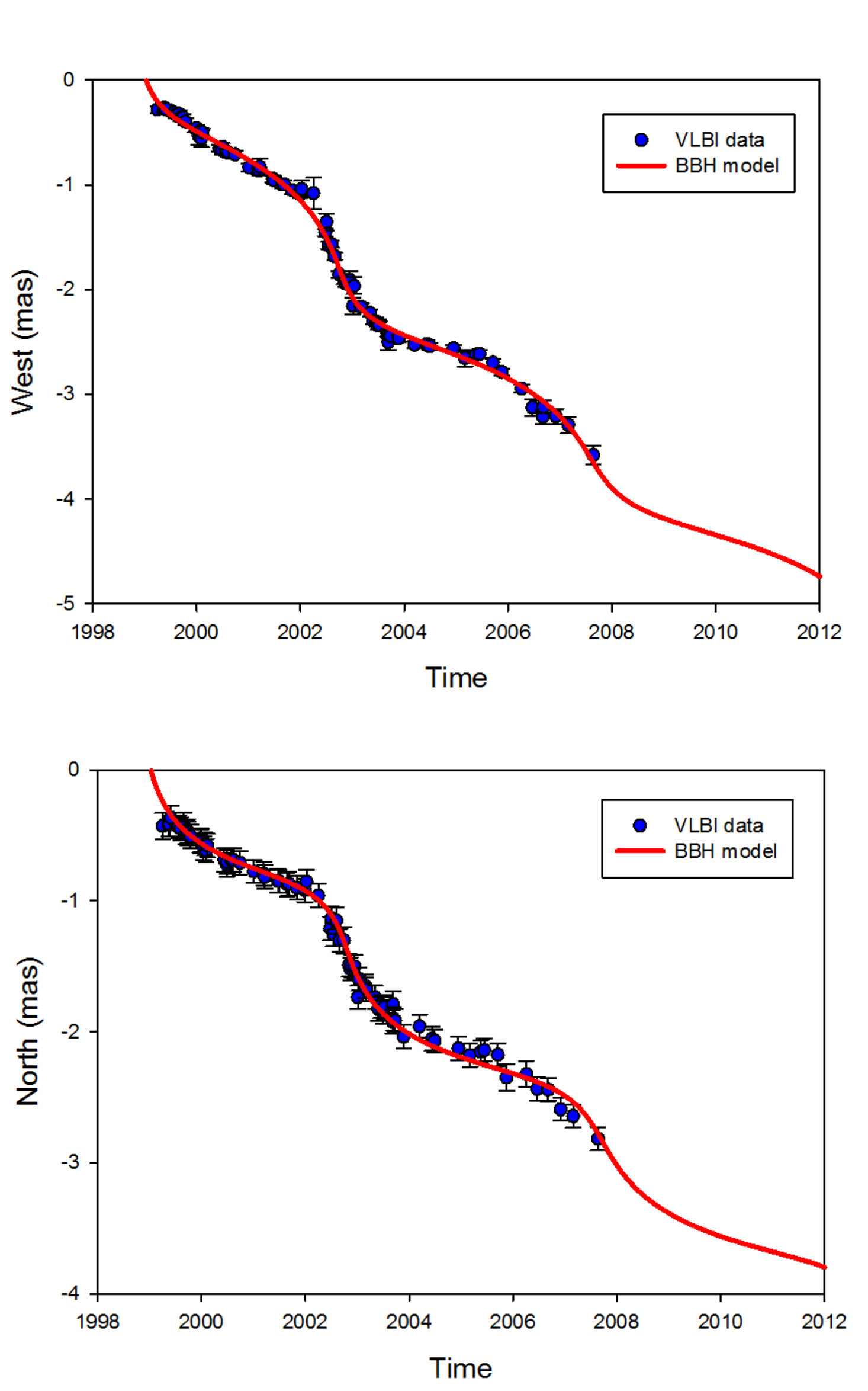}}
\caption{Fit of the two coordinates $W(t)$ and $N(t)$ of component C5 of 3C 279. 
They correspond to the solution with $T_{p}/T_{b} \approx 140$, 
$M_{1}/ M_{2} \approx 2.75$, and $i_{o} \approx 10.4^{\circ}$. 
The points are the observed coordinates of component C5 that were  
corrected for the offsets $\Delta W \approx +405$ $\mu as$ and $\Delta N \approx +110$ $\mu as$.  
VLBI coordinates are taken from \citet{LiAl+:09}. 
The red lines are the coordinates of the component trajectory calculated using the BBH model.}
\label{fig:3C279_C5_Xt+Yt_Ofb12-}
\end{figure}

The fit of both coordinates $W(t)$ and $N(t)$ of component C5 of 3C 279 are 
shown in Figure \ref{fig:3C279_C5_Xt+Yt_Ofb12-}. 
The points are the observed coordinates of component C5 that were corrected 
for the offsets $\Delta W \approx +405$ $\mu as$ and $\Delta N \approx +110$ $\mu as$, 
the red lines are the coordinates of the component trajectory calculated using the BBH model 
assuming the solution parameters, i.e., $T_{p}/T_{b} \approx 140$, $M_{1}/ M_{2} \approx 2.75$, 
 $i_{o} \approx 10.4^{\circ}$, etc.

Finally, we compared this solution with the solution obtained using 
the precession model. The $\chi^{2}_{min}(i_{o})$ is about 151.4 for the fit using 
the BBH system and $> 1000$ for the precession model (see section \ref{sec:Prec_3C279}).
To fit the ejection of component C5 we used 152 observations (76 epochs), so the reduced 
$\chi^{2}$ is $\chi^{2}_{r} = \chi^{2}_{min} /152 \approx 0.996$.

\subsubsection{Determining the family of solutions}

The solution is not unique, but there exists a family of solutions. 
For the inclination angle previously found, i.e., 
$i_{o}\approx 10.4^{\circ}$ and using the parameters of the corresponding solution, 
i.e., $T_{p}/T_{b} \approx 140$, $M_{1}/M_{2} \approx 2.75$ 
and $R_{bin} \approx 420$ $\mu as$, 
we gradually varied $V_{a}$ between $0.01$ c and $0.45$ c. 
The function $\chi^{2}(V_{a})$ remains constant, indicating a degeneracy 
of the solution, and we deduced the range of variation of 
the BBH system parameters. They are given in Table 2.

\begin{center}
Table 2 : Ranges for the BBH system parameters\medskip

\begin{tabular}
[c]{c||c|c}\hline
$V_{a}$                  & $0.01 \; c$                           & $0.45 \; c$                            \\\hline
$T_{p}(V_{a})$           & $\approx 3.12 \; 10^{6}$ yr           & $\approx 38500$ yr                     \\\hline
$T_{b}(V_{a})$           & $\approx 22300$ yr                    & $\approx 275$ yr                       \\\hline
($M_{1}+ M_{2})(V_{a})$  & $\approx 3.2 \; 10^{8}$ $M_{\odot}$   & $\approx 213. \; 10^{10}$ $M_{\odot}$  \\\hline
\end{tabular}
\end{center}

\subsubsection{Determining the size of the accretion disk}

From the knowledge of the mass ratio $M_{1}/M_{2} \approx 2.75$ and 
the ratio $T_{p}/T_{b} \approx 140$, we calculated in the previous section 
the mass of the ejecting black hole $M_{1}$, the orbital period $T_{b}$, 
and the precession period $T_{p}$ for each value of $V_{a}$.

We calculated the rotation period of the accretion disk, $T_{disk}$, using (\ref{eq:Tdisk}). 
Assuming that the mass of the accretion disk is $M_{disk} \ll M_{1}$, 
the size of the accretion disk $R_{disk}$ is calculated using (\ref{eq:Rdisk}).

We found that the size of the accretion disk does not depend on $V_{a}$ and is 
$R_{disk}  \approx 0.26 \; pc \approx 0.041 \; mas$.

\subsubsection{Comparing of the trajectories of C5 and C10}

We see from figure \ref{fig:3C279_sep_time} that
\begin{itemize}
	\item components C5 and C6 probably follow probably the same trajectories,
	\item component C10 follows a different trajectory than C5 and C6.
\end{itemize}

\begin{figure}[ht]
\centerline{
\includegraphics[scale=0.5, width=8cm,height=6cm]{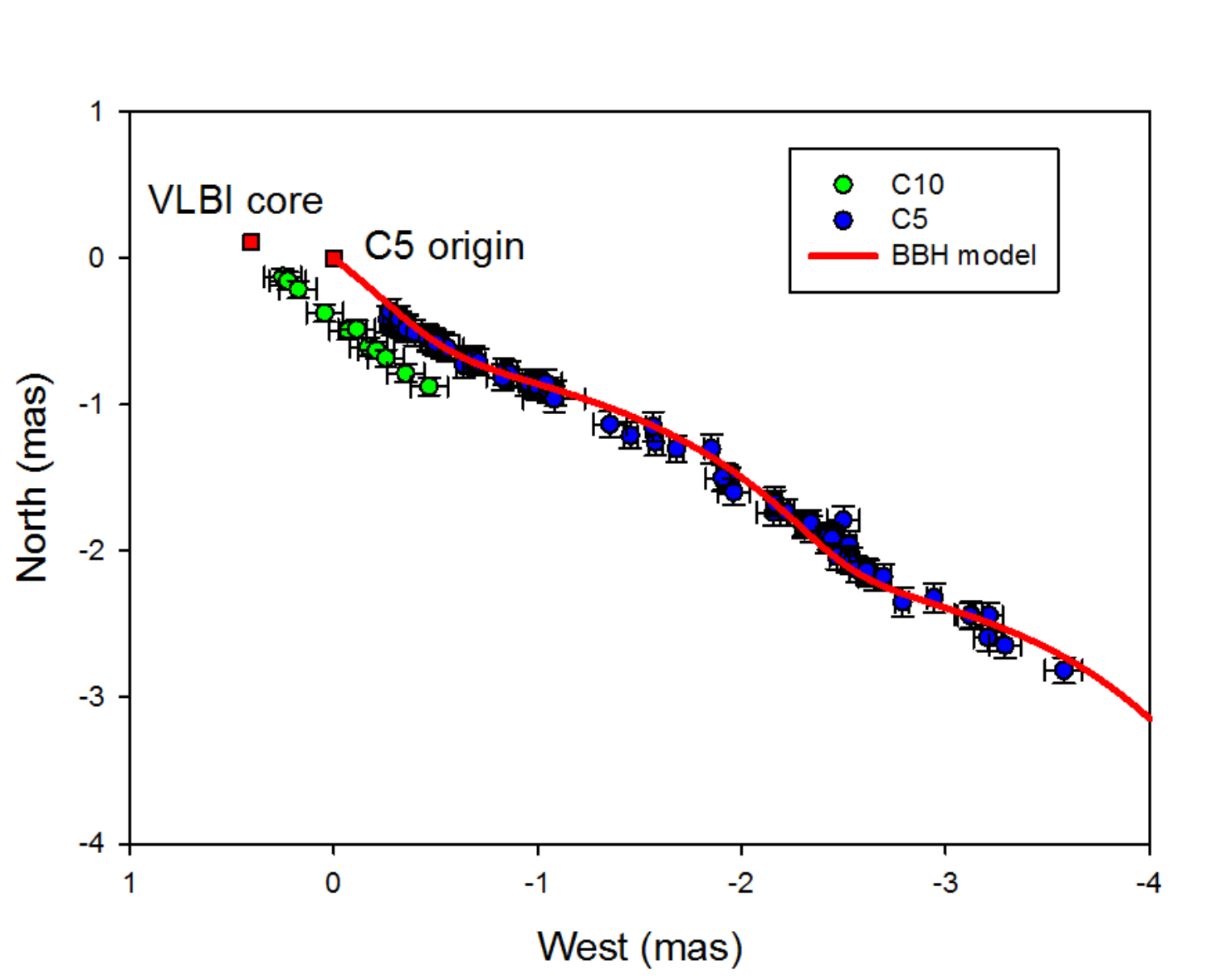}}
\caption{Using the MOJAVE data \citep{LiCo+:09}, we plot the trajectories of C5 and C10. 
Component C10 is probably ejected by the VLBI core and component C5 is ejected 
with a large offset from the VLBI core. Components C5 and C10 follow two different 
trajectories and are ejected from different origins, indicating that the 
nucleus of 3C 279 contains a BBH system. Note that the origin of this caption corresponds to the origin 
of the ejection of component C5, thus all MOJAVE coordinates have been corrected 
for the offsets $\Delta W \approx +405$ $\mu as$ and $\Delta N \approx +110$ $\mu as$.}
\label{fig:3C279_C5+C10_Trajectories}
\end{figure}

Thus, using the MOJAVE data \citep{LiCo+:09}, we plot in 
figure \ref{fig:3C279_C5+C10_Trajectories} the trajectories of C5 and C10. 
We found that
\begin{itemize}
	\item component C10 is probably ejected by the VLBI core, 
	\item component C5 is ejected with a large offset from the VLBI core, and
	\item components C5 and C10 follow two different trajectories and are not ejected 
	from the same origins, indicating that the nucleus of 3C 279 contains a BBH system.
\end{itemize}

\section{Discussion and conclusion}

We show how from the knowledge of the coordinates 
$West(t)$ and $North(t)$ of the ejected VLBI component one can find the characteristics 
of the BBH system in both cases. To illustrate case I, we fitted component S1 of 
1823+568, and to illustrate case II, we fitted component C5 of 3C 279. 

From the fit of the coordinates of component S1 of 1823+568, 
the main characteristics of the final solution of the BBH system 
associated with 1823+568 are that
\begin{itemize}
    \item the radius of the BBH system is $R_{bin} \approx 60$ $\mu as$ $\approx 0.42$ $pc$,
    \item the VLBI component S1 is not ejected by the VLBI core, and the offsets  
    of the observed coordinates are $\Delta W \approx +5$ $\mu as$ 
    and $\Delta N \approx 60$ $\mu as$,
    \item the ratio $T_{p}/T_{b}$ is $8.88 \leq T_{p}/T_{b} \leq 9.88$,
    \item the ratio $M_{1}/M_{2}$ is $0.095 \leq M_{1}/M_{2} \leq 0.25$,
		\item the inclination angle is $i_{o} \approx 4.0^{\circ}$,
		\item the bulk Lorentz factor of the VLBI component is $\gamma_{c} \approx 17.7$, and
		\item the origin of the ejection of the VLBI component is $t_{o} \approx 1995.7$.
\end{itemize}

From the fit of the coordinates of component C5 of 3C 279, 
the main characteristics of the final solution of the BBH system 
associated with 3C 279 are that
\begin{itemize}
    \item the radius of the BBH system is $R_{bin} \approx 420$ $\mu as$ $\approx 2.7$ $pc$,
    \item the VLBI component C5 is not ejected by the VLBI core and the offsets  
    of the observed coordinates are $\Delta W \approx +405$ $\mu as$ 
    and $\Delta N \approx +110$ $\mu as$,
    \item the ratio $T_{p}/T_{b}$ is $T_{p}/T_{b} \approx 140$,
    \item the ratio $M_{1}/M_{2}$ is $M_{1}/M_{2} \approx 2.75$,
		\item the inclination angle is $i_{o} \approx 10.4^{\circ}$,
		\item the bulk Lorentz factor of the VLBI component is $\gamma_{c} \approx 16.7$, and
		\item the origin of the ejection of the VLBI component is $t_{o} \approx 1999.0$.
\end{itemize}

If, in addition to the radio observations, one can obtain optical, X-ray, or 
$\gamma$-ray observations that show a light curve with peaks, 
the simultaneous fit of the VLBI coordinates and this light curve put 
stronger constraints on the characteristics of the BBH system. 
The high-frequency emission can be modeled as the synchrotron emission or the 
inverse Compton emission of a point source ejected in the perturbed beam, 
see \citet{BrRo+:01} for PKS 0420-014 and 
\citet{LoRo:05} for 3C 345. This short burst of very energetic
relativistic $e^{\pm}$ is followed immediately by a very long burst
of less energetic relativistic $e^{\pm}$. This long burst is 
modeled as an extended structure along the beam and is responsible for
the VLBI radio emission. 
The simultaneous fit of the VLBI coordinates and the optical light curve 
using the same method as the one developed in this article has to be achieved. 

Observations of compact radio sources in the first $mas$ show that the VLBI 
ejections do not follow a straight line, and modeling the ejection 
shows in each case studied that the nucleus contains a BBH system. 
Accordingly, \citet{BrRo+:01} assumed that all radio sources contain a BBH system. 
If extragalactic radio sources are associated with galaxies formed after the
merging of galaxies and if the formation of extragalactic
radio sources is related to the presence of binary black
hole systems in their nuclei, we can explain

\begin{itemize}
	\item why extragalactic radio sources are associated with elliptical galaxies, 
	\item why more than 90\% of the quasars are radio-quiet quasars, e.g., 
	\citet{KeSr+:89} and \citet{MiPe+:90}.
\end{itemize}

Radio-quiet quasars are active nuclei that contain a single
black hole and can be associated with spiral galaxies
\citep{PeMi+:86}. Although it has not been proven
yet that radio-quiet quasars only contain a single black
hole, the hypothesis for distinguishing between radio-loud
and radio-quiet quasars on the basis of the binarity of
the central engine is supported by comparing 
the optical properties of the two classes \citep{GoKu+:99}. 
Recent observations of the central parts of radio galaxies 
and radio-quiet galaxies show a systematic difference between 
the two classes \citep{KhCa+:12}.

Because GAIA will provide positions of extragalactic radio sources within 
$\approx 25$ $\mu as$, the link between the GAIA reference frame from 
optical observations of extragalactic radio sources and the reference 
frame obtained from VLBI observations will have to take into account 
the complex structure of the nuclei of extragalactic radio sources, because 
with a resolution of $\approx 25$ $\mu as$, probably all these sources 
will appear as double sources, and the radio core, obtained from VLBI 
observations and the optical core obtained by GAIA will not necessarily be 
 the same.

We conclude, remarking that if the inner parts of the accretion disk 
contain a warp  or precess faster than the precession of the outer part, 
this will produce a very small perturbation that will produce a day-to-month 
variability of the core flux \citep{RoBi+:09}.

\begin{acknowledgements}
JR thanks Alain Lecavelier des Etangs and Simon Prunet for useful 
discussions and comments. This research has made use of data from the MOJAVE 
database that is maintained by the MOJAVE team \citep{LiAl+:09} and part of 
this work was supported by the COST Action MP0905 Black Holes in a Violent Universe.

\end{acknowledgements}

\bibliographystyle{aa} 
\bibliography{JRoland_BIB}

\newpage
\appendix
\section{Fit of component S1 of 1823+568}
\label{Fit_1823+568}
\subsection{Fit of S1 using the precession model}
\label{sec:Prec_1823+568}

To fit the ejection of component S1, we used 56 observations (28 epochs).

We studied the two cases $\pm \omega_{p}(t - z/V_{a})$. The final solution 
of the fit of component S1 of 1823+568 using a BBH system corresponds 
to $+\omega_{p}(t - z/V_{a})$, therefore we discuss only this case in this appendix. 
In this section, we assume that $V_{a} = 0.1$ c.

The range of inclination we explore is $0.5^{\circ} \leq i_{o} \leq 10^{\circ}$.  

\begin{figure}[ht]
\centerline{
\includegraphics[scale=0.5, width=8cm,height=6cm]{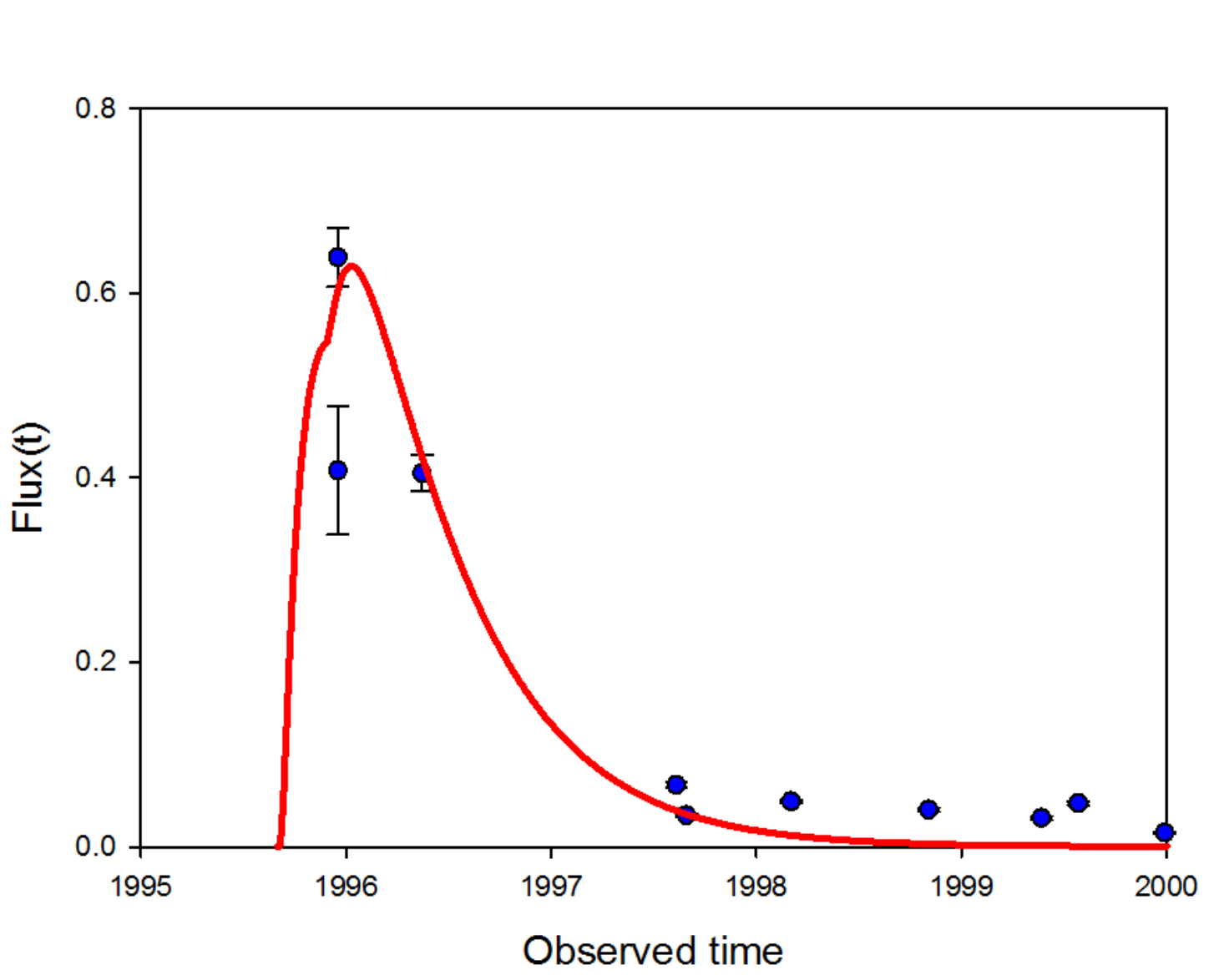}}
\caption{Fit of the first peak flux of component S1 of 1823+568 using the precession model. 
The time origin deduced from the fit of the peak flux is $t_{o} \geq 1995.65$ and is comparable with the time 
origin deduced from VLBI observations interpolation, i.e., $t_{o} \approx 1995.60$.}
\label{fig:Flux_1823_S1_Precession}
\end{figure}

An important parameter for the fit  is the time origin 
of the ejection of the VLBI component. We model the flux using equation

\begin{eqnarray}
S_{c} & = & \delta_{c}(t)^{2+\alpha_{r}}/ Fac \nonumber \\
        &&{} * \exp(- T_{opacity} /t) * \exp(-t /T_{decay}) \ ,
\label{eq:flux_model}
\end{eqnarray}
where $Fac$ is a scaling factor, $T_{opacity}$ is the characteristic time 
to describe the synchrotron opacity, and $T_{decay}$ is the characteristic 
time to describe the losses. This is the simplest way to 
model the flux and does not take into account in situ re-acceleration of 
the relativistic particles along the beam and synchrotron, inverse Compton, 
or expansion losses. We do not aim to fit the flux light curve, 
but we wish to compare the time position of the modeled first peak flux 
with the observed first peak flux (Figure \ref{fig:Flux_1823_S1_Precession}). 
This provides the minimum for the time origin of the ejection of the 
VLBI component. For S1 we find $t_{o}\geq 1995.65$. 
This value agrees well with the time origin obtained from VLBI data 
interpolation, which is $t_{o} \approx 1995.60$ \citep{Gl:10}. This case corresponds to case I, 
i.e., if there is an offset of the VLBI ejection, it is smaller than or on the order of 
the smallest error bars of the VLBI component coordinates. 
In the following we keep $t_{o}$ as a free parameter in the range 
$1995.65 \leq t_{o} \leq 1995.90$.

Because the function $\chi^{2}_{t}(i_{o})$ is mostly flat between 4 and 10 degrees, 
to continue we abitrarily adopted the inclination angle $i_{o} \approx 6^{\circ}$. 
The main results of the fit for the precession model are that
\begin{enumerate}
    \item the opening angle fo the precession cone is $\Omega \approx 0.46^{\circ}$,
    \item the bulk Lorentz factor of S1 is $\gamma_{c} \approx 20$,
    \item the origin of S1 is $t_{o} \approx 1995.7$, and
    \item $\chi^{2}(i_{o}\approx 6^{\circ}) \approx 67.4$.
\end{enumerate}

\subsection{Determining the family of solutions}
The solution is not unique. 
For the inclination angle previously found, i.e., 
$i_{o}\approx 6^{\circ}$ and using the parameters of the corresponding solution,
we gradually varied $V_{a}$ between $0.01$ c and $0.45$ c. 
At each step of $V_{a}$, we minimized the function 
$\chi^{2}_{t}(\lambda)$, where $\lambda$ are the free parameters. 
The function $\chi^{2}(V_{a})$ remained constant, indicating a degeneracy 
of the solution, and we obtained the range of possible values for 
the precession period given in Table 3.

\begin{center}
Table 3 : Range for the precession period \medskip

\begin{tabular}
[c]{c||c|c}\hline
$V_{a}$            & $0.01 \; c$           & $0.45 \; c$       \\\hline
$T_{p}(V_{a})$     & $\approx 570000$ yr   & $\approx 7700$ yr \\\hline
\end{tabular}
\end{center}

\subsection{Determining the BBH system parameters}
\label{sec:BBHparam1823}
Because the precession is defined by $+\omega_{p}(t - z/V_{a})$, the 
BBH system rotation is defined by $-\omega_{b}(t - z/V_{a})$.
In this section, we kept the inclination angle previously found, 
i.e., $i_{o} \approx 6^{\circ}$ and $V_{a} = 0.1$ c.

To determine the BBH system parameters corresponding to a value of 
$T_{p}/T_{b}$, we minimized $\chi^{2}_{t}(M_{1})$ 
when the mass of the ejecting black hole $M_{1}$ varied gradually between 
1 $M_{\odot}$ to a value corresponding to $M_{1} /M_{2} = 2$ with a starting 
value of $M_{2}$, such that $10^{6} \leq M_{2} \leq 10^{9}$. 
During the minimization $M_{2}$ is a free parameter, 
and at each step of $M_{1}$, we minimized the function 
$\chi^{2}_{t}(\lambda)$, where $\lambda$ are the free parameters. 
Thus we constrained the parameters of the BBH system 
when the two black holes have the same masses, i.e., $M_{1} = M_{2}$.  

We determined the parameters of the BBH system 
model for different values of the parameter $ T_{p}/T_{b}$, 
namely $ T_{p}/T_{b} = 4.6$, $10$, $22$, $46$, $100$, and $220$. 

For a given value of $T_{p}/T_{b}$, we found the radius of the BBH system 
defined by Equation (\ref{eq:Rbin_mas}). Note that the radius of the BBH 
system does not depend on the starting value of $M_{2}$.

\subsection{$\chi^{2}(T_{p}/T_{b})$ - diagram}
\label{sec:Chi2TpTb1823}

In this section, we kept the inclination angle previously found, 
i.e., $i_{o} \approx 6^{\circ}$, $V_{a} = 0.1$ c and assumed $M_{1} = M_{2}$.

The diagram $\chi^{2}(T_{p}/T_{b})$ provides the 
possible solutions at a given inclination angle. Some of the solutions can 
be mirage solutions when $i_{o}$  varies.

We calculated $\chi^{2}(T_{p}/T_{b})$ for $1 \leq T_{p}/T_{b} \leq 300$. 
We started for each value of the BBH system parameters found in the previous section, i.e., 
corresponding to the values of $ T_{p}/T_{b} = 4.6$, $10$, $22$, $46$, 
$100$, and $220$, and covered the complete interval $1 \leq T_{p}/T_{b} \leq 300$. 
For instance, if we started at $T_{p}/T_{b} = 22$, 
we covered the ranges varying $T_{p}/T_{b}$ from 22 to 1 and from 22 to 300.
We found the possible solutions of the BBH system, i.e,. the solutions that correspond 
to the minima of $\chi^{2}(T_{p}/T_{b})$. They are given in Table 4.

\begin{center}
Table 4 : Main Solutions found for $i_{o} \approx 5.98^{\circ}$ \medskip

\begin{tabular}
[c]{l||l|l|l}\hline
Solution      & $ (T_{p}/T_{b})_{min} $  & $\chi^{2}(min)$   &Remark                \\\hline
Sol 1         & $\approx 3.3$            & $\approx 50.8$    &$\gamma_{c} > 30$     \\\hline
Sol 2         & $\approx 11.45$          & $\approx 53.7$    &                      \\\hline
Sol 3         & $\approx 23.0$           & $\approx 63.9$    &                      \\\hline
Sol 4         & $\approx 76$             & $\approx 63.6$    &                      \\\hline
Sol 5         & $\approx 109$            & $\approx 62.2$    &$\gamma_{c} > 30$     \\\hline
\end{tabular}
\end{center}

Solutions 1 and 5 are  excluded because they have $\gamma_{c} > 30$ . There are three possible 
solutions, the best one is Solution 2, which corresponds to a BBH system whose radius is 
$R_{bin} \approx 60$ $\mu as$ . In the following, we continue with Solution 2.

\subsection{Possible offset of the origin of the ejection}
In this section, we kept the inclination angle previously adopted, 
i.e., $i_{o} \approx 6^{\circ}$. We assumed that $V_{a} = 0.1$ c, $M_{1} = M_{2}$, 
$T_{p}/T_{b} = 11.45$ and used the parameters of Solution 2 previously found.

\begin{figure}[ht]
\centerline{
\includegraphics[scale=0.5, width=8cm,height=8cm]{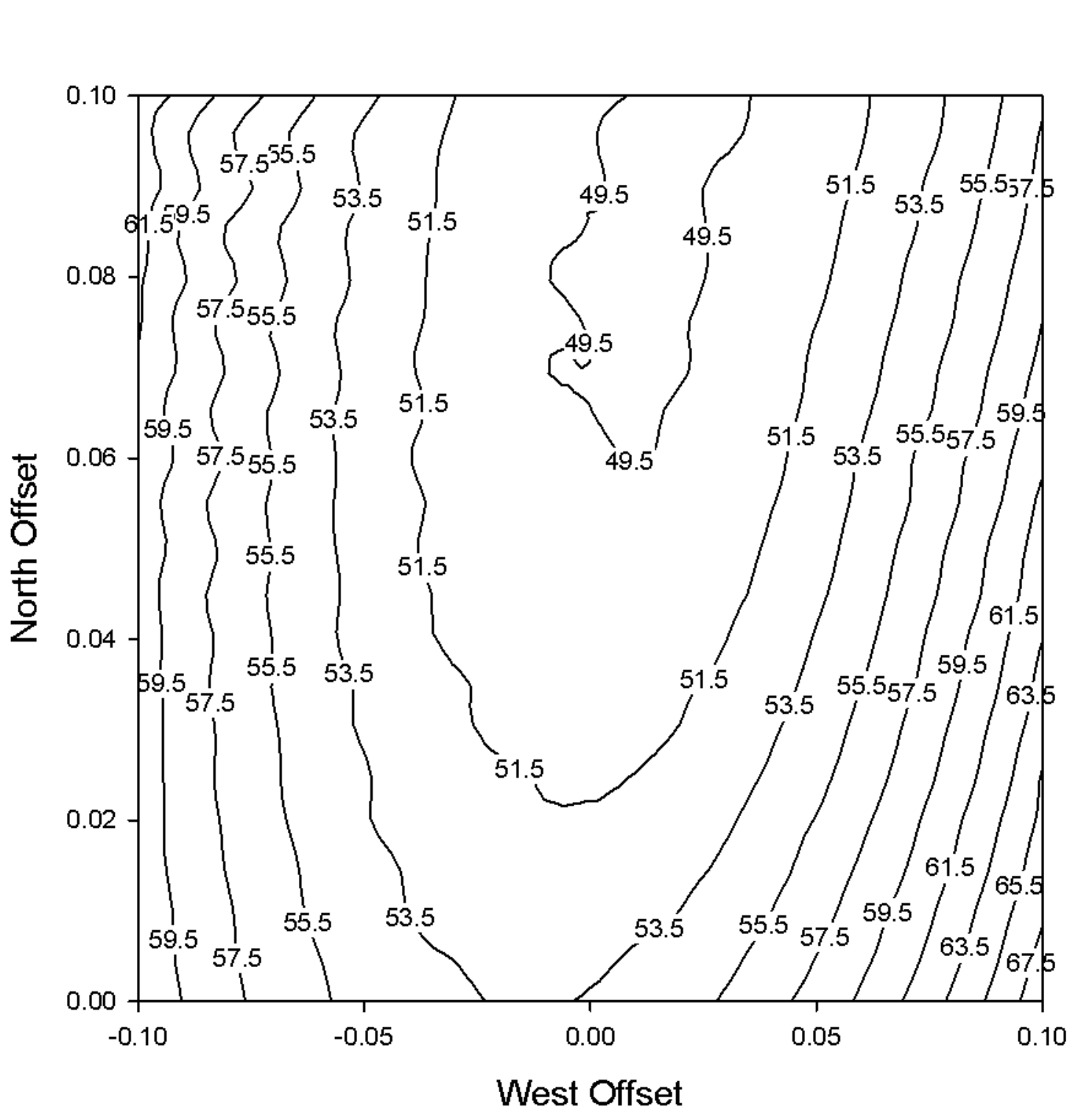}}
\caption{Calculation of $\chi^{2}(\Delta W, \Delta N)$ using the BBH model. 
Non-zero offsets are possible, but the size of the offset 
must be the same as the radius of the BBH system calculated at this point. 
This is the case if the offsets are $\Delta W_{1} \approx 0.010$ $mas$ and 
$\Delta N_{1} \approx 0.070$ $mas$. We determined the offset 
 at $i_{o} \approx 6^{\circ}$; it does not depend on the value of the adopted inclination angle.}
\label{fig:Chi2_Off_XY_1823}
\end{figure}

To test whether the VLBI component is ejected from the VLBI core 
or from the second black hole, we calculated $\chi^{2}(\Delta W, \Delta N)$,
where $\Delta W$ and $\Delta N$ are offsets in the west and north directions. 
The step used in the west and north directions is $5 \; \mu as$. 
At each step of $\Delta W$ and $\Delta N$, we minimized the function 
$\chi^{2}_{t}(\lambda)$, where $\lambda$ are the free parameters 
(Figure \ref{fig:Chi2_Off_XY_1823}). The radius of the BBH system was left  
free to vary during the minimization.\\

The minimum of $\chi^{2}(\Delta W, \Delta N)$ is $\approx 49.5$, and we see 
from Figure \ref{fig:Chi2_Off_XY_1823} that the corresponding non-zero offsets are 
with $\Delta N \geq 0.060$ $mas$. However, all points 
with the smallest $\chi^{2}(\Delta W, \Delta N)$ are not possible. 
Indeed, for a point with the smallest $\chi^{2}$, the size of the offset
offset must be equal to the radius of the BBH system calculated at this point. 
This is the case if the offsets are $\Delta W_{1} \approx +0.010$ $mas$ and 
$\Delta N_{1} \approx +0.070$ $mas$.
The radius of the BBH system at this point is $R_{bin} \approx 70$ $\mu as$ 
and the offset size is $\approx 71$ $\mu as$, i.e., the offset and the 
radius of the BBH system are the same at this point. \\

Therefore we conclude that 
\begin{itemize}
  \item the VLBI component S1 is not ejected from the VLBI core, but from 
  the second black hole of the BBH system,
  \item the radius of the BBH system is $R_{bin} \approx 71$ $\mu as$. 
  It is about twice the smallest error bars of the observed VLBI component 
	coordinates (the component positions), but it is significantly detected 
	($2 \; \sigma$ from Figure \ref{fig:Chi2_Off_XY_1823}).
\end{itemize}

We must correct the VLBI coordinates from the offset 
before we continue.

Note that determining the offset of the origin does 
not depend on the value adopted for the inclination angle. This was shown by 
calculating the offset at different inclination angles, i.e., $i_{o} \approx 4^{\circ}$, 
$5^{\circ}$, $7^{\circ}$.

\subsection{Determining $T_{p}/T_{b}$}
From this point onward, the original coordinates of the VLBI component S1 are corrected 
for the offsets $\Delta W_{1}$ and $\Delta N_{1}$ found in the previous section. 
In this section, we assumed that $R_{bin} = 71$ $\mu as$, $M_{1} = M_{2}$ and 
$V_{a} = 0.1$ c. 

Previously, we found that Solution 2 is characterized by $T_{p}/T_{b} \approx 11.45$ 
for $i_{o} \approx 6^{\circ}$. In this section we obtain the 
range of possible values of $T_{p}/T_{b}$ when $i_{o}$ varies.

\begin{figure}[ht]
\centerline{
\includegraphics[scale=0.5, width=8cm,height=6cm]{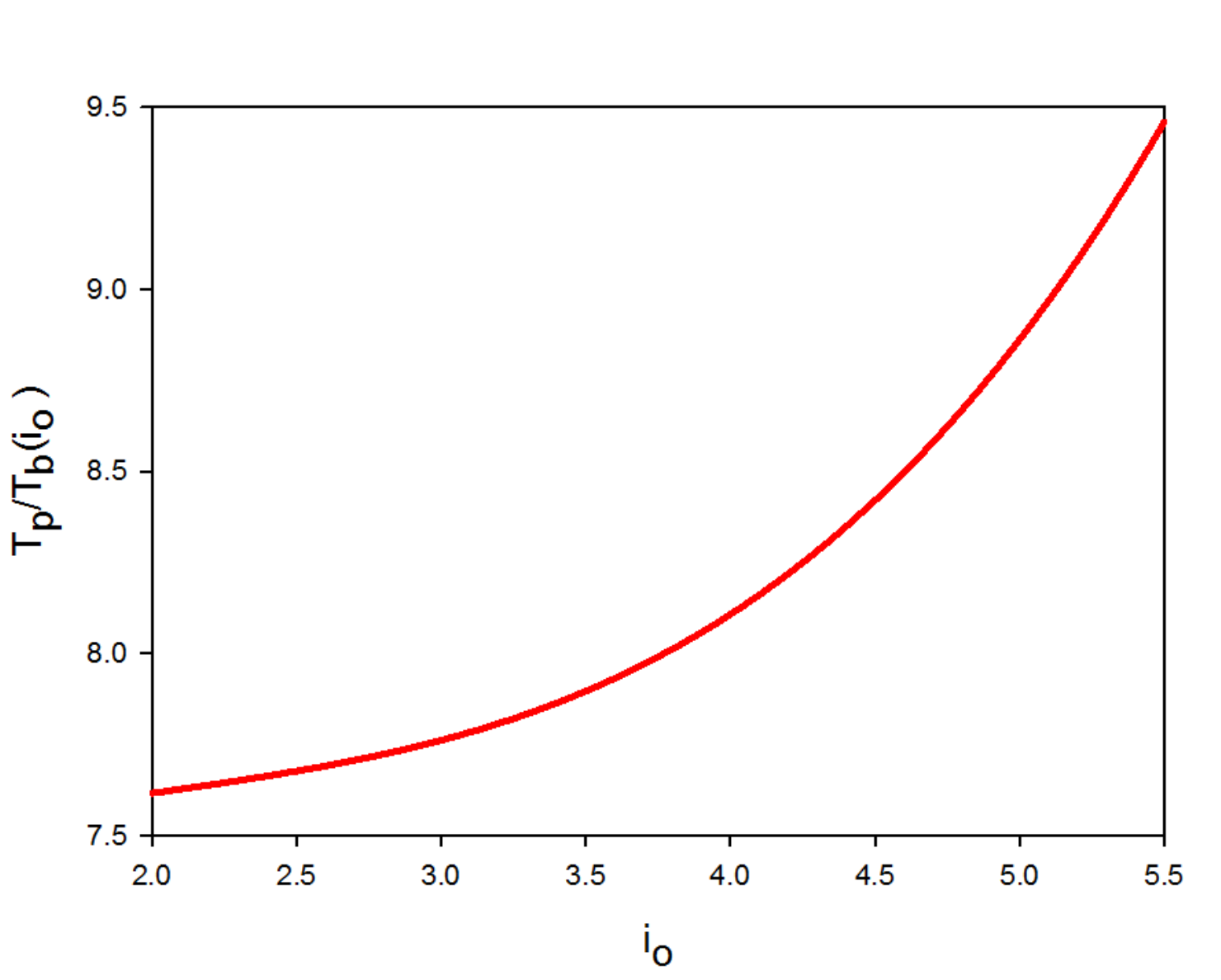}}
\caption{$T_{p}/T_{b}$ as function of $i_{o}$ obtained by minimizing $\chi^{2}(i_{o})$.}
\label{fig:TpTb_io_1823_Rb071}
\end{figure}

We calculated the funtion $\chi^{2}(i_{o})$  
in the inteval $2^{\circ} \leq i_{o} \leq 7^{\circ}$, assuming that the ratio 
$T_{p}/T_{b}$ is free. The relation between $T_{p}/T_{b}$ and $i_{o}$ 
is plotted in Figure \ref{fig:TpTb_io_1823_Rb071}.

Knowing of the possible values of the ratio $T_{p}/T_{b}$ allows us to 
calculate in the next section the function $\chi^{2}(i_{o})$ for various values 
of the ratios $T_{p}/T_{b}$ and $M_{1}/ M_{2}$ and then estimate the  
mass ratio $M_{1}/ M_{2}$ of the BBH system.

\subsection{Preliminary determination of $i_{o}$, $T_{p}/T_{b}$ and $M_{1}/M_{2}$.}
\label{sec:Mass_ratio_case_I}

In this section, we assumed that $V_{a} = 0.1$ c and the radius of the BBH system 
is $R_{bin} = 71$ $\mu as$.

\begin{figure}[ht]
\centerline{
\includegraphics[scale=0.5, bb =-275 0 700 660,clip=true]{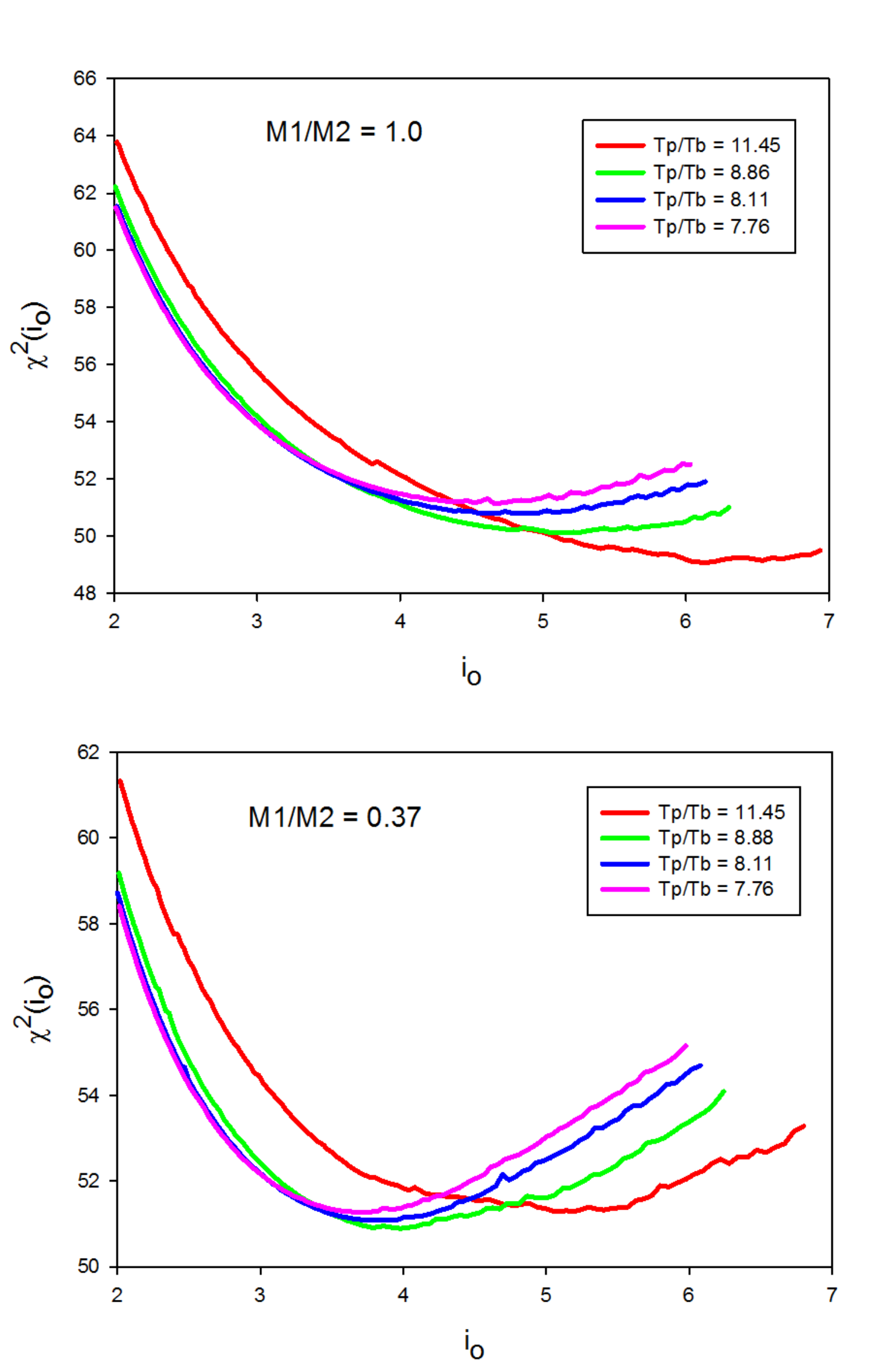}}
\caption{Function $\chi^{2}(i_{o})$. It stops at large inclination angles 
because the bulk Lorentz factor becomes greater than 30. 
\textit{Top figure:} The ratio $M_{1}/ M_{2}$ is $M_{1}/ M_{2} = 1$. When $T_{p}/T_{b}$ 
increases, the minimum decreases and disappears. This means that the function $\chi^{2}(i_{o})$ does not 
show a minimum and is a mirage solution. 
\textit{Bottom figure:} The ratio $M_{1}/ M_{2}$ $ = 0.37$. The function 
$\chi^{2}(i_{o})$ has a minimum for $T_{p}/T_{b} \approx 8.88$ and $i_{o} \approx 4$. The 
robustness of this solution, defined as the square root of the difference $\chi^{2}(\gamma = 30) -  \chi^{2}(min)$, 
is $\approx 1.8 \times \sigma$.}
\label{fig:Chi2_io_OS_Rb071_TpTb=f_M1M2=037+100}
\end{figure}

We varied $i_{o}$  between 2 and 7 degrees and calculated $\chi^{2}(i_{o})$ for 
various values of $T_{p}/T_{b}$ and $M_{1}/ M_{2}$. The values of  $T_{p}/T_{b}$ 
investigated are $T_{p}/T_{b}$ = 11.45, 8.88, 8.11, 7.76, and 7.62. The values of 
$M_{1}/ M_{2}$ investigated are $M_{1}/ M_{2}$= 1.0, 0.5, 0.37, 0.25, and 0.1. For each 
value of $M_{1}/ M_{2}$ we calculated $\chi^{2}(i_{o})$ for all values of $T_{p}/T_{b}$. 
The plots $\chi^{2}(i_{o})$ for $M_{1}/ M_{2}$= 1.0 and 0.37 are shown in Figure 
\ref{fig:Chi2_io_OS_Rb071_TpTb=f_M1M2=037+100}.\\

The main results are that
\begin{itemize}
  \item when $i_{o}$ is larger than about 6 degrees, the bulk Lorentz factor increases and 
  becomes greater than 30, which is excluded, 
  \item the critical value of $M_{1}/ M_{2}$  $ \approx 0.5$,
  \item if $M_{1}/ M_{2} > 0.5$, the solution $\chi^{2}(i_{o})$ is a mirage solution, 
  \item if $M_{1}/ M_{2} < 0.5$, the solution $\chi^{2}(i_{o})$ has a minimum,
  \item the solutions with a robustness larger than 1.7 $\sigma$ are those with 
  $M_{1}/ M_{2} < 0.37$ (see Table 8),
  \item when $M_{1}/ M_{2}$ decreases, the solutions are more robust, but they are of 
  lower quality, i.e., their $\chi^{2}(min)$ increases (see Table 8), and
  \item when $M_{1}/ M_{2} < 0.5$, the value of $T_{p}/T_{b}$ that produces the best fit 
  is $T_{p}/T_{b} \approx 8.88$, independently of the value of $M_{1}/ M_{2}$.
\end{itemize}

We present in Table 5 the results of solutions corresponding to $T_{p}/T_{b} \approx 8.88$ 
and $M_{1}/ M_{2} =$ 0.1, 0.25 and 0.37.\\

\begin{center}
Table 5 : Solutions found for $T_{p}/T_{b} \approx 8.88$ \medskip

\begin{tabular}
[c]{l||l|l|l}\hline
$M_{1}/ M_{2}$           & 0.1                      & 0.25                     & 0.37                     \\\hline
$i_{o}$                  & $\approx 4.10^{\circ}$   & $\approx 3.64^{\circ}$   & $\approx 3.98^{\circ}$   \\\hline
$\chi^{2}(min)$          & $\approx 53.7$           & $\approx 51.5$           & $\approx 50.9$           \\\hline
Robustness($i_{o}$)      & $\approx 2.4 \; \sigma$  & $\approx 1.9 \; \sigma$  & $\approx 1.8 \; \sigma$  \\\hline
\end{tabular}
\end{center}

\subsection{Determining a possible new offset correction}

In this section, we assumed $V_{a} = 0.1$ c. Using the solution found 
in the previous section (Table 8), we can verify whether if 
there is an additional correction to the offset of the origin of the VLBI 
component. For this, we calculated $\chi^{2}(\Delta W, \Delta N)$,
where $\Delta W$ and $\Delta N$ are offsets in the west and north directions. 
We assumed the radius of the BBH system to be free to vary. We found 
that a small additional correction is needed $\Delta W_{2} \approx - 0.005$ $mas$ 
and $\Delta N_{2} \approx -0.010$ $mas$.

Finally, we found that the total offset is $\approx 60$ $\mu as$ and the radius 
of the BBH system is also $R_{bin} \approx 60$ $\mu as$. 

\subsection{Final fit of component S1 of 1823+568}
\label{sec:Final_Fit_1823+568}
From this point onward, the coordinates of the VLBI component S1 are corrected 
for the new offsets $\Delta W_{2}$ and $\Delta N_{2}$ found in the previous section. 
In this section, we assumed $V_{a} = 0.1$ c and $R_{bin} = 60$ $\mu as$. 

We can now find the final solution for S1. We calculated $\chi^{2}(i_{o})$ 
for various values of $T_{p}/T_{b}$ assuming $M_{1}/ M_{2} \approx 0.25$. 
We found that the best range for $T_{p}/T_{b}$ is: 
$8.88 \leq T_{p}/T_{b} \leq 9.88$. With this we can estimate 
the range of the mass ratio 
assuming $T_{p}/T_{b} \approx 8.88$ and $T_{p}/T_{b} \approx 9.88$. 
We defined the range of the mass ratio in the following way:
\begin{enumerate}
    \item we found the mass ratio that produces a solution of at least 
    1.7 $\sigma$ robustness, and
    \item we found the mass ratio that produces a solution that is poorer by 1 $\sigma$ 
    than the previous one, but that is more robust.
\end{enumerate}

The results of the fit are presented in Tables 6 and 7. The improvement of the 
solutions of Tables 6 and 7 compared to the solutions of Table 5 is due to the 
new offset and the new value of the BBH system radius.

\begin{center}
Table 6 : The range of $M_{1}/ M_{2}$ when $T_{p}/T_{b} \approx 8.88$ \medskip

\begin{tabular}
[c]{l||l|l|l}\hline
$M_{1}/ M_{2}$           & 0.09                     & 0.17                     & 0.29                     \\\hline
$i_{o}$                  & $\approx 3.79^{\circ}$   & $\approx 3.98^{\circ}$   & $\approx 4.27^{\circ}$   \\\hline
$\chi^{2}(min)$          & $\approx 51.7$           & $\approx 51.2$           & $\approx 50.7$           \\\hline
Robustness($i_{o}$)      & $\approx 2.4 \; \sigma$  & $\approx 2.2 \; \sigma$  & $\approx 1.8 \; \sigma$  \\\hline
\end{tabular}
\end{center}

\begin{center}
Table 7 : The range of $M_{1}/ M_{2}$ when $T_{p}/T_{b} \approx 9.88$ \medskip

\begin{tabular}
[c]{l||l|l|l}\hline
$M_{1}/ M_{2}$           & 0.095                    & 0.16                     & 0.25                       \\\hline
$i_{o}$                  & $\approx 4.06^{\circ}$   & $\approx 4.22^{\circ}$   & $\approx 4.53^{\circ}$     \\\hline
$\chi^{2}(min)$          & $\approx 51.7$           & $\approx 51.2$           & $\approx 50.7$             \\\hline
Robustness($i_{o}$)      & $\approx 2.2 \; \sigma$  & $\approx 1.8 \; \sigma$  & $\approx 1.7   \; \sigma$  \\\hline
\end{tabular}
\end{center}

We see that the solutions found with $T_{p}/T_{b} \approx 8.88$ are 
slightly more robust, but both solutions can be used.

The characteristics of the final solution of the BBH system 
associated with 1823+568 are given in section \ref{sec:solution_1823+568}.

\section{Fit of component C5 of 3C 279}
\label{Fit_3C279}
\subsection{Fit of C5 using the precession model }
\label{sec:Prec_3C279}

To fit the ejection of component C5 we used 152 observations (76 epochs).

We studied the two cases $\pm \omega_{p}(t - z/V_{a})$. The final solution 
of the fit of component C5 of 3C 279 using a BBH system corresponds 
to $-\omega_{p}(t - z/V_{a})$, therefore we discuss only this case in this appendix. 

To fit the component C5, we assumed $T_{d} \leq 2000$ yr.

In this section, we assume that $V_{a} = 0.1$ c.

The range of inclination explored is $0.5^{\circ} \leq i_{o} \leq 10^{\circ}$.

\begin{figure}[ht]
\centerline{
\includegraphics[scale=0.5, width=8cm,height=6cm]{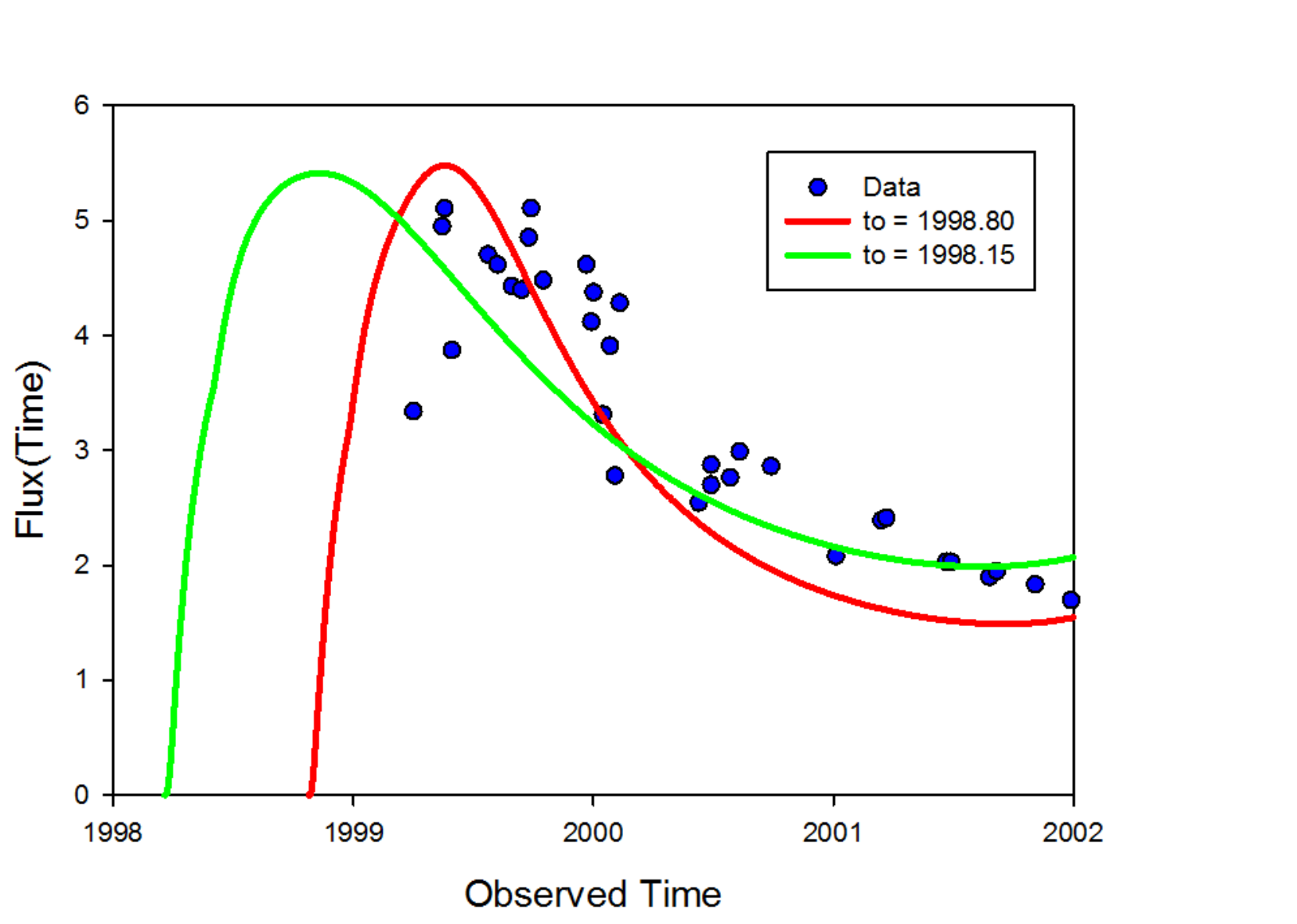}}
\caption{Determination of the minimum time for the ejection of component C5.
We fit the first peak flux of component C5 of 3C 279 
using the precession model. The peak flux corresponding to an ejection 
at $t_{o} \approx 1998.15$ is shown (green line), it is too early by at least 8 months.
The precession model can fit the position of the first peak flux for $t_{o} \geq 1998.80$ 
(red curve). }
\label{fig:C5_3C279_Flux_Precession}
\end{figure}

To begin, we allowed the time origin of the VLBI component to be a free parameter. 
We assumed $1997.0 \leq t_{o} \leq 1998.5$.
We found that the function $\chi^{2}_{t}(i_{o})$ is characteristic of a 
function corresponding to case II (see Section \ref{sec:IntrocaseII}), 
and the possible range for the inclination angle is $\left[ 0.5, 5.5 \right]$.
The time origin is $t_{o} \approx 1997.52$ 
when $i_{o} \rightarrow 0.55$, and the time origin is $t_{o} \approx 1998.15$ 
when $i_{o} \rightarrow 5.5$. We plotted the first peak flux corresponding to the solution 
$t_{o} \approx 1998.15$ and $i_{o} \rightarrow 5.5$ (solution with the smallest $\chi^{2}$) 
and found that it is too early by at least eight months (green curve in Figure 
\ref{fig:C5_3C279_Flux_Precession}). As indicated in Section \ref{sec:Prec_1823+568}, 
we do not aim to fit the flux light curve, but we wish to compare the time position of the 
modeled first peak flux with the observed first peak flux (Figue \ref{fig:C5_3C279_Flux_Precession}). 

\begin{figure}[ht]
\centerline{
\includegraphics[scale=0.5, bb =-275 0 700 660,clip=true]{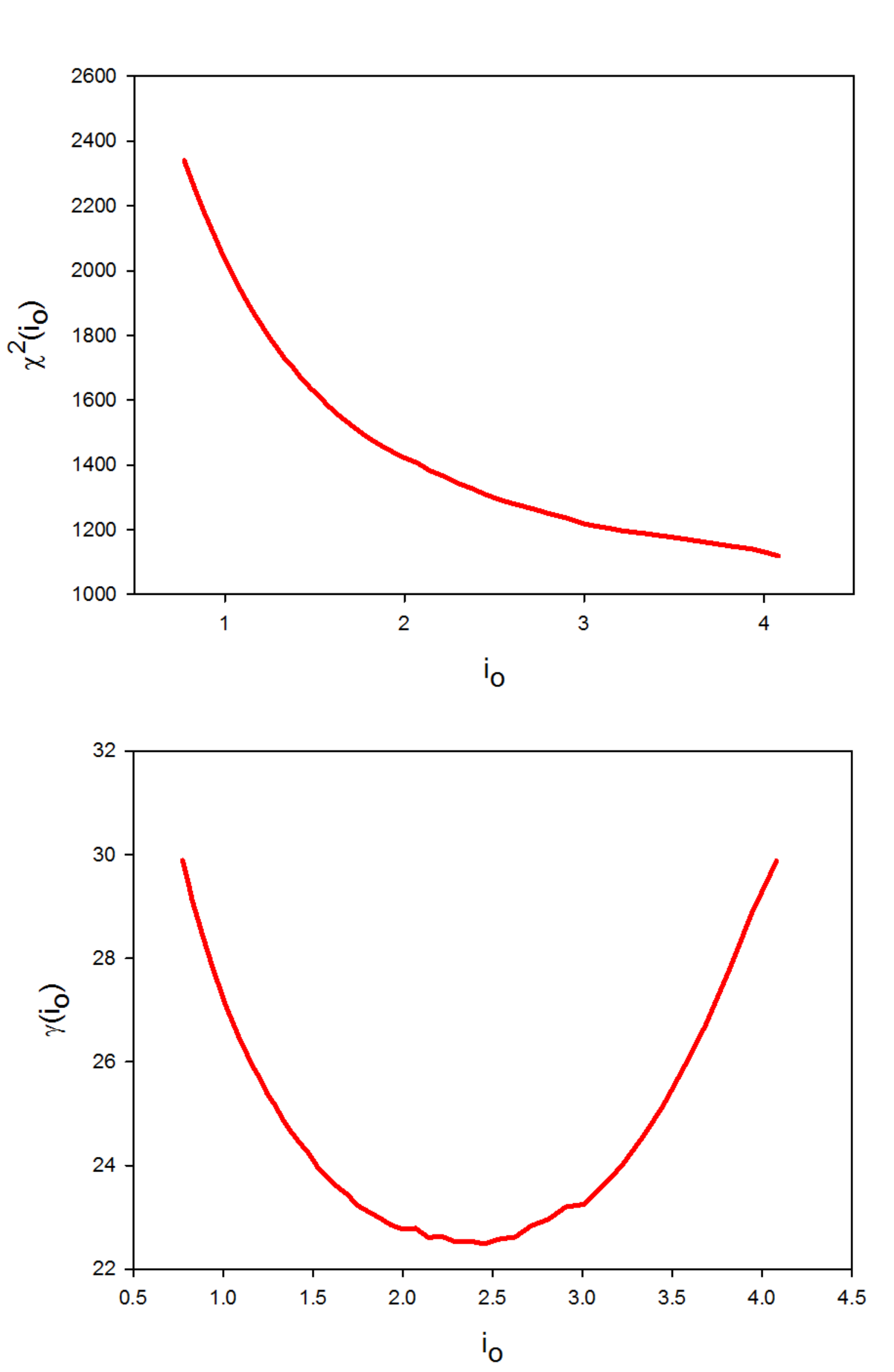}}
\caption{Precession model applied to the component C5 of 3C 279. We assumed that 
the time origin is $1998.80 \leq t_{o} \leq 1999.10$. 
\textit{Top figure:} The function $\chi^{2}(i_{o})$ is limited to $0.8^{\circ} \leq i_{o} \leq 4.3^{\circ}$ and 
has no minimum in this interval. It stops at $i_{o} \approx 4.3^{\circ}$ and $i_{o} \approx 0.8^{\circ}$ because 
at these points the bulk Lorentz factor becomes larger than 30. 
\textit{Bottom figure:} The bulk Lorentz factor diverges when 
$i_{o} \rightarrow 4.3^{\circ}$ and $i_{o} \rightarrow 0.8^{\circ}$.}
\label{fig:Chi2+Gamma_C5_3C279_Precession}
\end{figure}

Next, we allowed $t_{o}$ to be a free parameter in the range 
$1998.80 \leq t_{o} \leq 1999.10$ and calculated the new function $\chi^{2}_{t}(i_{o})$. 
The possible range for the inclination angle is reduced to $0.8^{\circ} \leq i_{o} \leq 4.3^{\circ}$. 
The plots of $\chi^{2}_{t}(i_{o})$ and $\gamma(i_{o})$ are presented in 
Figure \ref{fig:Chi2+Gamma_C5_3C279_Precession}. We plotted the first peak flux corresponding to the solution 
$t_{o} \approx 1998.80$ and $i_{o} \rightarrow 4.3$ (red curve in Figure 
\ref{fig:C5_3C279_Flux_Precession}). From Figure \ref{fig:C5_3C279_Flux_Precession}, we conclude that 
the minimun time for the ejection of C5 is $t_{o} \geq 1998.80$.

The behavior of the functions $\chi^{2}(i_{o})$ and $\gamma_{c}(i_{o})$ are the 
second signature of case II, i.e.,  
the offset is larger than the smallest error bars of the VLBI component 
coordinates.

We see from Figure \ref{fig:Chi2+Gamma_C5_3C279_Precession} that the bulk 
Lorentz factor is $\gamma_{c} \geq 22$. Because the function $\chi^{2}(i_{o})$ 
does not show a minimum, we arbitrarily chose an inclination angle such that 
$22 \leq \gamma \leq 26$ and the corresponding $\chi^{2}(i_{o})$ is the smallest. To 
continue, we chose $i_{o} \approx 2.98^{\circ}$ and the corresponding parameters of 
the precession solution (the $\chi^{2}$ of this solution is $\chi^{2} \approx 1211$ 
and its bulk Lorentz factor is $\gamma \approx 22.6$). We used this 
solution to apply the method explained in section \ref{sec:method} and 
we will see in the following how the BBH system model allows us 
to find the concave part of the funtion $\chi^{2}(i_{o})$.

\subsection{Determining the family of solutions (precession model)}
The solution is not unique. 
For the inclination angle previously found, i.e., 
$i_{o}\approx 2.98^{\circ}$ and using the parameters of the corresponding solution, 
we gradually varied $V_{a}$ between $0.01$ c and $0.45$ c. 
The function $\chi^{2}(V_{a})$ remains constant, indicating a degeneracy 
of the solution, and we obtained the range of possible values for  
the precession period given in Table 8.

\begin{center}
Table 8 : Range for the precession period\medskip%

\begin{tabular}
[c]{c||c|c}\hline
$V_{a}$            & $0.01 \; c$         & $0.45 \; c$       \\\hline
$T_{p}(V_{a})$     & $\approx 91900$ yr  & $\approx 1150$ yr  \\\hline
\end{tabular}
\end{center}

\subsection{Possible offset of the origin of the ejection (precession model)}
\label{sec:Offset_prec_3C279}

In this section, we kept the inclination angle previously found, 
i.e., $i_{o} \approx 2.98^{\circ}$. We assumed that $V_{a} = 0.1$ and used 
the parameters of the solution previously found.

To test whether the VLBI component is ejected from the VLBI core 
or if it is ejected with an offset of the origin, we calculated $\chi^{2}(\Delta W, \Delta N)$,
where $\Delta W$ and $\Delta N$ are offsets in the west and north directions, 
using the precession model. The step used in West and North directions 
is $10 \; \mu as$. \\

\begin{figure}[ht]
\centerline{
\includegraphics[scale=0.5, width=8cm,height=8cm]{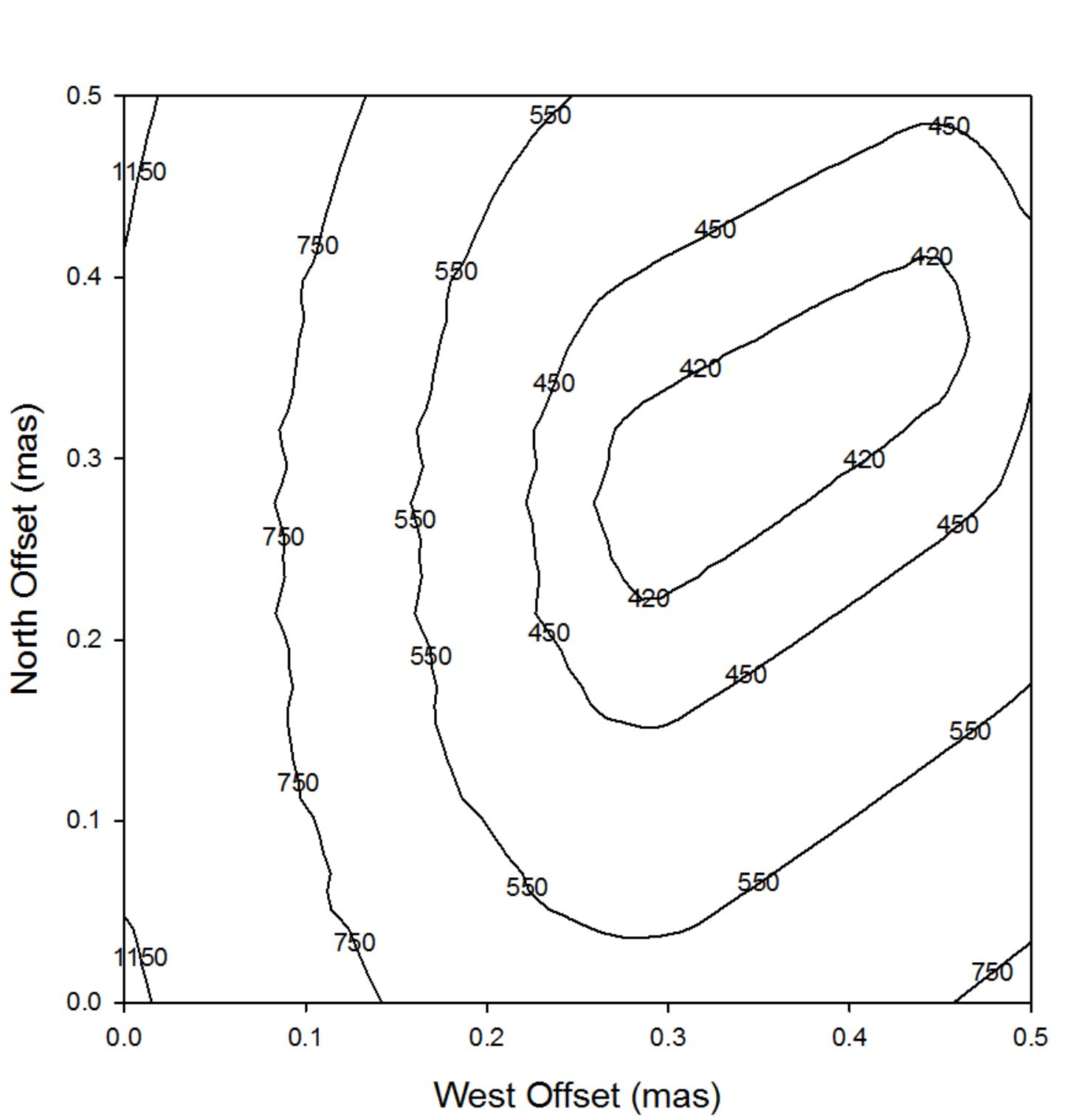}}
\caption{Using the precession model, we calcuated $\chi^{2}(\Delta W, \Delta N)$ corresponding 
to the solution with $i_{o} \approx 2.98^{\circ}$. 
Non-zero offsets are possible and the smallest offsets are 
$\Delta W \approx 0.300$ $mas$ and $\Delta N \approx 0.280$ $mas$, 
which corresponds to a BBH system of radius $R_{bin} \geq 410$ $\mu as$.}
\label{fig:Chi2_Off_YX_3C279_C5_Prec}
\end{figure}

We see from Figure \ref{fig:Chi2_Off_YX_3C279_C5_Prec}, that 
non-zero offsets are possible and the smallest offsets of the coordinates are 
$\Delta W \approx +0.300$ $mas$ and $\Delta N \approx +0.280$ $mas$,
which, a priori, corresponds to an offset of the space origin of 
$\geq 410$ $\mu as$ or to a BBH system of radius 
$R_{bin} \geq 410$ $\mu as$. This minimum offset corresponds to an improvement 
of about $28 \; \sigma$. If the offset of the space origin can be estimated 
using the precession model, it cannot be explained if we assume that the nucleus contains 
a single black hole, but it can be explained if we assume that 
the nucleus contains a BBH system.\\

It is important to note that the offset does not depend on the 
inclination angle chosen in section \ref{sec:Prec_3C279}. Indeed, we took 
the solution corresponding to $i_{o} \approx 1.5^{\circ}$, whose $\chi^{2}$ is 
$\chi^{2} \approx 1610$ and whose bulk Lorentz factor is $\gamma \approx 24$, and we 
calculated $\chi^{2}(\Delta W, \Delta N)$, which yielded the same result.

\begin{figure}[ht]
\centerline{
\includegraphics[scale=0.5, width=8cm,height=8cm]{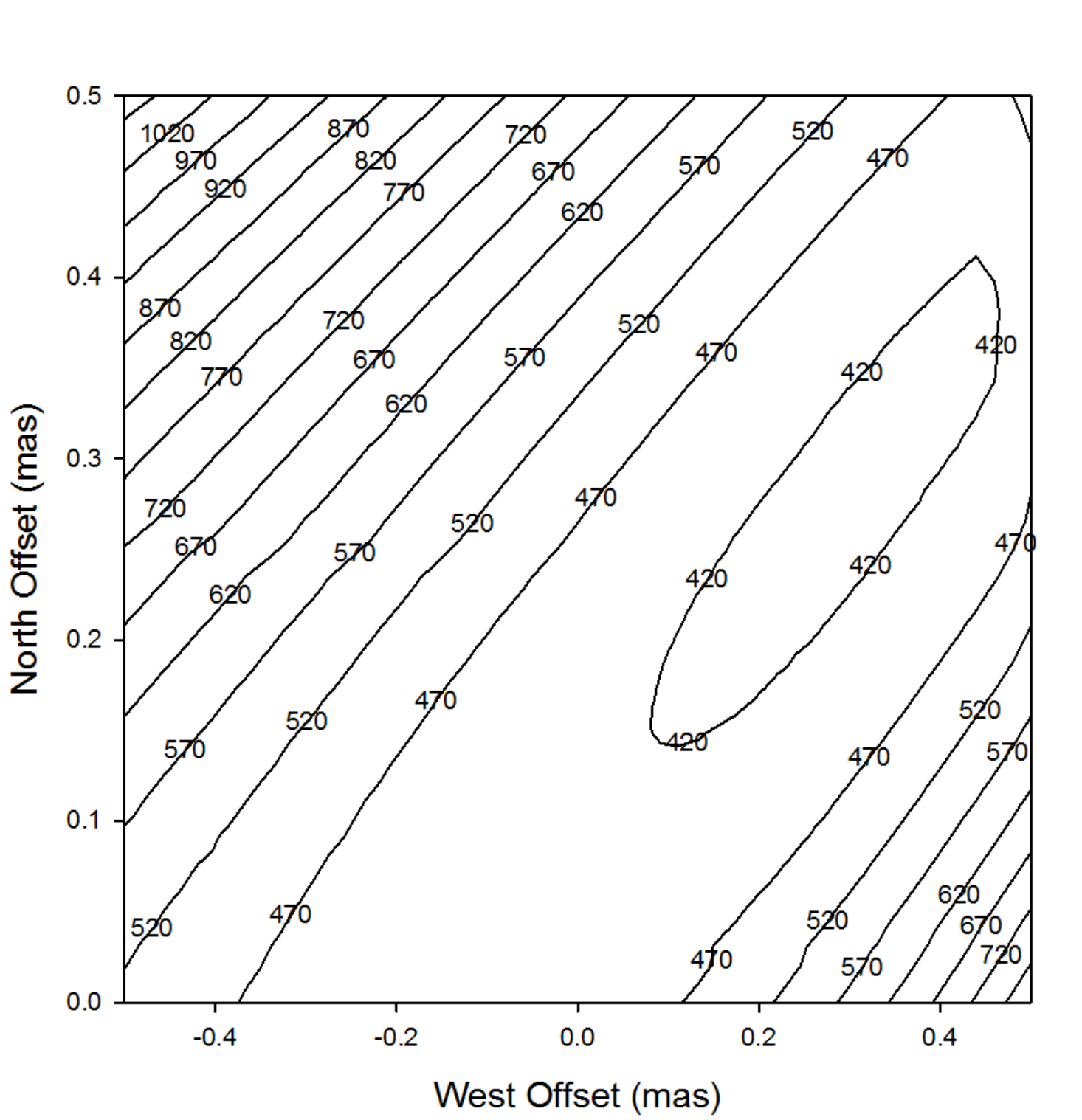}}
\caption{Calculation of $\chi^{2}(\Delta W, \Delta N)$ using the precession model and 
assuming that the time origin is a free parameter. We find that 
non-offset are possible and the smallest offset corresponds to 
the point $\Delta W \approx 100$ $\mu as $ and $\Delta N \approx +150$ $\mu as $. 
At this point, the time origin is $t_{o} \approx 1998.45$ which is $\approx 7$ months 
too early.}
\label{fig:Chi2_Off_YX_3C279_C5_Prec_to=f}
\end{figure}

It is easy to prove that the value of the offset of the space 
origin is related to the time origin problem. Indeed, 
Figure \ref{fig:Chi2_Off_YX_3C279_C5_Prec} shows that there is a 
significant offset of the space origin when we assume 
that the time origin of component C5 of 3C 279 is $t_{o} \geq 1998.80$. 
Now, using again the precession model, we calculated the possible offset of the space 
origin assuming that the time origin is a free parameter 
(Figure \ref{fig:Chi2_Off_YX_3C279_C5_Prec_to=f}). We see from 
Figure \ref{fig:Chi2_Off_YX_3C279_C5_Prec_to=f} that non-zero offsets are possible 
and the smallest offsets of the coordinates are $\Delta W \approx 100$ $\mu as $ 
and $\Delta N \approx +150$ $\mu as $, 
at this point, the time origin is $t_{o} \approx 1998.45$. This time origin corresponds 
to an ejection that is about seven months too early.\\

To continue, two possibilities arise:
\begin{enumerate}
  \item either we keep the original VLBI coordinates and determine the parameters 
	of the BBH system and the  $\chi^{2}(T_{p}/T_{b})$ - diagram. Then, 
	we determine a first offset correction using the BBH model, and after a preliminary determination 
	of $T_{p}/ T_{b}$ and  $M_{1}/ M_{2}$, we determine a second offset correction using 
	the BBH model;
  \item or we apply the precession offset correction to the VLBI coordinates and then we determine the 
  parameters of the BBH system and the $\chi^{2}(T_{p}/T_{b})$ - diagram. 
  Then we determine a first offset correction using the BBH model, and after a preliminary determination 
	of $T_{p}/ T_{b}$ and  $M_{1}/ M_{2}$, we determine a second offset correction using 
	the BBH model.
\end{enumerate}

For component C5 of 3C 279, the two possibilities 
were followed. We found that they provide the same result in the end. In 
this article, we present the first one.

The determination of the offsets of the origin of the ejection does 
not depend on the inclination angle.

\subsection{Determining the BBH system parameters}
\label{sec:BBHparam3C279}
Because the precession is defined by $-\omega_{p}(t - z/V_{a})$, the 
BBH system rotation is defined by $+\omega_{b}(t - z/V_{a})$.
In this section we kept the inclination angle previously found, 
i.e., $i_{o} \approx 2.98^{\circ}$ and $V_{a} = 0.1$\:~c.

In the previous section, we saw that the BBH system has a large radius, 
i.e., $R_{bin} \geq 410$ $\mu as$. Therefore, we determined the parameters 
of a BBH system with small $T_{p}/T_{b}$ and 
a radius for the BBH system that is a free parameter (solutions with small $T_{p}/T_{b}$ 
have large radii), i.e., we determined the parameters of a BBH system with $T_{p}/T_{b} = 1.01$ 
and calculated the corresponding $\chi^{2}(T_{p}/T_{b})$ - diagram.

\subsection{$\chi^{2}(T_{p}/T_{b})$ - diagram}
\label{sec:Chi2TpTb3C279}
In this section, we kept the inclination angle previously found, 
i.e., $i_{o} \approx 2.98^{\circ}$, $V_{a} = 0.1$ c and assumed $M_{1} = M_{2}$. 
Furthermore we assumed that the radius of the BBH system is a free parameter.

\begin{figure}[ht]
\centerline{
\includegraphics[scale=0.5, width=8cm,height=6cm]{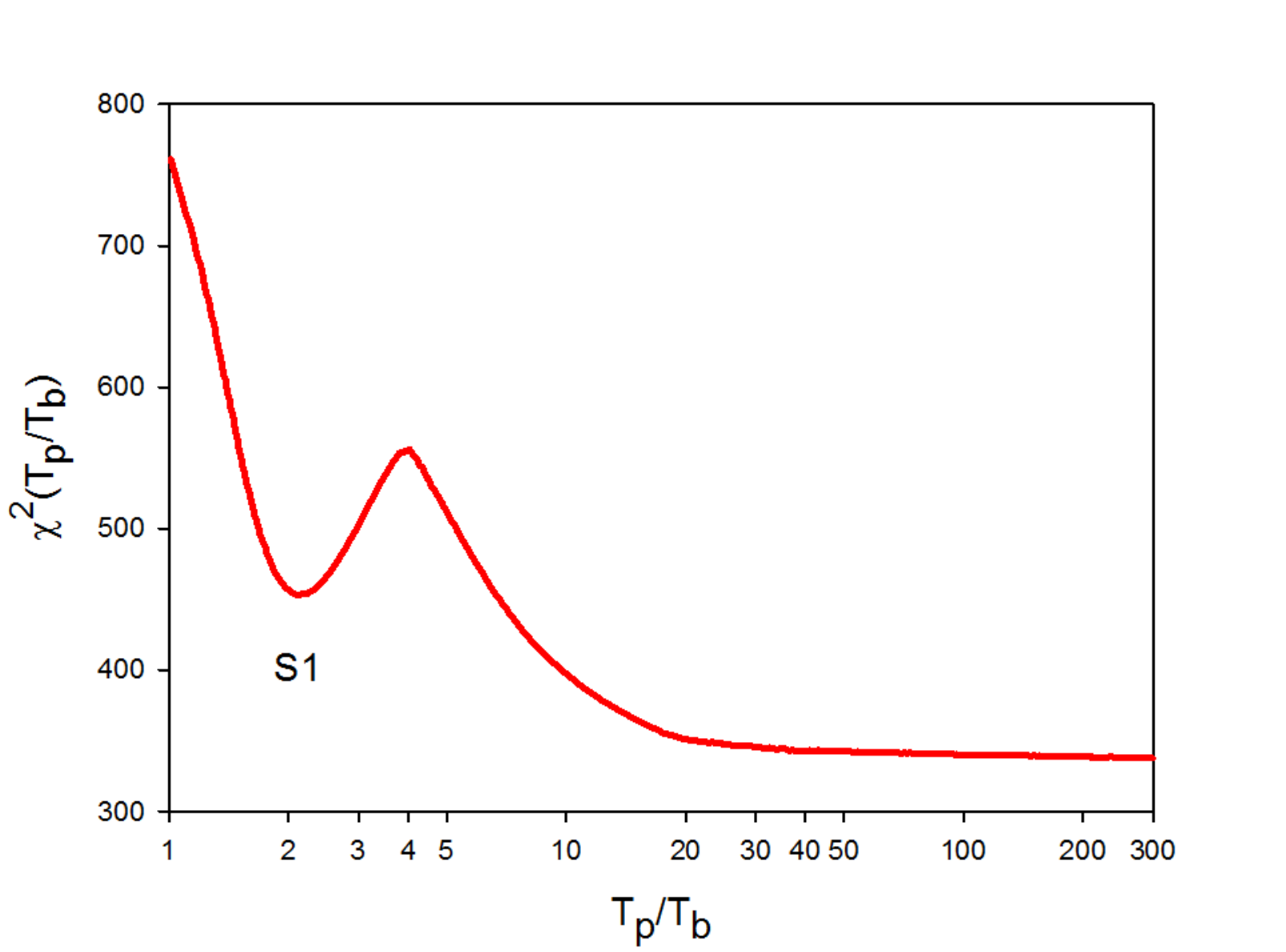}}
\caption{Calculation of $\chi^{2}(T_{p}/T_{b})$. The curve corresponds 
to the minimization when $T_{p}/T_{b}$ varies from 1 to 300. 
There is one solution S1.}
\label{fig:Chi2_TpTb_3C279_C5}
\end{figure}

We calculated $\chi^{2}(T_{p}/T_{b})$ for $1 \leq T_{p}/T_{b} \leq 300$.
We started for BBH system parameters corresponding to the 
values of $ T_{p}/T_{b} = 1.01$ and cover the complete 
interval of $T_{p}/T_{b}$. 
The result is shown in Figure \ref{fig:Chi2_TpTb_3C279_C5}.

We found the one solution given in Table 9.

\begin{center}
Table 9 : Solution found for $i_{o} \approx 2.98^{\circ}$\medskip%

\begin{tabular}
[c]{l||l|l|l}\hline
Solution      & $ (T_{p}/T_{b})_{min} $  & $\chi^{2}(min)$   &$R_{bin}$                  \\\hline
S1            & $\approx 2.13 $          & $\approx 453$     &$\approx 389$ $\mu as$     \\\hline
\end{tabular}
\end{center}

\subsection{Determining the offset of the origin of the ejection (BBH model)}
\label{sec:BBH_Offset1_3C279}
In this section, we kept the inclination angle previously found, 
i.e., $i_{o} \approx 2.98^{\circ}$. We assumed that $V_{a} = 0.1$ c, $M_{1} = M_{2}$.

We calculated $\chi^{2}(\Delta W, \Delta N)$,
where $\Delta W$ and $\Delta N$ are offsets in the west and north directions. 
The step used in the west and north directions is $5$ $\mu as$. 
The radius of the BBH system and $T_{p}/T_{b}$ are free parameters 
during the minimization.

\begin{figure}[ht]
\centerline{
\includegraphics[scale=0.5, width=8cm,height=8cm]{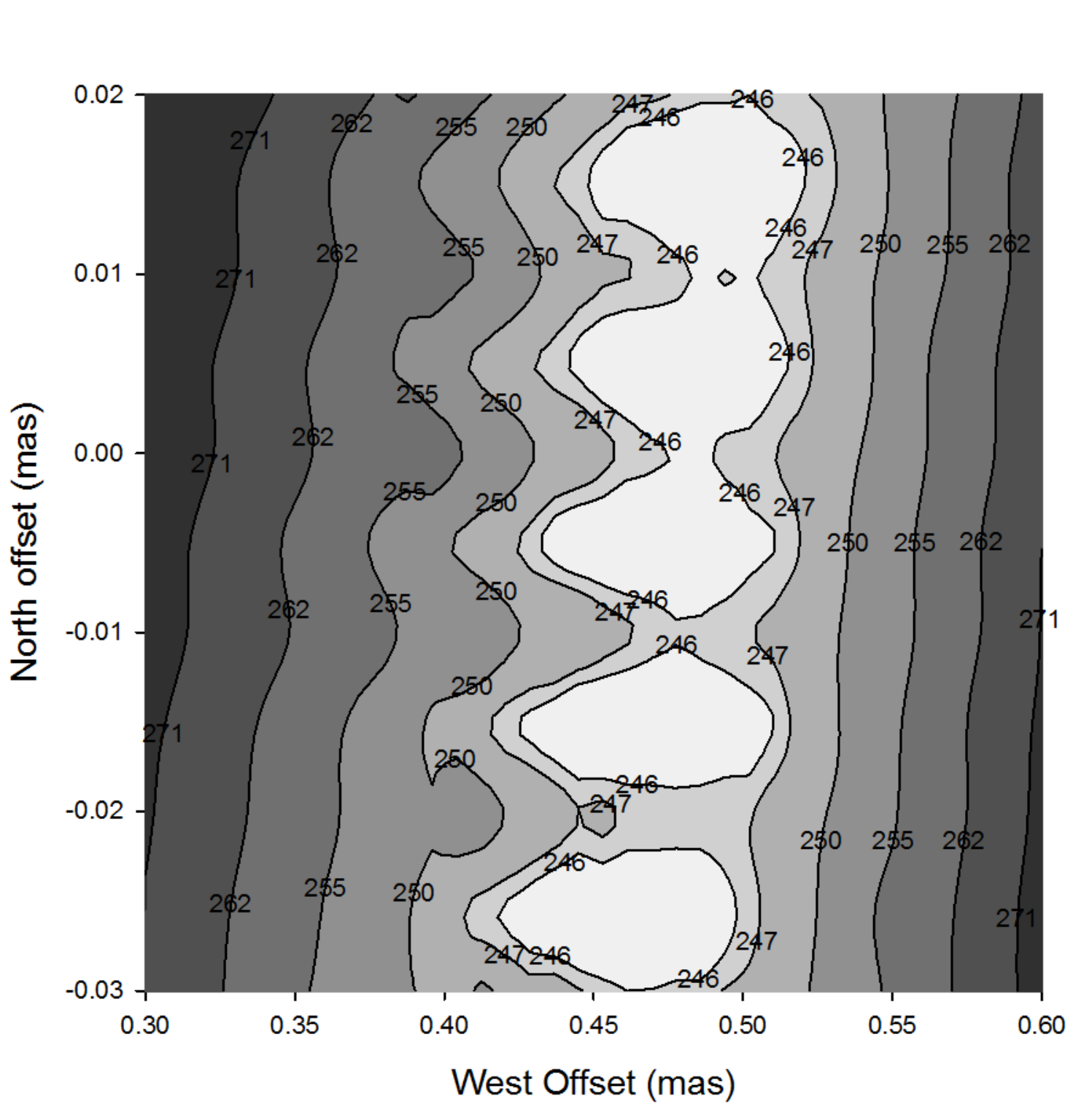}}
\caption{Calculation of $\chi^{2}(\Delta W, \Delta N)$ using the BBH model. 
Contour levels are 246, 247, 250, 255, etc corresponding to the minimum, $1 \;\sigma$, 
$2 \;\sigma$, $3 \;\sigma$, etc
There is a valley of possible offsets, but the size of the offset 
must be the same as the radius of the BBH system. This is true when the 
offsets are $\Delta W_{1} \approx +0.490$ $mas$ and $\Delta N_{1} \approx 0.005$ $mas$.}
\label{fig:Chi2_Off_YX_3C279_C5_BBH_1}
\end{figure}

We calculated $\chi^{2}(\Delta W, \Delta N)$ starting with the parameters of 
solution S1 found in the previous section. The result is shown in 
Figure \ref{fig:Chi2_Off_YX_3C279_C5_BBH_1}.

We see from Figure \ref{fig:Chi2_Off_YX_3C279_C5_BBH_1} that non-zero offsets are possible. 
However, all points with the smallest $\chi^{2}(\Delta W, \Delta N)$ are not possible. Indeed, 
for a point with the smallest $\chi^{2}$, the offset size must be equal to the 
radius of the BBH system calculated at this point. 
This is the case if the offsets are $\Delta W_{1} \approx +0.490$ $mas$ and 
$\Delta N_{1} \approx +0.005$ $mas$. 

The radius of the BBH system at this point is $R_{bin} \approx 487$ $\mu as$ 
and the offset size is $\approx 490$ $\mu as$, i.e., the offset and the 
radius of the BBH system are the same at this point. \\

Therefore we conclude that
\begin{itemize}
  \item the VLBI component C5 is not ejected from the VLBI core, but from 
  the second black hole of the BBH system, and
  \item the radius of the BBH system is $R_{bin} \approx 490$ $\mu as$. 
  It is more than ten times the smallest error bars of the VLBI component coordinates.
\end{itemize}

Note that if the size of the offset found with the BBH model is the same as 
the size of the offset found with the precession model, the first offsets 
are not the same for the coordinates. However, after the preliminary determination 
of the ratios $T_{p}/T_{b}$ and $M_{1}/M_{2}$, the second and third offset corrections 
provide the same final offset corrections (the two methods indicated in section 
\ref{sec:Offset_prec_3C279} provide the same corrections in the end).

\subsection{Preliminary determination of $i_{o}$, $T_{p}/T_{b}$ and $M_{1}/M_{2}$}
\label{sec:Mass_ratio_case_II}

From this point onward, the original coordinates of the VLBI component C5 are corrected 
for the offsets $\Delta W_{1}$ and $\Delta N_{1}$ found in the previous section. 
In this section, we assumed that $V_{a} = 0.1$ c and the radius of the BBH system 
is $R_{bin} = 490$ $\mu as$.

For given values of the ratio $M_{1}/ M_{2}$ = 1.0, 1.25, 1.50, 1.75, and 2.0, 
we varied $i_{o}$  between 3.0 and 10 degrees and calculated $\chi^{2}(i_{o})$ 
assuming that the ratio $T_{p}/T_{b}$ is variable. 

We found that $\chi^{2}(i_{o})$ is minimum for the parameters

\begin{itemize}
	\item $i_{o} \approx 5.9^{\circ}$,
	\item $M_{1}/ M_{2} \approx 1.75$, and
	\item $T_{p}/T_{b} \approx 14.6$.
\end{itemize}

\subsection{Detemining a possible new offset correction}

In this section, we assumed $V_{a} = 0.1$ c.

\begin{figure}[ht]
\centerline{
\includegraphics[scale=0.5, width=8cm,height=8cm]{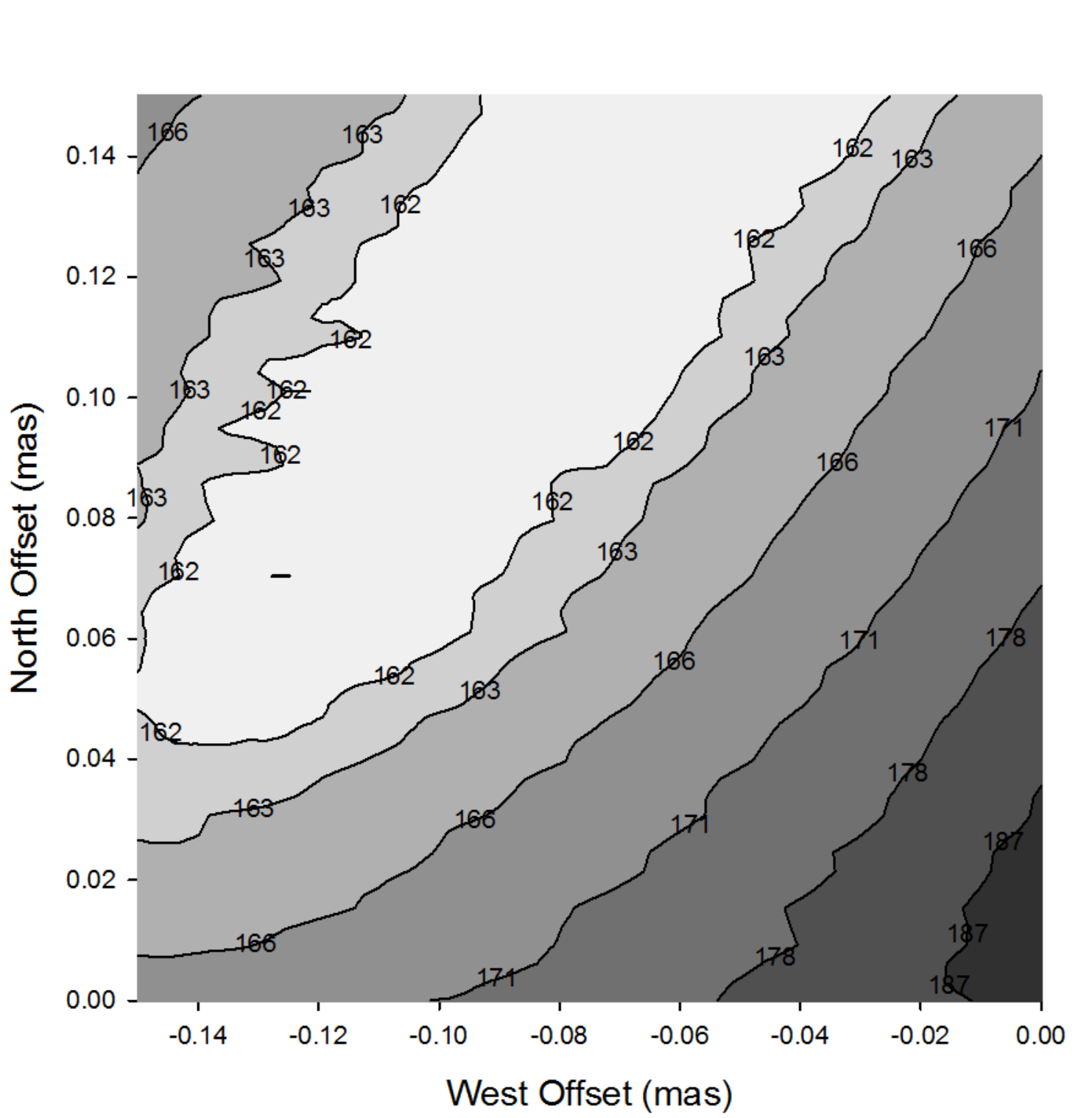}}
\caption{Calculation of $\chi^{2}(\Delta W, \Delta N)$ using the BBH model. 
Contour levels are 162, 163, 166, 171, etc corresponding to the minimum, $1 \;\sigma$, 
$2 \;\sigma$, $3 \;\sigma$, etc.
There is a valley of possible offsets, but the size of the offset 
must be the same as the radius of the BBH system. This is the case when the 
offsets are $\Delta W_{2} \approx -0.085$ $mas$ and $\Delta N_{2} \approx +0.105$ $mas$.}
\label{fig:Chi2_Off_YX_3C279_C5_BBH_2}
\end{figure}

With $i_{o} \approx 5.9^{\circ}$, with a variable ratio $T_{p}/T_{b}$,  
$M_{1}/ M_{2} \approx 1.75$ and the parameters of the solution found in the previous 
section, we can verify whether 
there is an additional correction to the offset of the origin of the VLBI 
component. We calculated $\chi^{2}(\Delta W, \Delta N)$,
where $\Delta W$ and $\Delta N$ are offsets in the west and north directions. 
We assumed that the radius of the BBH system is left free to vary. The result is shown in 
Figure \ref{fig:Chi2_Off_YX_3C279_C5_BBH_2}. We found 
that an additional correction is needed, namely $\Delta W_{2} \approx - 0.085$ $mas$ and 
$\Delta N_{2} \approx + 0.105$ $mas$.

At this point the total offset is $\approx 418$ $\mu as$ and the radius 
of the BBH system is $R_{bin} \approx 420$ $\mu as$.

\subsection{Final fit of component C5 of 3C 279}
\label{sec:Final_fit_3C279}

The coordinates of the VLBI component C5 
are corrected for the new offsets $\Delta W_{2}$ and $\Delta N_{2}$. 
In this section, we assumed $V_{a} = 0.1$ c and $R_{bin} = 420$ $\mu as$.

We can now find the final solution for the fit of C5. We calculated 
$\chi^{2}(i_{o})$ for various values of $T_{p}/T_{b}$ and $M_{1}/ M_{2}$, 
namely $T_{p}/T_{b} \leq 1000$ and $M_{1}/ M_{2} \leq 3.5$ with a typical 
step $\Delta(M_{1}/ M_{2}) = 0.25$.

\begin{figure}[ht]
\centerline{
\includegraphics[scale=0.5, width=8cm,height=6cm]{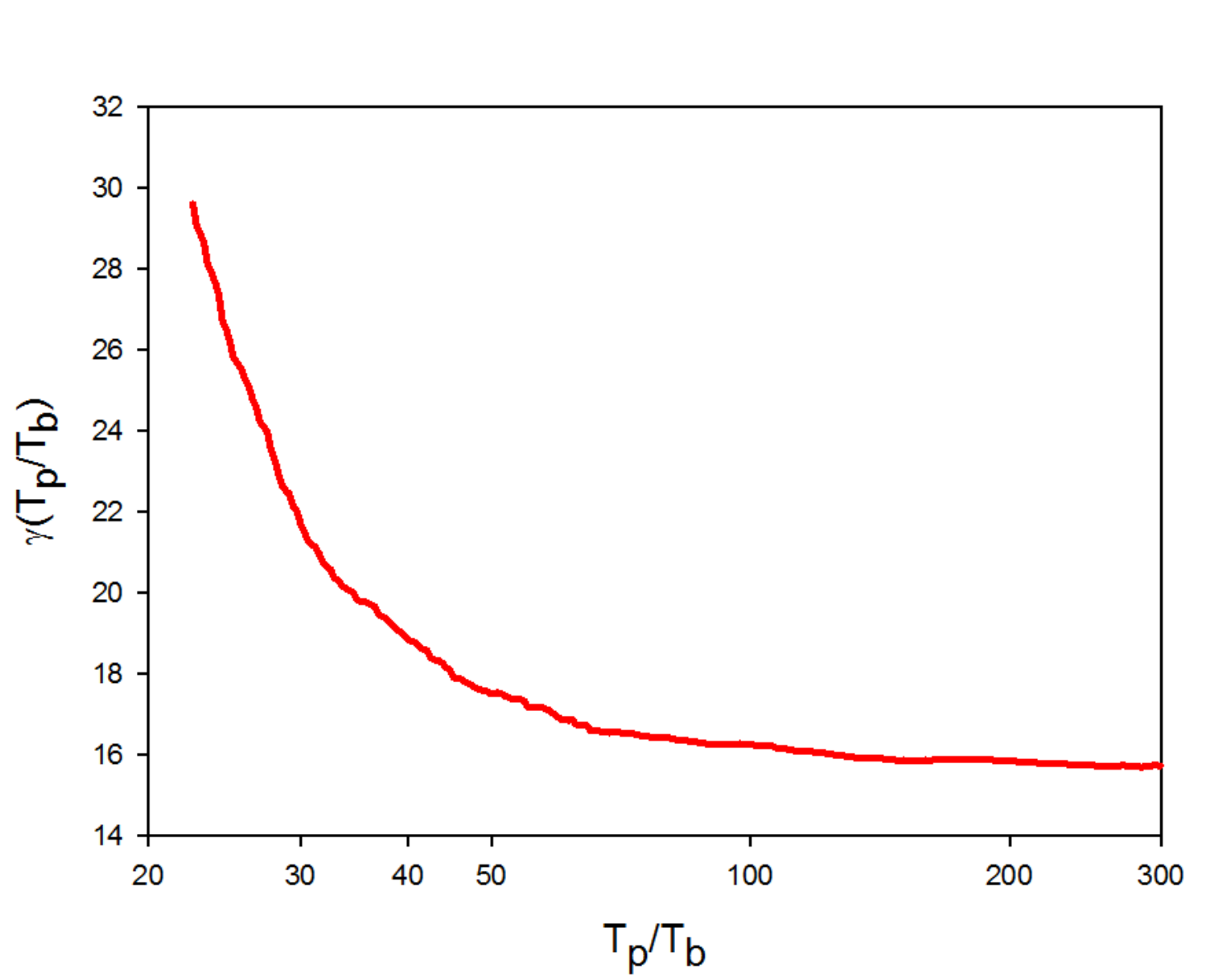}}
\caption{Calculation of $\gamma(T_{p}/T_{b})$. For a given value of $M_{1}/ M_{2}$, the 
bulk Lorentz factor decreases when $T_{p}/T_{b}$ increases, showing that if the ratio $T_{p}/T_{b}$ 
is high enough, we can find solutions that are not mirage solutions in relation to the
variable $\gamma$.}
\label{fig:3C279_C5_Gamma_TpTb_M1M2_175}
\end{figure}

The first important result is that when the ratio $T_{p}/T_{b}$ is high enough, we can 
find non-mirage solutions in relation to the variable $\gamma$. To illustrate this 
result, we plot in Figure \ref{fig:3C279_C5_Gamma_TpTb_M1M2_175} the function 
$\gamma(T_{p}/T_{b})$ corresponding to $M_{1}/ M_{2} = 1.75$.

The determination of the solution used an iterative method. 
Starting with a given value of $T_{p}/T_{b}$, we calculated $\chi^{2}(i_{o})$ for various values 
of $M_{1}/ M_{2}$. Then we calculated for the parameters corresponding to the solution found  
the function $\chi^{2}(T_{p}/T_{b})$ to determine the new value of $T_{p}/T_{b}$ that 
minimizes the function $\chi^{2}(T_{p}/T_{b})$. Starting with the new value of $T_{p}/T_{b}$, 
we repeated the procedure.

At each step of the procedure, we calculated $\chi^{2}(\gamma)$ to check that the solution corresponds to 
the concave part and is not a mirage solution.

The best fit is obtained for $T_{p}/T_{b} \approx 140$ and $M_{1}/ M_{2} \approx 2.75$.  
The results of the fits are presented in Table 10.

\begin{center}
Table 10 : Solutions found for $T_{p}/T_{b} = 140$ \medskip

\begin{tabular}
[c]{l||l|l|l}\hline
$M_{1}/ M_{2}$            & 2.50                     & 2.75                     & 3.00                     \\\hline
$i_{o}$                   & $\approx 10.0^{\circ}$   & $\approx 10.4^{\circ}$   & $\approx 11.2^{\circ}$   \\\hline
$\chi^{2}(min)$           & $\approx 151.7$          & $\approx 151.4$          & $\approx 152.0$          \\\hline
$\gamma_{min}$            & $\approx 15.0$           & $\approx 16.7$           & $\approx 21.4$           \\\hline
\end{tabular}
\end{center}

\begin{figure}[ht]
\centerline{
\includegraphics[scale=0.5, width=8cm,height=6cm]{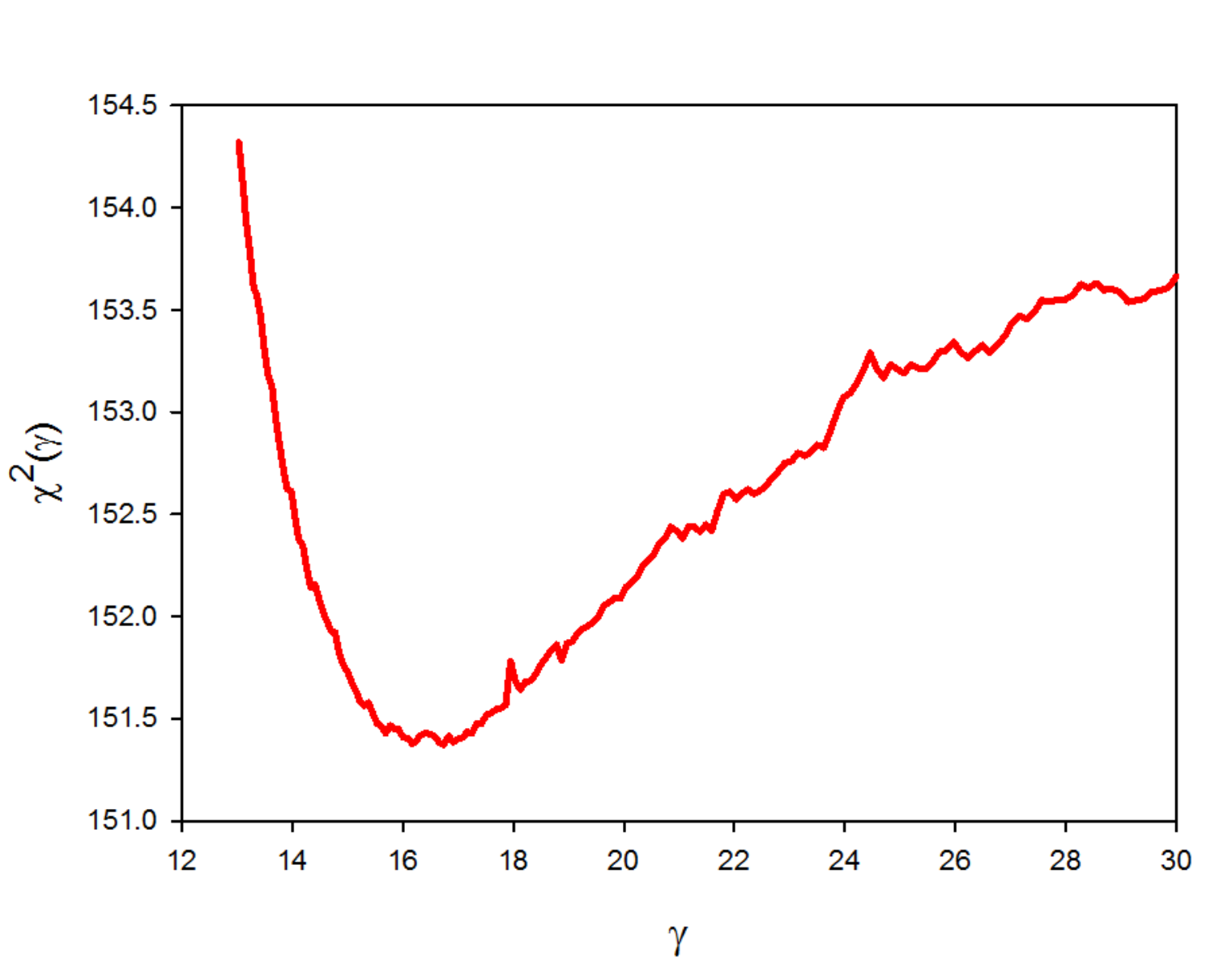}}
\caption{Calculation of $\chi^{2}(\gamma)$ for the solution with  $T_{p}/T_{b} = 140$ and 
$M_{1}/ M_{2} = 2.75$. It shows that the solution is not a mirage solution in ralation to 
$\gamma$. The minimum corresponds to $\gamma \approx 16.7$.}
\label{fig:3C279_C5_Chi2_Gam_S3_Rb418_TpTb140_M1M2_275}
\end{figure}

We plot in Figure \ref{fig:3C279_C5_Chi2_Gam_S3_Rb418_TpTb140_M1M2_275} the calculation of 
$\chi^{2}(\gamma)$ corresponding to the solution characterized by $T_{p}/T_{b} = 140$ 
and $M_{1}/ M_{2} = 2.75$. It shows that the solution is not a mirage solution in relation to 
$\gamma$.

\begin{figure}[ht]
\centerline{
\includegraphics[scale=0.5, width=8cm,height=6cm]{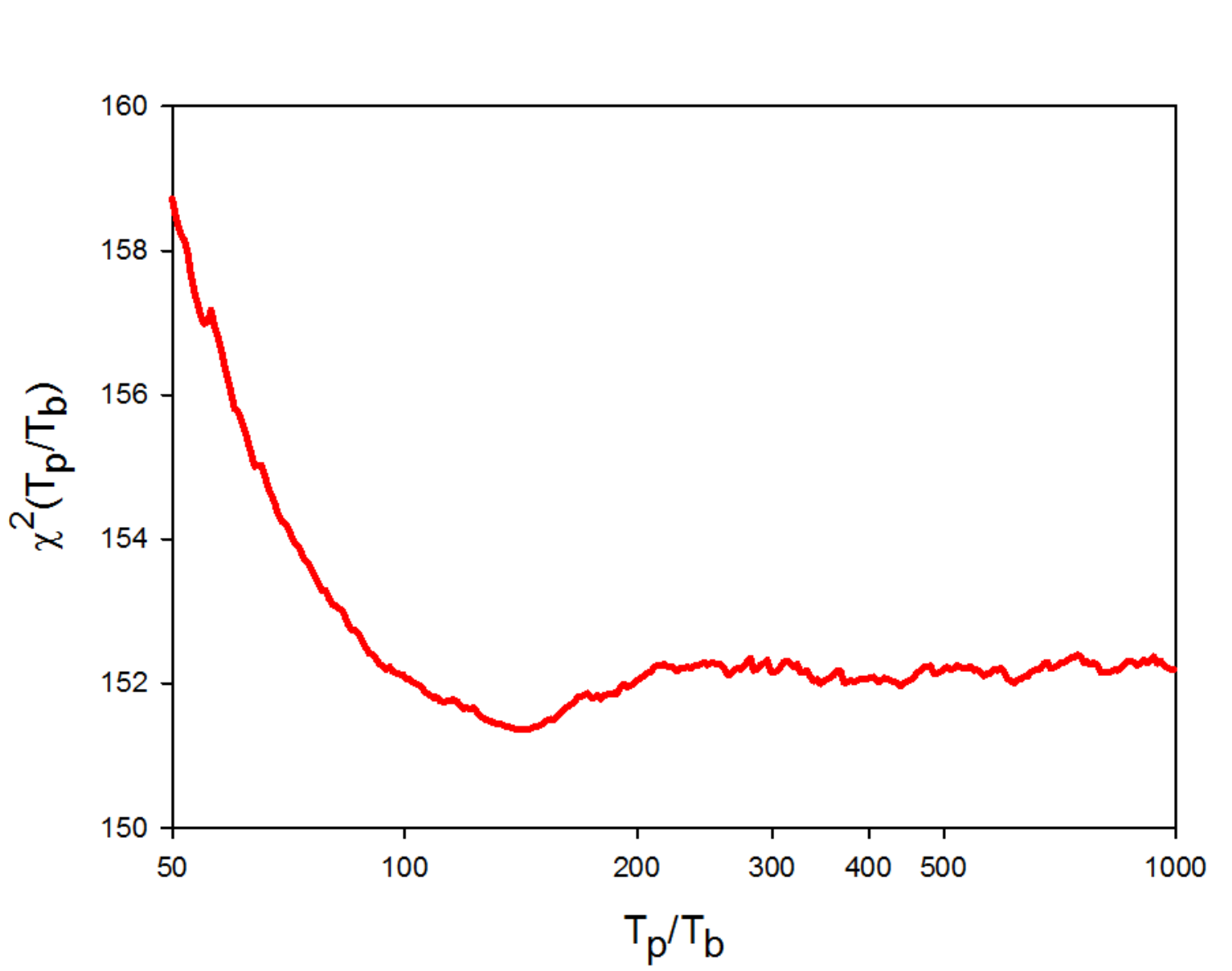}}
\caption{Calculation of $\chi^{2}(T_{p}/T_{b})$ for the solution with $M_{1}/ M_{2} = 2.75$.}
\label{fig:3C279_C5_Chi2_TpTb_S3_Rb418_M1M2_275}
\end{figure}

The best fit is obtained for $T_{p}/T_{b} \approx 140$ (see Figure 
\ref{fig:3C279_C5_Chi2_TpTb_S3_Rb418_M1M2_275}). When the ratio $T_{p}/T_{b}$ increases,  
the $\chi^{2}$ remains mostly constant but the robustness of the solution in relation 
to $\gamma$ increases.

\begin{figure}[ht]
\centerline{
\includegraphics[scale=0.5, width=8cm,height=6cm]{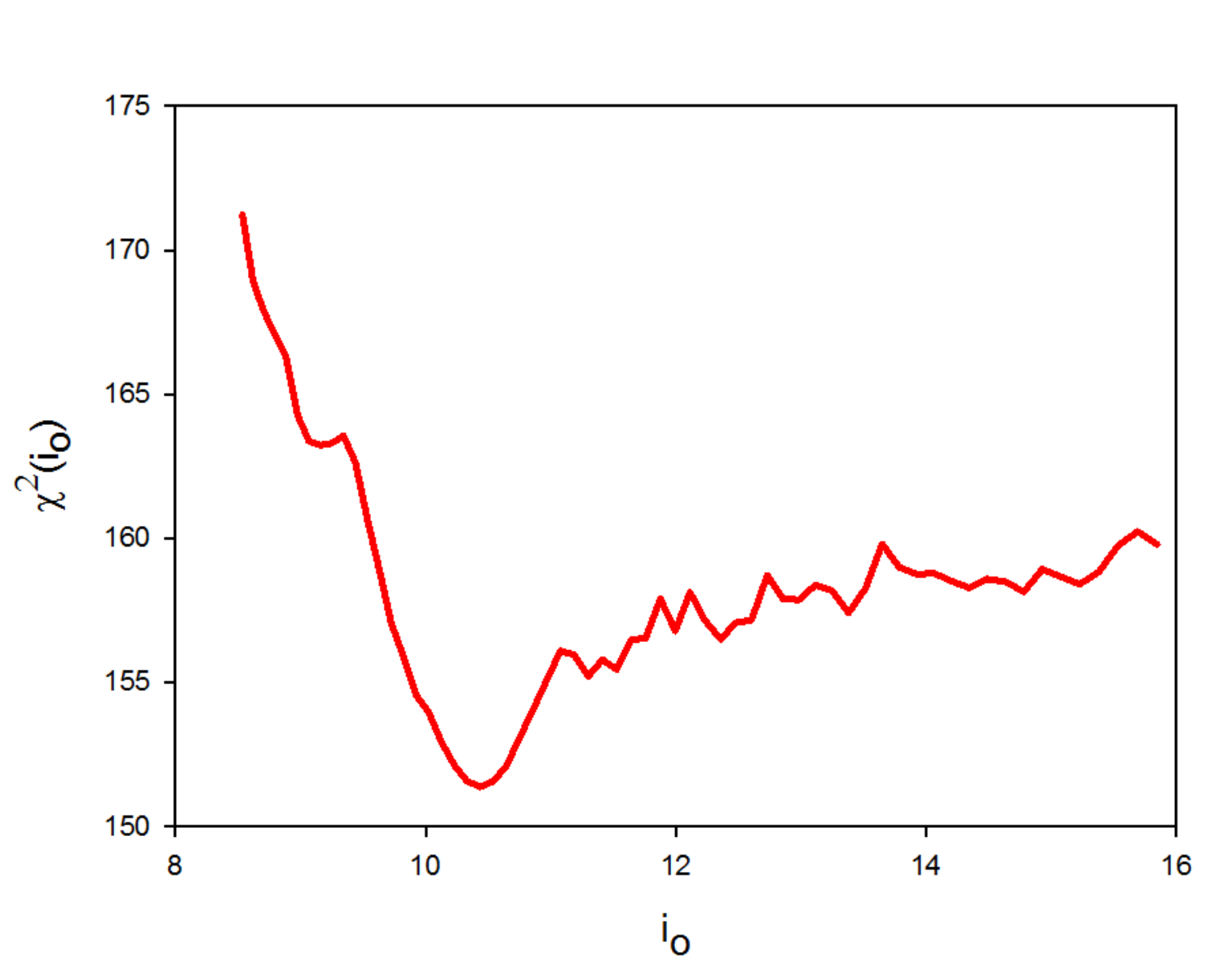}}
\caption{Calculation of $\chi^{2}(i_{o})$ for the solution with $T_{p}/T_{b} = 140$ and $M_{1}/ M_{2} = 2.75$.}
\label{fig:3C279_C5_Chi2_io_Fin}
\end{figure}

Finally, we plot in Figure \ref{fig:3C279_C5_Chi2_io_Fin} the function $\chi^{2}(i_{o})$.

The characteristics of final solution of the BBH system 
associated with 3C 279 are given in section \ref{sec:solution_3C279}.

\section{Error bars}
\label{sec:appendix_III}

\subsection{Minimum error bar values}
\label{sec:Intro_appenIII}

Observations used to fit the components S1 of 1823+568 and C5 of 3C279 were performed 
at 15 GHz. We adopted for the minimum values of the error bars, $\Delta_{min}$, 
values in the range $Beam /15 \leq \Delta_{min} \leq Beam /12$.

There are three important points concerning the minimum values used for the error bars:
\begin{enumerate}
  \item The minimum values are chosen empirically, but the adopted values are justified 
	a posteriori by comparing of the value of $\chi^{2}$ of the final solution and the 
	number of constraints used to make the fit. Indeed, the reduced $\chi^{2}$ has to be close to 1.

	\item The minimum value of the error bars used at 15 GHz produces a value of 
	$(\chi^{2})_{final}$ concistent with the value of the \textit{realistic error} obtained from the VLBI Service 
	for Geodesy and Astrometry \citep{ScBe:07}, which is a permanent geodetic and astrometric 
	VLBI program. It has been monitoring the position of thousands of extragalactic radio sources 
	for more than 30 years. In 2009, the second realization of the International Celestial Reference 
	Frame (ICRF2) was released \citep{FeGo+:10}, obtained after the treatment of about 6.5 millions 
	of ionosphere-corrected VLBI group delay measurements at 2 and 8~GHz. This catalog is currently 
	the most accurate astrometric catalog, giving absolute positions of 3414 extragalactic bodies at 8~GHz. 
	The observations at 2~GHz are used for the ionospheric correction only. Therefore, 
	the positions at 2~GHz are not provided. The ICFR2 is found to have a noise floor of only 
	40 microseconds of arc ($\mu$as), which is five to six times better than the previous ICRF realization 
	\citep{MaAr+:98}. The positions of more than 200 radio sources are known with a precision 
	(inflated error, or ``realistic'' error) better than 0.1~mas. 
	
	Since the ICRF2 release, the positional accuracy of the sources has increased, 
	and it is likely that the next VLBI realization of the ICRF will have a noise floor 
	lower than 40~$\mu$as.
	
	\item The adopted minimum value of the error bars also includes typical errors 
	due to opacity effects, which shift the measured position at different frequencies \citep{Lo:98}.
	
\end{enumerate}

Thus the minimum values for the error bars adopted at 15 GHz, using equation (\ref{eq:min_error}), are correct.
The fit of VLBI coordinates of components of 3C 345 (work in progress) indicates that the adopted values 
for the minimun values of the error bars, using equation (\ref{eq:min_error}), are correct for frequencies 
between 8 GHz and 22 GHz. At lower frequencies, the minimum values may be higher than $Beam /12$ 
due to strng opacity effects and at 43 GHz, the minimum values are also probably higher ($\approx 20$ $\mu$as).\\

It has been suggested by \citet{LiHo:05} that the positional error bars should be about 
$1/5$ of the beam size. To study the influence of the minimum values of the error 
bars on the characteristics of the solution, we calculate in the next sections the 
solution of the fit of the component C5 assuming for the minimum values of the error bars the value suggested 
by \citet{LiHo:05}, i.e. the value $\Delta_{min} = Beam /5$ , or $(\Delta W)_{min} \approx 102$ $\mu as$ 
and $(\Delta N)_{min} \approx 267$ $\mu as$.

\subsection{Fit of C5 using the precession model }
\label{sec:Prec_3C279_LaEr}

We look for a solution with $-\omega_{p}(t - z/V_{a})$ and $t_{o} \geq 1980.80$ 
(see Section \ref{sec:Prec_3C279}).

In this section, we assumed that $V_{a} = 0.1$ c.

The range of inclination explored is $0.5^{\circ} \leq i_{o} \leq 10^{\circ}$.

\begin{figure}[ht]
\centerline{
\includegraphics[scale=0.5, bb =-275 0 700 660,clip=true]{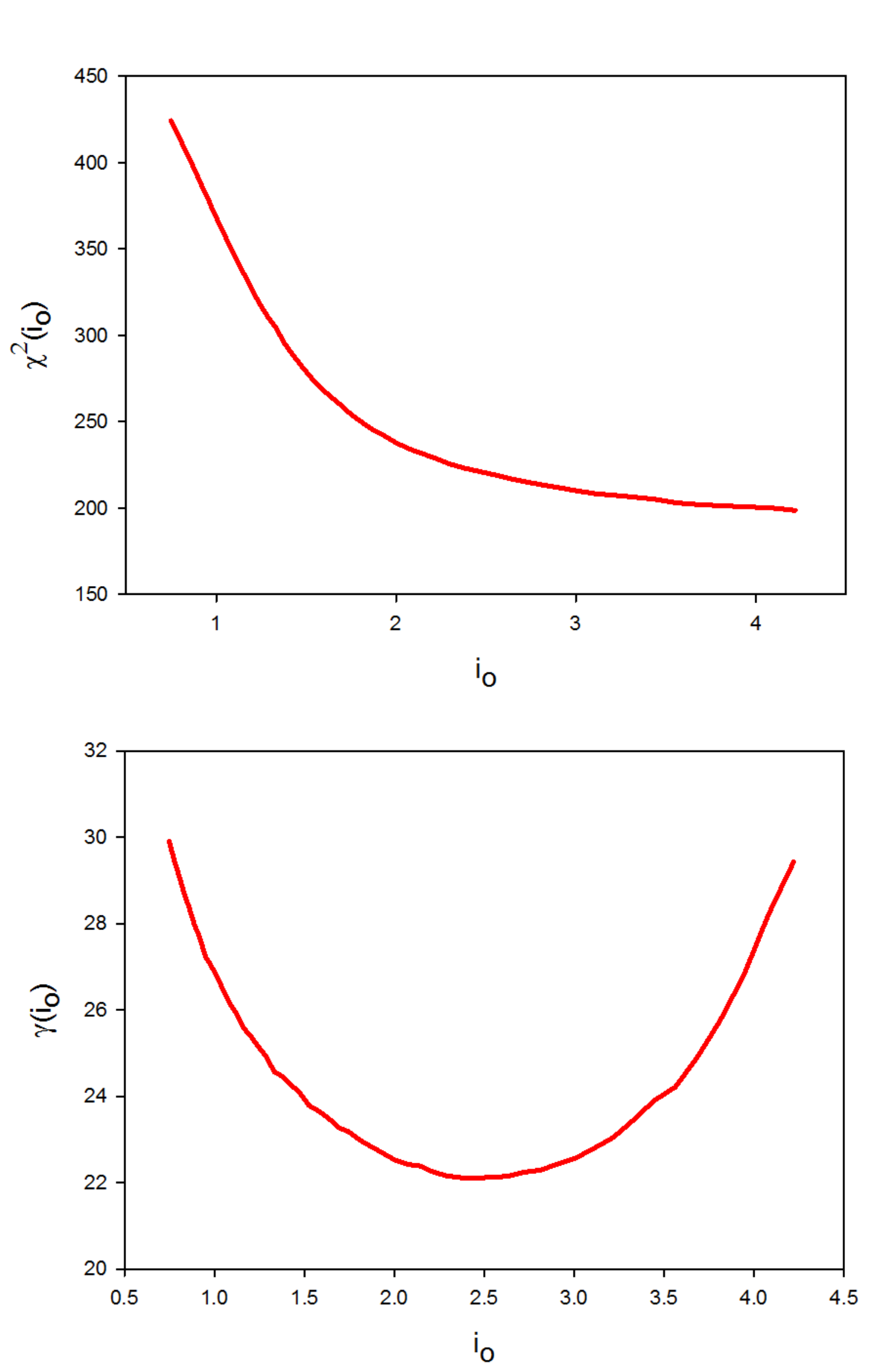}}
\caption{Precession model applied to the component C5 of 3C 279 assuming 
high values for the minimum error bars. We assumed that 
the time origin is $1998.80 \leq t_{o} \leq 1999.10$. 
\textit{Top figure:} The function $\chi^{2}(i_{o})$ is limited to $0.8^{\circ} \leq i_{o} \leq 4.3^{\circ}$ and 
has no minimum in this interval. It stops at $i_{o} \approx 4.3^{\circ}$ and $i_{o} \approx 0.8^{\circ}$ because 
at these points the bulk Lorentz factor becomes larger than 30. 
\textit{Bottom figure:} The bulk Lorentz factor diverges when 
$i_{o} \rightarrow 4.3^{\circ}$ and $i_{o} \rightarrow 0.8^{\circ}$.}
\label{fig:Chi2+Gamma_C5_3C279_Precession_LaEr}
\end{figure}

We allowed $t_{o}$ to be a free parameter in the range 
$1998.80 \leq t_{o} \leq 1999.10$ and calculated the function $\chi^{2}_{t}(i_{o})$. 
The possible range for the inclination angle is  $0.74^{\circ} \leq i_{o} \leq 4.3^{\circ}$. 
The plots of $\chi^{2}_{t}(i_{o})$ and $\gamma(i_{o})$ are presented in 
Figure \ref{fig:Chi2+Gamma_C5_3C279_Precession_LaEr}.

The behavior of the functions $\chi^{2}(i_{o})$ and $\gamma_{c}(i_{o})$ are the 
second signature of case II.\\

Comparison of Figure \ref{fig:Chi2+Gamma_C5_3C279_Precession_LaEr} and 
Figure \ref{fig:Chi2+Gamma_C5_3C279_Precession} shows that 
the range of the inclination angle, and the values of the bulk Lorentz factor in this range,
are the same for the different values of the minimum error bars used.

\subsection{Possible offset of the origin of the ejection (precession model)}
\label{sec:Offset_prec_3C279_La_Er}

In this section, we kept the inclination angle used in section 
\ref{sec:Offset_prec_3C279}, i.e., $i_{o} \approx 3.3^{\circ}$. 
We assumed that $V_{a} = 0.1$ and used the parameters of the solution found 
in section \ref{sec:Prec_3C279_LaEr}.

\begin{figure}[ht]
\centerline{
\includegraphics[scale=0.5, width=8cm,height=8cm]{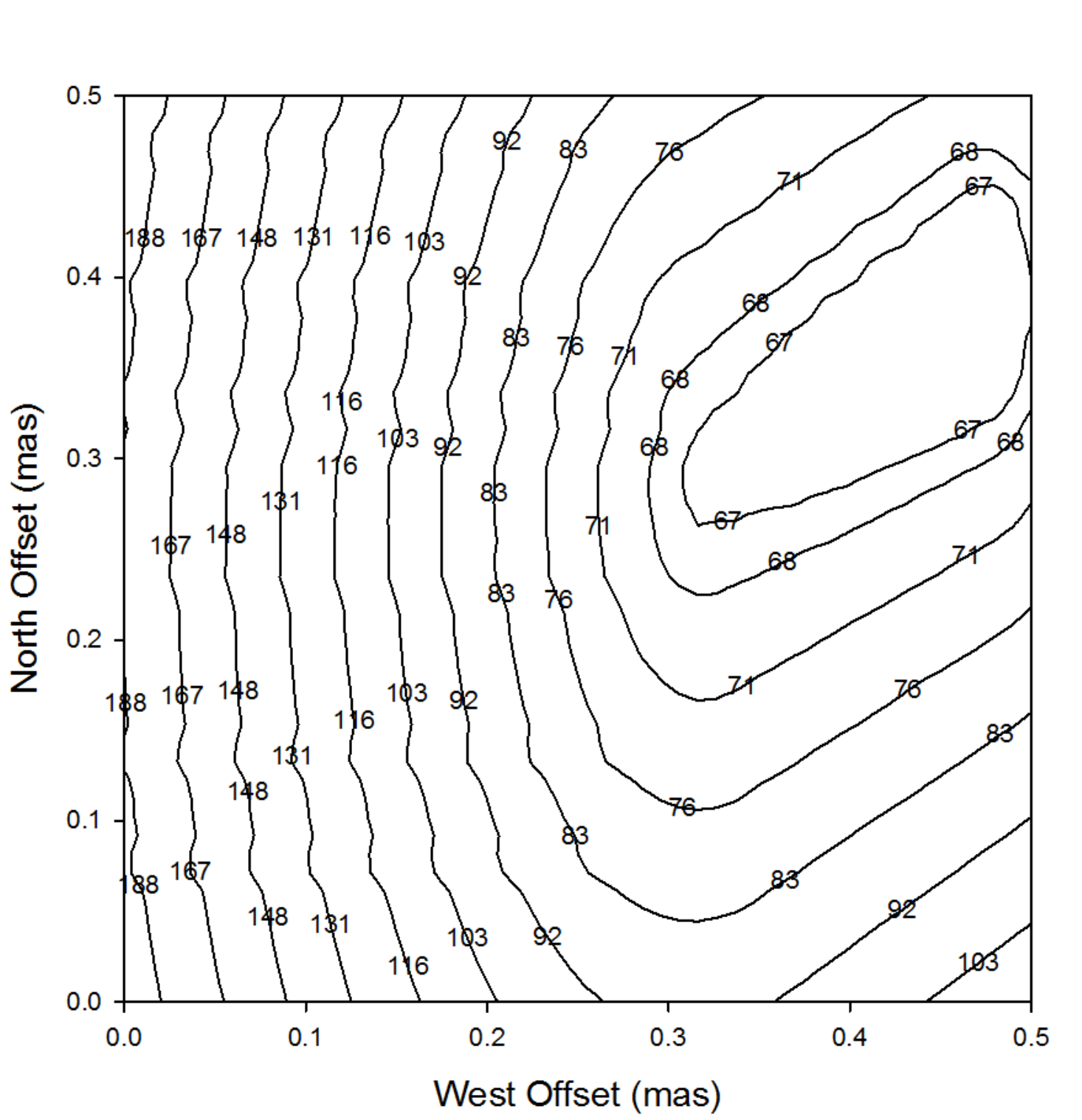}}
\caption{Using the precession model, we calcuated $\chi^{2}(\Delta W, \Delta N)$. 
The contour levels 68, 71, 76, etc correspond to 1 $\sigma$, 2 $\sigma$, 3 $\sigma$, etc. 
Non-zero offsets are possible and the smallest offsets are 
$\Delta W \approx 0.320$ $mas$ and $\Delta N \approx 0.280$ $mas$, 
which corresponds to a BBH system of radius $R_{bin} \geq 425$ $\mu as$.}
\label{fig:Chi2_Off_YX_3C279_C5_Prec_LE}
\end{figure}

We calculated $\chi^{2}(\Delta W, \Delta N)$,
where $\Delta W$ and $\Delta N$ are offsets in the west and north directions, 
using the precession model. The step used in the west and north directions 
is $10 \; \mu as$. The result of the calculation is plotted in Figure 
\ref{fig:Chi2_Off_YX_3C279_C5_Prec_LE}.\\

Comparison of Figure \ref{fig:Chi2_Off_YX_3C279_C5_Prec_LE} with 
Figure \ref{fig:Chi2_Off_YX_3C279_C5_Prec} shows that 

\begin{itemize}
  \item the smallest offsets of the coordinates are  
	$\Delta W \approx +0.320$ $mas$ and $\Delta N \approx +0.280$ $mas$. They are 
	similar to the offsets found assuming that the minimum error bars are 
	$\Delta_{min} = Beam /15$ (see Section \ref{sec:Offset_prec_3C279}),
	\item the value of $\chi^{2}$ at the minimum is $\chi^{2}_{min} \approx 67$ 
	instead of $\chi^{2}_{min} \approx 400$ when the minimum error bars are 
	$\Delta_{min} = Beam /15$. The reduced $\chi^{2}_{r}$ is $\chi^{2}_{r} \approx 67 /152 \approx 0.44$, 
	indicating that the minimum error bars are too large.
\end{itemize}

Accordingly, with high values for the minimum values of the error bars, we find using the 
precession model that the component C5 is ejected with an offset of the space origin 
of at least $0.425$ $mas$ with a robustness higher than 11 $\sigma$. 
The offset of the space origin can be estimated using 
a single black hole and the precession of the accretion disk.  
It cannot be explained when we assume that the nucleus contains 
a single black hole, but it can be explained when we assume that 
the nucleus contains a BBH system.\\

\subsection{$\chi^{2}(T_{p}/T_{b})$ - diagram}
\label{sec:Chi2TpTb3C279_La_Er}

Because the precession is defined by $-\omega_{p}(t - z/V_{a})$, the 
BBH system rotation is defined by $+\omega_{b}(t - z/V_{a})$. As in Section 
\ref{sec:BBHparam3C279}, we calculated the BBH parameters for the inclination angle 
$i_{o} \approx 2.98^{\circ}$ and the ratio $T_{p}/T_{b} = 1.01$ and calculated the 
corresponding $\chi^{2}(T_{p}/T_{b})$ - diagram assuming $M_{1} = M_{2}$ and $V_{a} = 0.1$ c.

\begin{figure}[ht]
\centerline{
\includegraphics[scale=0.5, width=8cm,height=6cm]{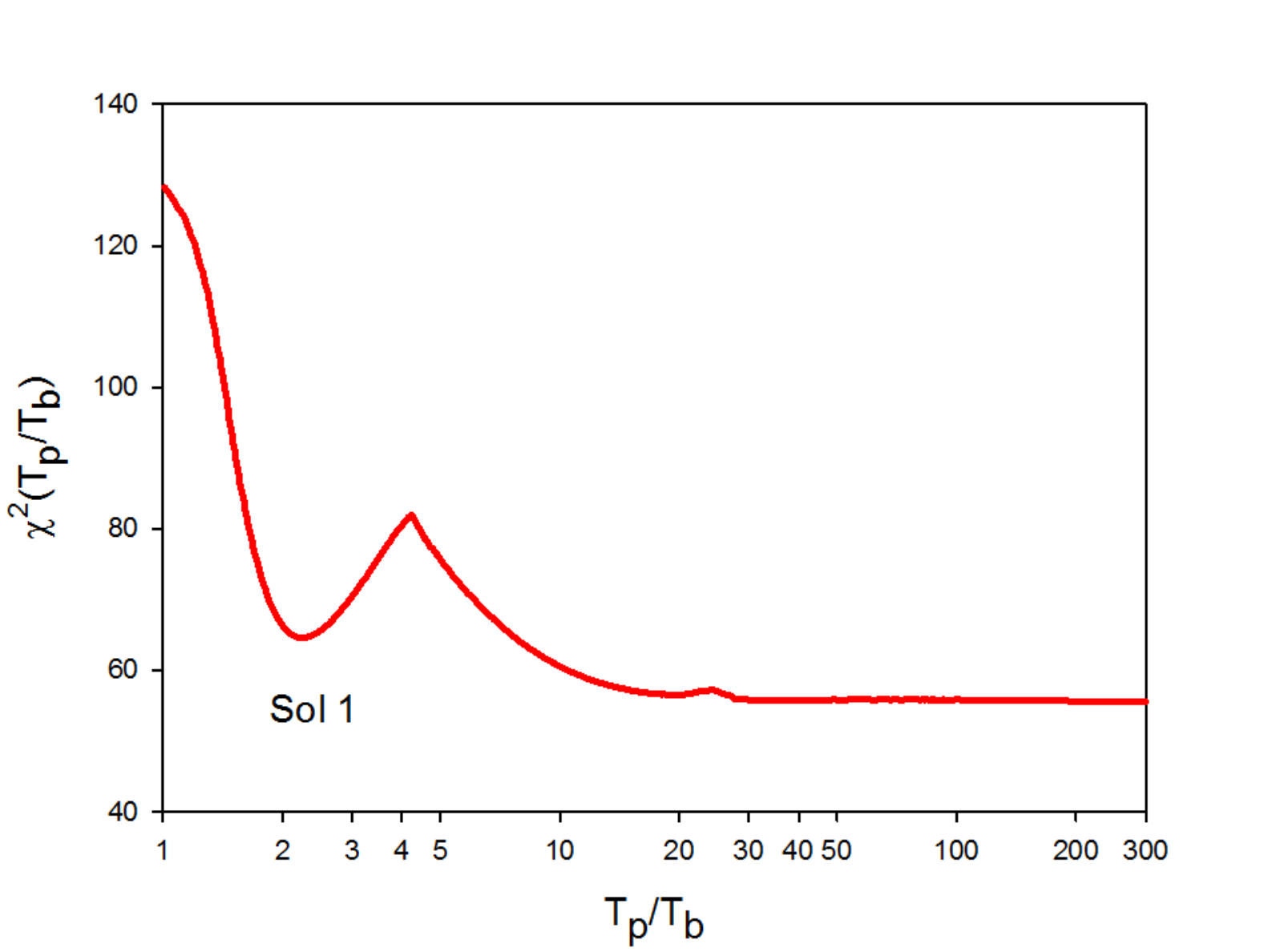}}
\caption{Calculation of $\chi^{2}(T_{p}/T_{b})$. There is one solution Sol 1.}
\label{fig:Chi2_TpTb_3C279_C5_LE}
\end{figure}

We calculated $\chi^{2}(T_{p}/T_{b})$ for $1 \leq T_{p}/T_{b} \leq 300$.
The result is shown in Figure \ref{fig:Chi2_TpTb_3C279_C5_LE}.

We found a possible solution of the BBH system given in Table 11.

\begin{center}
Table 11 : Solution found for $i_{o} \approx 2.98^{\circ}$ \medskip%

\begin{tabular}
[c]{l||l|l|l}\hline
Solution      & $ (T_{p}/T_{b})_{min} $  & $\chi^{2}(min)$   &$R_{bin}$                  \\\hline
Sol 1         & $\approx 2.24 $          & $\approx 64.6$     &$\approx 398$ $\mu as$     \\\hline
\end{tabular}
\end{center}

Comparison of Tables 9 and 11 and of Figures \ref{fig:Chi2_TpTb_3C279_C5} and \ref{fig:Chi2_TpTb_3C279_C5_LE} 
shows that the solutions S1 and Sol1 are mostly identical.

\subsection{Determining the offset of the origin of the ejection (BBH model)}

In this section, we kept the inclination angle previously found, 
i.e., $i_{o} \approx 2.98^{\circ}$. We assumed that $V_{a} = 0.1$ c, $M_{1} = M_{2}$.

We calculated $\chi^{2}(\Delta W, \Delta N)$,
where $\Delta W$ and $\Delta N$ are offsets in the west and north directions. 
The step used in the west and north directions is $5$ $\mu as$. 
The radius of the BBH system and $T_{p}/T_{b}$ are free parameters 
during the minimization.

\begin{figure}[ht]
\centerline{
\includegraphics[scale=0.5, width=8cm,height=8cm]{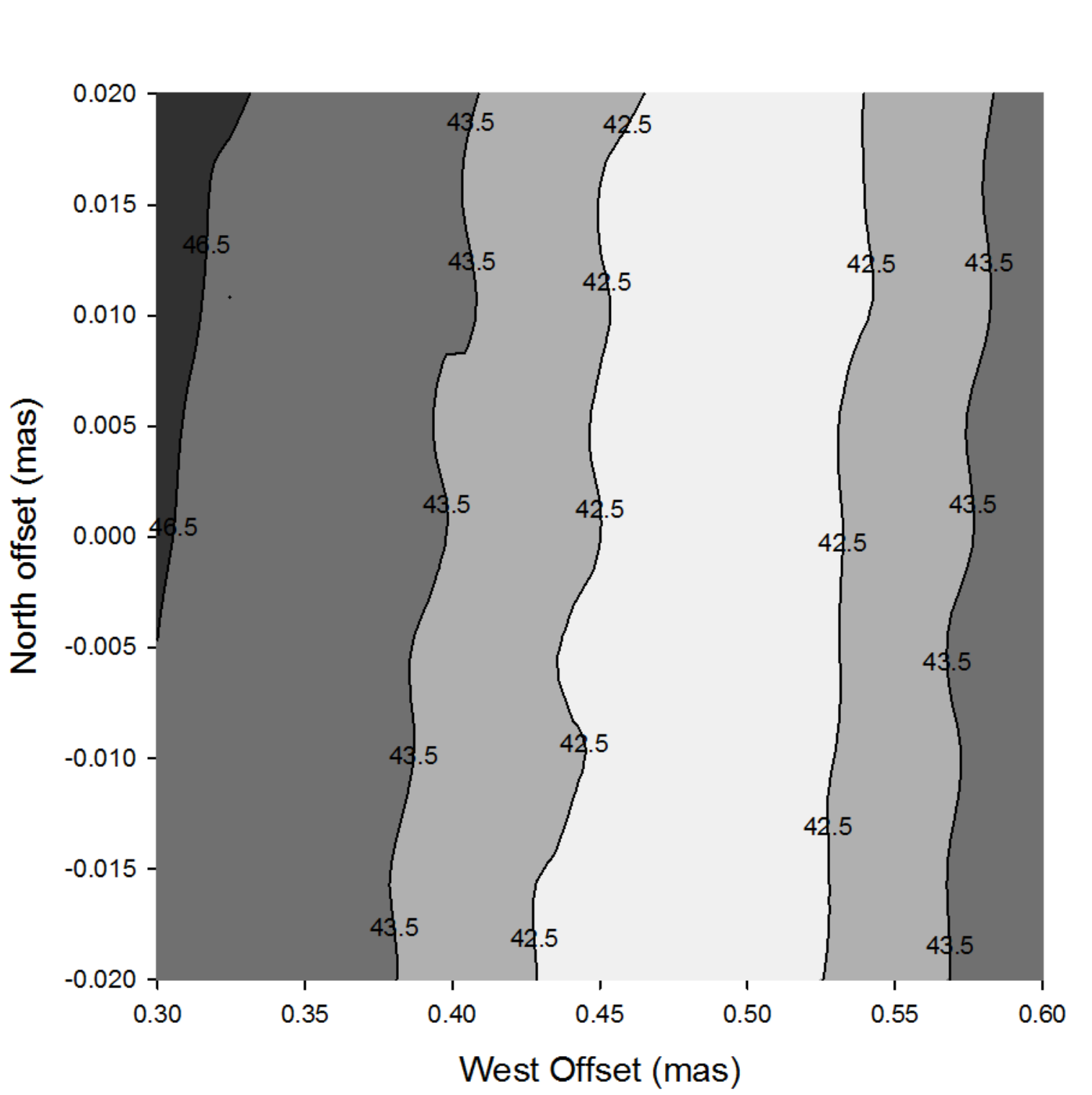}}
\caption{Calculation of $\chi^{2}(\Delta W, \Delta N)$ using the BBH model. 
There is a valley of possible offsets, but the size of the offset 
must be the same as the radius of the BBH system. This is the case if the 
offsets are $\Delta W_{1} \approx +0.495$ $mas$ and $\Delta N_{1} \approx 0.005$ $mas$. 
The corresponding radius of the BBH system is $R_{bin} \approx 495$ $\mu as$. }
\label{fig:Chi2_Off_YX_3C279_C5_BBH_1_LE}
\end{figure}

We calculated $\chi^{2}(\Delta W, \Delta N)$ starting with the parameters of 
solution Sol 1 found in the previous section. 
The result is shown in Figure \ref{fig:Chi2_Off_YX_3C279_C5_BBH_1_LE}.

We see from Figure \ref{fig:Chi2_Off_YX_3C279_C5_BBH_1_LE} that non-zero offsets are possible. 
However, all points with the smallest $\chi^{2}(\Delta W, \Delta N)$ are not possible. Indeed, 
for a point with the smallest $\chi^{2}$, the size offset must be equal to the 
radius of the BBH system calculated at this point. 
This is the case if the offsets are $\Delta W_{1} \approx +0.495$ $mas$ and $\Delta N_{1} \approx +0.005$ $mas$.

The radius of the BBH system at this point is $R_{bin} \approx 495$ $\mu as$ 
and the offset size is $\approx 495$ $\mu as$, i.e. the offset and the 
radius of the BBH system are the same at this point. \\

Comparison of the result found in Section \ref{sec:BBH_Offset1_3C279} and of 
Figure \ref{fig:Chi2_Off_YX_3C279_C5_BBH_1_LE} and Figure \ref{fig:Chi2_Off_YX_3C279_C5_BBH_1} 
show that the offset determined using large error bars is the same as the offset calculated with the 
small error bars.\\

Therefore we conclude that the VLBI component C5 is not ejected from the VLBI core but from 
the second black hole of the BBH system.

\subsection{Determining a possible new offset correction}

From this point onward, the original coordinates of the VLBI component C5 are corrected 
for the offsets $\Delta W_{1}$ and $\Delta N_{1}$ found in the previous section.

In this section, we assumed $V_{a} = 0.1$ c.

As in Section \ref{sec:Mass_ratio_case_II}, we preliminarily determined  
the parameters $T_{p}/T_{b}$, $M_{1}/ M_{2}$ and $i_{o}$.

\begin{figure}[ht]
\centerline{
\includegraphics[scale=0.5, width=8cm,height=8cm]{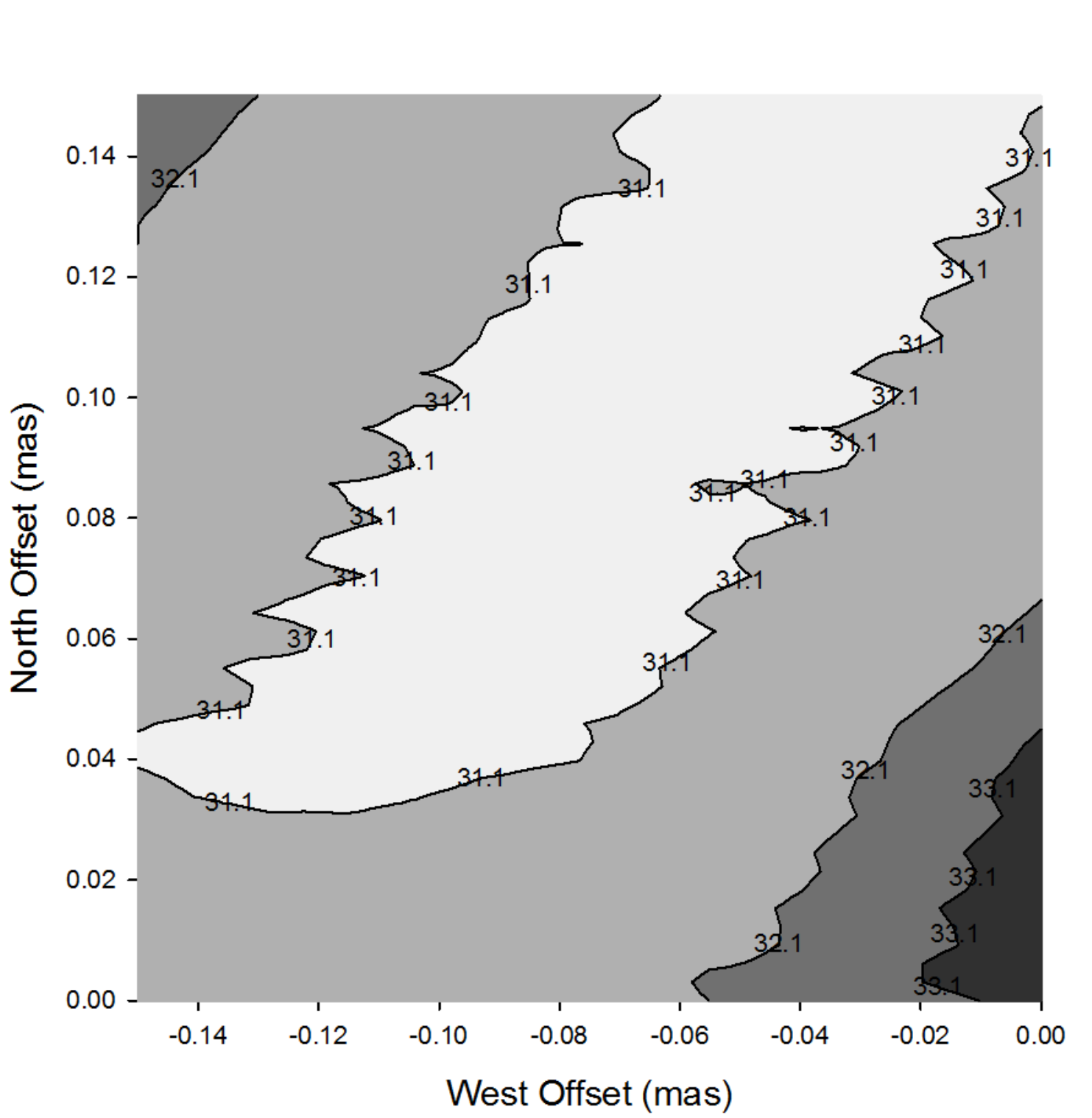}}
\caption{Calculation of $\chi^{2}(\Delta W, \Delta N)$ using the BBH model. 
There is a valley of possible offsets, but the size of the offset 
must be the same as the radius of the BBH system. This is the case if the 
offsets are $\Delta W_{2} \approx -0.085$ $mas$ and $\Delta N_{2} \approx 0.085$ $mas$. 
The corresponding radius of the BBH system is $R_{bin} \approx 419$ $\mu as$. }
\label{fig:Chi2_Off_YX_3C279_C5_BBH_2_LE}
\end{figure}

With $i_{o} \approx 5.9^{\circ}$, using the ratios $T_{p}/T_{b}$ free and  
$M_{1}/ M_{2} \approx 1.75$, we can verify whether  
there is an additional correction to the offset of the origin of the VLBI 
component. We calculated $\chi^{2}(\Delta W, \Delta N)$,
where $\Delta W$ and $\Delta N$ are offsets in the west and north directions. 
We assumed that the radius of the BBH system is let free to vary. The result is shown in Figure 
\ref{fig:Chi2_Off_YX_3C279_C5_BBH_2_LE}. We found 
that an additional correction is needed, namely $\Delta W_{2} \approx - 0.085$ $mas$ 
and $\Delta N_{2} \approx + 0.085$ $mas$.

At this point the total offset is $\approx 419$ $\mu as$ and the radius 
of the BBH system is $R_{bin} \approx 419$ $\mu as$.

Thus, using for the highest values of the error bars the values $\Delta_{min} = Beam /5$, 
we found that the final offset is $\Delta W_{t} \approx + 0.410$ $mas$, and
$\Delta N_{t} \approx + 0.090$ $mas$, and the radius of the BBH system is 
$R_{bin} \approx 0.419$ $mas$.

These values have to be compared with the values obtained assuming for the lowest error bars  
the value used $\Delta_{min} = Beam /15$, which are $\Delta W_{t} \approx + 0.405$ $mas$, and 
$\Delta N_{t} \approx + 0.110$ $mas$, and the radius of the BBH system is 
$R_{bin} \approx 0.420$ $mas$.

\subsection{Final solution}

The characteristics of the final solution determined assuming for the minimum value of the 
error bars the value $\Delta_{min} = Beam /5$ are 

\begin{itemize}
	\item $T_{p}/T_{b} \approx 140$, 
	\item $M_{1}/ M_{2} \approx 2.75$ ,
	\item $i_{o} \approx 11.0^{\circ}$, and
	\item $\chi^{2}_{min} \approx 28.9$.
\end{itemize}
Thus the reduced $\chi^{2}$ at the minimum is $\chi^{2}_{r} \approx 0.19$, 
indicating that the minimum error bars are too large.

\subsection{Conclusion}

We determined the characteristics of the solution, assuming for the minimum value of the 
error bars the value $\Delta_{min} = Beam /5$ suggested by \citet{LiHo:05}. We found that

\begin{enumerate}
	\item the characteristics of the solution are the same as those of the solution 
	determined assuming for the minimum value of the error bars the value $\Delta_{min} = Beam /15$ and
	\item the corresponding reduced $\chi^{2}$ is $\chi^{2}_{r} \approx 28.9 / 152 \approx 0.19$, 
	indicating that the minimum error bars are too large.
\end{enumerate}

The correct value for the minimum error bars at 15 GHz is $Beam /15 \leq \Delta_{min} \leq Beam /12$.

This value
\begin{enumerate}
	\item produces a reduced $\chi^{2}$, $\chi^{2}_{r} \approx 1$,
	\item the minimum value agrees with the value of the \textit{realistic error} obtained from the VLBI Service 
	for Geodesy and Astrometry \citep{ScBe:07} and
	\item the fit of VLBI coordinates of components of 3C 345 (work in progress) indicates that the adopted values 
	for the minimun values of the error bars, i.e., $Beam /15 \leq \Delta_{min} \leq Beam /12$, 
	are correct for frequencies between 8 GHz and 22 GHz.
\end{enumerate}

\end{document}